\documentclass[fleqn,usenatbib]{mnras}

\usepackage{newtxtext,newtxmath}
\usepackage[T1]{fontenc}
\usepackage{ae,aecompl}

\usepackage{etoolbox}
\makeatletter
\patchcmd\@combinedblfloats{\box\@outputbox}{\unvbox\@outputbox}{}{\errmessage{\noexpand patch failed}}
\makeatother

\usepackage{graphicx}	% Including figure files
\usepackage{amsmath}	% Advanced maths commands
\usepackage{amssymb}	% Extra maths symbols
\usepackage[figure,figure*]{hypcap}
\usepackage{float}

\newcommand{\comm}[1]{}
\defcitealias{2019MNRAS.490.1539R}{R19}

\title[The {\sl HERON} Project II]{The Halos and Environments of Nearby Galaxies ({\sl HERON}) II: The outer structure of edge-on galaxies}

\author[A. Mosenkov et al.]{
Aleksandr Mosenkov,$^{1,2}$\thanks{E-mail: aleksandr.mosenkov@tdtu.edu.vn}
R. Michael Rich,$^{3}$ 
Andreas Koch,$^{4}$
Noah Brosch,$^{5}$
\newauthor
David Thilker,$^{6}$
Javier Rom\'an,$^{7}$
Oliver M\"uller,$^{8}$
Anton Smirnov,$^{9,10}$
\newauthor
and Pavel Usachev$^{9,11}$
\\
$^{1}$Department for Management of Science and Technology Development, Ton Duc Thang University, Ho Chi Minh City, Vietnam\\
$^{2}$Faculty of Applied Sciences, Ton Duc Thang University, Ho Chi Minh City, Vietnam\\
$^{3}$Department of Physics \& Astronomy, Univ. of California Los Angeles, 430 Portola Plaza, Los Angeles, CA 90095-1547, USA\\
$^{4}$Zentrum f\"ur Astronomie der Universit\"at Heidelberg, Astronomisches Rechen-Institut, 69120 Heidelberg, Germany\\
$^{5}$Wise Observatory, Tel Aviv University, 69978 Tel Aviv, Israel\\
$^{6}$Department of Physics and Astronomy, Johns Hopkins University, 3400 N. Charles Street, Baltimore, MD 21218, USA\\
$^{7}$Instituto de Astrof\'isica de Andaluc\'ia (CSIC), Glorieta de la Astronom\'ia, 18008 Granada, Spain\\
$^{8}$Observatoire Astronomique de Strasbourg (ObAS), Universite de Strasbourg - CNRS, UMR 7550 Strasbourg, France\\
$^{9}$St. Petersburg State University, Universitetskij pr. 28, 198504 St. Petersburg, Stary Peterhof, Russia\\
$^{10}$Central (Pulkovo) Astronomical Observatory, Russian Academy of Sciences, Pulkovskoye Chaussee 65/1, 196140 St. Petersburg, Russia\\
$^{11}$Special Astrophysical Observatory, Russian Academy of Sciences, 369167 Nizhnij Arkhyz, Russia\\
}

\date{Accepted XXX. Received YYY; in original form ZZZ}

\pubyear{2019}

\begin{document}
\label{firstpage}
\pagerange{\pageref{firstpage}--\pageref{lastpage}}
\maketitle

% Abstract of the paper
\begin{abstract}
The {\sl HERON} project is aimed at studying halos and low surface brightness details near galaxies. In this second {\sl HERON} paper we consider in detail deep imaging (down to surface brightness of $\sim28$\,mag/arcsec$^2$ in the $r$ band) for 35 galaxies, viewed edge-on. We confirm a range of low surface brightness features previously described in the literature but also report new ones.  We classify the observed outer shapes of the galaxies into three main types (and their prototypes): disc/diamond-like (NGC\,891), oval (NGC\,4302), and boxy (NGC\,3628). We show that the shape of the outer disc in galaxies does not often follow the general 3D model of an exponential disc: 17 galaxies in our sample exhibit oval or even boxy isophotes at the periphery. Also, we show that the less flattened the outer disc is, the more oval or boxy its structure. Many galaxies in our sample have an asymmetric outer structure. We propose that the observed diversity of the galaxy outer shapes is defined by the merger history and its intensity: if no recent multiple minor or single major merging took place, the outer shape is diamond-like or discy. On the contrary, interacting galaxies show oval outer shapes, whereas recent merging appears to transform the outer shape to boxy.
\end{abstract}


%%%%%%%%%%%%%%%%%%%%%%%%%%%%%%%%%%%%%%%%%%%%%%%%%%

%%%%%%%%%%%%%%%%% BODY OF PAPER %%%%%%%%%%%%%%%%%%

\section{Introduction}
% Importance of studying edge-on galaxies
Edge-on galaxies are unique targets in the sense that they allow us to directly study their vertical structure without contamination from unknown inclinations \citep{1981A&A....95..105V,1994A&AS..103..475B,1996A&AS..117...19D,2002A&A...389..795B,2010MNRAS.401..559M,2014ApJ...787...24B} and to uniquely decompose them into different structural components such as thin and thick discs \citep{1983MNRAS.202.1025G,2008ApJ...673..864J,2011ApJ...738L..17C,2018A&A...610A...5C}, bulges \citep{ 2010ApJ...715L.176K,2012MNRAS.423..877G,2012A&AT...27..325S}, bars \citep{2015MNRAS.446.3749Y}, and stellar halos \citep{2005Ap.....48..221T,2018JKAS...51...73A}. Also different structural details can be well-distinguished in edge-on galaxies: disc warps \citep{1976A&A....53..159S,1998A&A...337....9R,2016MNRAS.461.4233R}, orthogonal structures such as polar rings \citep{1990AJ....100.1489W,1997A&A...325..933R,2019MNRAS.483.1470R} and polar bulges \citep{2012MNRAS.423L..79C,2015AstL...41..748R}, disc truncations \citep{2000A&A...357L...1P,2002MNRAS.334..646K,2012ApJ...759...98C,2019MNRAS.483..664M}, disc flaring \citep{2011ApJ...741...28C,2014A&A...567A.106L}, X-shaped or boxy structure of a bar \citep{2006MNRAS.370..753B,2017MNRAS.471.3261S}. The disc flattening, which can be measured in the case of the edge-on view \citep{2015MNRAS.451.2376M}, is related to the ratio of dark-matter halo to the total galaxy mass \citep{2002AstL...28..527Z,2010MNRAS.401..559M} and the minor-merger history \citep{2008MNRAS.389.1041R}. In particular, stellar discs can be effectively heated up in the vertical direction by accretion of small galaxies. Recently, \citet{2019A&A...632A..13S} studied the radio halo of the edge-on spiral galaxy NGC\,4013, along with radio scaleheights and the linear polarization. Other features, such as tidal streams and tails \citep{1998ApJ...504L..23S, 2009ApJ...692..955M,2011A&A...536A..66M}, galactic fountains \citep{1980ApJ...236..577B,1997ApJ...485..159B,2012ARA&A..50..491P}, and extraplanar gas \citep{1997ApJ...490..247L,2008A&ARv..15..189S,2019A&A...631A..50M} and dust structures \citep{1999AJ....117.2077H,2018ApJS..239...21S} can be studied as well. In summary, edge-on galaxies 
come in a variety of flavours \citep{2006A&A...445..765K} and this wealth of information, which can be obtained from an analysis of the vertical structure of edge-on galaxies, is extremely important for probing existing models of galaxy formation and evolution within the dominant $\Lambda$CDM paradigm.

% Model of discs. Shape of the outermost isophotes in galaxies.
Usually, the surface brightness (SB) distribution in galactic discs is well-represented by an exponential law in the radial direction \citep{1940BHarO.914....9P,1970ApJ...160..811F} and by an exponential \citep{1989ApJ...337..163W} or $\mathrm{sech}^2$-profile, as a physically motivated isothermal case \citep{1942ApJ....95..329S,1950MNRAS.110..305C,1981A&A....95..105V,1981A&A....95..116V,1982A&A...110...61V}, in the vertical direction (see also the review by \citealt{2011ARA&A..49..301V}). However, a general law to describe the vertical distribution is also used \citep{1988A&A...192..117V,1994A&AS..103..475B,1997A&A...327..966D}. Starting with \citet{2011ApJ...741...28C}, \citet{ 2012ApJ...759...98C,2018A&A...610A...5C} argue that these simplistic functions do not describe well a thin+thick disc structure because the sum of two isothermal sheets in hydrostatic equilibrium does not result into the sum of two $\mathrm{sech}^2$-profiles.
An edge-on disc in all cases would be characterised by diamond-like isophotes (we re-evaluate this in Sect.~\ref{sec:correlations}). However, real observations of edge-on galaxies (see, for example, the EGIS catalogue by \citealt{2014ApJ...787...24B}\footnote{\url{http://users.apo.nmsu.edu/~dmbiz/EGIS/}}) reveal that the outer shape of their discs is much more complicated and often does not follow a general perception that edge-on galaxy discs always have diamond-like isophotes with sharp tips at the periphery (see Sect.~\ref{sec:examples} and Fig.~\ref{fig:Example_galaxies} where we compare three different shapes of galaxies).  

% Boxy/penut-shape structures. No works on the outer shape of edge-on galaxies
The shape of galaxy isophotes is a valuable source of information about the structural composition of a galaxy which may point to some physical processes responsible for the observed structure. For example, giant elliptical galaxies have boxy-shaped isophotes which are explained by anisotropic random motion of their stars, whereas less luminous, mid-sized rotationally supported ellipticals show discy isophotes \citep[see e.g. ][]{1988A&A...193L...7B}. Boxy/peanut-shaped isophotes, which are observed in the central region of an edge-on galaxy, give us a clue that this galaxy harbours a bar which photometrically manifests itself in this way \citep[see e.g.][]{1990A&A...233...82C,2000A&A...362..435L}. \citet{2004A&A...417..527L} introduced the term ``thick boxy bulges'', which are boxy in shape and large in respect to the galaxy diameter. They proposed that this particular structure of large boxy bulges can be a result of accreted material of merging satellites or can be barlenses \citep[see e.g.][]{2017A&A...598A..10L,2018A&A...618A..34L}.
Similarly, the existence of galaxy stellar halos is naturally explained as a by-product of hierarchical galaxy formation by the accretion and disruption of infalling satellites \citep[see ][and references therein]{2018JKAS...51...73A}.

% HERON project and aims of this work
To our knowledge, the outer disc shape in edge-on galaxies has not up to now been studied in detail, though the observations show that the shape of the outermost isophotes in galaxies can be very different. In this study we address this issue using deep observations from the Halos and Environments of Nearby Galaxies ({\sl HERON}) survey \citep[][hereafter \citetalias{2019MNRAS.490.1539R}]{2019MNRAS.490.1539R}. The survey employs two dedicated 0.7-m telescopes with prime focus cameras dedicated to low surface brightness imaging. A description of the instrumentation is given in \citetalias{2019MNRAS.490.1539R}.

% Outline
This paper is structured as follows. In Sect.~\ref{sec:sample} we describe our sample of edge-on galaxies. In Sect.~\ref{sec:shape} we present our method for measuring the outer galaxy shape and apply it to the sample. We present our results in Sect.~\ref{sec:results}. In Sect.~\ref{sec:examples} we consider three typical galaxies of different outer shape and discuss our results in Sect.~\ref{sec:discussion}. We summarise our results in Sect.~\ref{sec:conlusions}.

\section{The sample of edge-on galaxies}
\label{sec:sample}

In the framework of the {\sl HERON} project we aim to study halos of galaxies in the Local Universe down to SB $\mu_r\sim28-30$~mag/arcsec$^2$, collected at the dedicated Jeanne Rich 0.7-m telescope at Lockwood Valley (see \citetalias{2019MNRAS.490.1539R}). It is a f/3.2 telescope with a prime focus imager behind a Ross doublet corrector which uses a luminance filter from 400 to 700~nm with an actual transmission equivalent very approximately to the full Sloan Digital Sky Survey \citep[SDSS,][]{2014ApJS..211...17A} $g$ and $r$ passbands.

At the moment, our sample consists of 119 galaxies and spans a substantial range of morphologies, including giant and dwarf elliptical galaxies, as well as spirals and irregular galaxies. From this sample we selected 35 galaxies which are oriented edge-on or close to it (most of these galaxies are labelled as edge-on in \citealt{2006A&A...445..765K} and \citealt{2014ApJ...787...24B}). To select these galaxies, we retrieved rough estimates of their inclinations from the HyperLeda database (the \textit{incl} parameter) and rejected galaxies with $incl<80\degr$. This yielded 43 objects. These galaxies were revisited by eye and 14 galaxies were rejected as obviously having a moderate inclination (in most cases the spiral arms are well seen, as, for example, in NGC\,3623, NGC\,4096 or NGC\,7052). However, we also revisited the remaining galaxies of the {\sl HERON} sample, and selected six more galaxies which look edge-on and have been extensively studied in the literature: NGC\,1055, NGC\,3034, NGC\,3556, NGC\,3628, NGC\,4244, and NGC\,4594 (for references, see Sect.~\ref{sec:description}). The 35 selected galaxies are listed in Table~\ref{tab:table1}, along with their morphology, visually determined envelope shape (`discy'/`diamond-like', `oval', `round', and `boxy'), and inclination taken from the literature.

The data reduction and preparation was described in detail in \citetalias{2019MNRAS.490.1539R}. Here we only note that from all images we have carefully subtracted the sky background and masked out all objects, which do not belong to the target galaxy. The calibration was done to the SDSS $r$ passband. Also, for each frame we measured a core of the point spread function (PSF) and combined it with the extended PSF (wings) from \citetalias{2019MNRAS.490.1539R}, up to the radius $R_\mathrm{PSF}\approx17.8$~arcmin (see fig.~10 in \citetalias{2019MNRAS.490.1539R}). Here we assume that the shape of the inner PSF is mainly governed by the seeing of the atmosphere and its outer wings (extended PSF) are shaped by the instrumental properties \citep[see e.g.][]{2009PASP..121.1267S}. Therefore, we combine the extended {\sl HERON} PSF from \citetalias{2019MNRAS.490.1539R} with a PSF built using the unsaturated PSFs (stars detected by SExtractor, \citealt{1996A&AS..117..393B}, the profiles of which were then averaged) in the vicinity of the target galaxy. This allows us to build the empirical inner PSF up to a radius of 7 arcsec. To merge the inner and outer PSFs, we normalise them in the overlapping region within 5 arcsec. By so doing, we are able to generate individual extended PSFs for each galaxy in our sample. All galaxies in our sample have the radii at the surface brightness level 28\,mag/arcsec$^2$ $R_{28}<1.5\,R_\mathrm{PSF}$, that satisfies the recommendation by \citet{2014A&A...567A..97S} to work with a PSF that is at least 1.5 times larger than the galaxy radius. The average PSF FWHM for our sample galaxies is $4.5\pm2.0$\,arcsec and the pixel scale in most cases is 0.83~arcsec/pix. 

The inclination values were estimated using different methods such as radiative transfer modelling \citep[e.g. ][]{2018A&A...616A.120M} or axisymmetric dynamical modelling \citep{2013MNRAS.432.1709C}. In some cases, when no estimate of the inclination angle was found, we estimated it using the position of the galaxy dust lane (or a stellar ring, if present) with respect to its geometric centre \citep[see ][]{2015MNRAS.451.2376M}. The provided inclinations should be reliable, with an error of several degrees at most (see the respecting uncertainties in the listed papers in Table~\ref{tab:table1}). The average inclination angle for our sample is $i=85.5\pm3.4\degr$. In fact, not all galaxies in our sample are viewed exactly edge-on. The least inclined galaxy in our sample is NGC\,4638 ($i=78\degr$).  We included such galaxies because the outer shape of highly inclined galaxies is much less affected by inclination than if we consider the inner structure. We note that the ratio of the scaleheights for the thick and thin disc is $4.6\pm1.6$ for a sample of 141 galaxies at 3.6~$\mu$m from \citet{2018A&A...610A...5C}.

Our sample includes lenticular galaxies as well as late-type spirals. Out of 35 galaxies, 18 have apparent boxy/peanut-shaped (B/PS) bulges (many have been studied by \citealt{2016MNRAS.459.1276C}, \citealt{2017MNRAS.471.3261S} and \citealt{2017A&A...598A..10L}), which are in fact bars viewed side-on. 15 galaxies harbour pseudo-bulges \citep[see for a review][]{2013seg..book....1K}.

The sample galaxies are distributed at an average distance of $\langle D \rangle=21.7\pm13.5$~Mpc (see table~B1 from \citetalias{2019MNRAS.490.1539R}) with average luminosities  $\langle M_V \rangle=-20.9\pm1.0$. Based on table~B2 from \citetalias{2019MNRAS.490.1539R}, the average depth of our images at the $3\sigma$ level within a box of $10 \times 10\,\mathrm{arcsec}^2$ is $27.8\pm0.5$~mag/arcsec$^2$.

Table~\ref{tab:table1} summarises some other properties of our sample galaxies, which are described in detail in the following sections.

\begin{table*}
	\centering
	\caption{The sub-sample of 35 edge-on galaxies selected out of the {\sl HERON} sample.}
	\label{tab:table1}
	\begin{tabular}{cccccccccccccc} % four columns, alignment for each
		\hline
		Name & Type & Shape & Outer   & $i$ & Ref. & $C_0$ & $b/a$ & Bulge & Ref. & B/PS? & Env. & $\rho$ & LSBF      \\
              &     &       &structure&(deg)&      &       &       & type  &      &       &       &  (Mpc$^{-3}$) &  \\
        (1)   & (2) &  (3)  &    (4)  &  (5)&  (6) &   (7) &  (8)  & (9)   &  (10)&  (11) &  (12) &   (13) & (14)       \\     
		\hline
NGC\,128 & S0 & oval/boxy & disc & 90 & 1 & -0.1 & 0.42 & P & 2 & yes & IntP,G & 0.17 &  +\\ 
NGC\,509 & S0 & oval/boxy & disc & 90 & 3 & 0.3 & 0.57 & P & 4 & yes & G & 0.50 & - \\ 
NGC\,518 & Sa & boxy/oval & disc & 83 & 1 & 0.6 & 0.35 & P & 4 & yes & G & 0.24 & +\\ 
NGC\,530 & SB0+ & oval & disc & 85 & 1 & 0.3 & 0.29 & P? & --- & yes & G & 0.61 & -\\ 
NGC\,891 & SA(s)b & diamond & disc & 89 & 5 & -0.8 & 0.25 & P & 6 & yes & G & 1.26 & -\\ 
NGC\,1055 & SBb & boxy & bulge & 86 & 7 & 0.8 & 0.59 & C & 7 & yes & G & 0.09 & +\\ 
NGC\,2481 & S0/a & oval & disc & 81 & 3 & 0.2 & 0.73 & P? & --- & no & IntP,G & 0.09 & +\\ 
NGC\,2549 & SA(r)0 & oval & disc & 89 & 3 & -0.1 & 0.31 & P? & 8 & yes & G & 0.06 & -\\ 
NGC\,2683 & SA(rs)b & diamond & disc & 83 & 9 & -1.0 & 0.37 & P? & --- & yes & I & 0.00 & -\\ 
NGC\,3034 & I0 & diamond & disc & 82 & 10 & -0.8 & 0.49 & NB & --- & no & IntP,G & 83.8 & -\\  %83.79
NGC\,3079 & SB(s)c & boxy/oval & disc & 82 & 11 & 0.5 & 0.18 & P & 12 & yes & IntP,G & 0.09 & -\\ 
NGC\,3115 & S0- & oval & halo & 86 & 13 & -0.0 & 0.54 & C & 14 & ? & G & 1279 & -\\ % 1278.68 
NGC\,3556 & SB(s)cd & oval & disc & 82 & 15 & -0.2 & 0.34 & P & 16 & ? & I & 0.05 & -\\ 
NGC\,3628 & SAb & boxy & disc & 88 & 17 & 5.5 & 0.29 & P? & --- & yes & G & 2.45 & +\\ 
NGC\,4206 & SA(s)bc & oval & disc & 80 & 1 & -0.2 & 0.17 & P? & --- & ? & G & 0.30 & -\\ 
NGC\,4216 & SAB(s)b & boxy & disc & 85 & 18 & 1.0 & 0.25 & P & 19 & ? & G & 0.24 & +\\ 
NGC\,4222 & Sc & disky & disc & 90 & 1 & -0.4 & 0.19 & P/NB & --- & ? & G & 0.19 & -\\ 
NGC\,4244 & SA(s)cd & disky & disc & 88 & 15 & -0.3 & 0.13 & NB & 20 & ? & G? & 0.01 & -\\ 
NGC\,4302 & Sc & oval & disc & 90 & 21 & -0.2 & 0.24 & P? & --- & yes & IntP,G & 5.87 & -\\ 
NGC\,4469 & SB(s)0/a & boxy/oval & disc & 88 & 1 & 0.4 & 0.31 & P & --- & yes & G & 1.24 & -\\ 
NGC\,4517 & SA(s)cd & oval/discy & disc & 86 & 1 & 0.1 & 0.17 & P/NB & --- & ? & IntP,G & 0.21 & -\\ 
NGC\,4550 & SB0 & diamond & disc & 81 & 3 & -0.7 & 0.29 & P & 22 & no & G & 0.24 & -\\ 
NGC\,4565 & SA(s)b & diamond & disc & 88 & 7 & -0.5 & 0.15 & P & 23 & yes & G & 1.45 & -\\ 
NGC\,4594 & SA(s)a & round & halo & 84 & 24 & 0.2 & 0.94 & C & 14 & yes & G & 0.19 & +\\ 
NGC\,4631 & SB(s)d & oval/boxy & disc & 86 & 25 & 0.4 & 0.23 & P & --- & ? & IntP,G & 0.03 & +\\ 
NGC\,4638 & S0- & boxy/oval & halo & 78 & 3 & 0.3 & 0.73 & C & 26 & ? & G & 88.42 & +\\ 
NGC\,4710 & SA(r)0+ & diamond & disc & 88 & 3 & -0.7 & 0.30 & P & 27 & yes & G & 0.68 & -\\ 
NGC\,4762 & SB(r)0 & boxy & disc & 90 & 3 & 0.2 & 0.29 & P & 28 & ? & IntP,G & 1.12 & +\\ 
NGC\,4866 & SB(rs)bc & oval/discy & disc & 82 & 1 & -0.3 & 0.22 & C & --- & yes & G & 0.37 & -\\ 
NGC\,5170 & SA(s)c & diamond/boxy & disc & 86 & 29 & -0.5 & 0.19 & P/NB & --- & ? & I & 0.14 & -\\ 
NGC\,5746 & SAB(rs)b & oval/discy & disc & 87 & 30 & -0.2 & 0.18 & P & 31 & yes & IntP,G & 0.64 & -\\ 
NGC\,5866 & S03 & oval/boxy & halo & 90 & 30 & 0.4 & 0.63 & C & 32 & ? & G & 0.53 & +\\ 
NGC\,5907 & SA(s)c & oval & disc & 85 & 33 & 0.2 & 0.13 & P & 33 & ? & G & 0.74 & +\\ 
NGC\,7332 & S0 & diamond/boxy & disc & 84 & 3 & -0.6 & 0.41 & P? & 34 & yes & IntP,G & 0.08 & -\\ 
UGC\,4872 & SBb & oval & disc & 82 & 1 & 0.0 & 0.19 & P? & --- & ? & G & 0.08 & -\\ 
		\hline
	\end{tabular}
   \parbox[t]{160mm}{ Columns: \\
   (1) Galaxy name, \\
   (2) morphological type from the NED database,  \\
   (3) envelope shape (note that for some galaxies it is slightly different than given in \citetalias{2019MNRAS.490.1539R}),  \\
   (4) outer structure for which we actually measure the shape, \\
   (5) inclination angle, \\
   (6) reference for inclination angle,  \\ 
   (7) ellipse shape from the S\'ersic fitting,  \\ 
   (8) apparent flattening of the outer structure from the S\'ersic fitting,  \\  
   (9) bulge type (`P' = pseudo-bulge, `C' = classical bulge, `NB' = bulgeless), \\
   (10) reference for bulge type, \\
   (11) presence of a boxy/peanut shape bulge (if it can be determined),\\
   (12) environment from the NED database: `IntP' stands for interacting pair, `G' -- group, `I' -- isolated,\\
   (13) volume number density calculated using the heliocentic velocities $cz$ and best distance moduli from the HyperLeda database (see text),\\
   (14) presence of LSB signatures of interaction (see Sect.\,\ref{sec:description}).\\
References:  (1) This work. (2) \citet{2011arXiv1102.0550B}. (3) \citet{2013MNRAS.432.1709C}. (4) \citet{2012AstBu..67..253S}. (5) \citet{1998A&A...331..894X}. (6) \citet{2013ApJ...773...45S}. (7) \citet{2014ApJ...795..136S}. (8) \citet{2017MNRAS.467.4540B}. (9) \citet{2016A&A...586A..98V}. (10) \citet{1963ApJ...137.1005L}. (11) \citet{1999AJ....118.2108V}. (12) \citet{2019ApJ...883..189S}. (13) \citet{1999MNRAS.303..495E}. (14) \citet{2004ARA&A..42..603K}. (15) \citet{1996PASJ...48..581K}. (16) \citet{2011ApJ...733L..47F}. (17) \citet{2015ApJ...815..133S}. (18) \citet{2013ApJ...767..133P}. (19) \citet{1999AJ....117..826S}. (20) \citet{2011ApJ...729...18C}. (21) \citet{2015ApJ...799...61Z}. (22) \citet{2002AJ....124..706A}. (23) \citet{2010ApJ...715L.176K}. (24) \citet{1996A&A...312..777E}. (25) \citet{2004A&A...414..475D}. (26) \citet{2012ApJS..198....2K}. (27) \citet{2016A&A...591A...7G}. (28) \citet{2013seg..book.....F}. (29) \citet{1987A&A...178...77B}. (30) \citet{MarjorieThesis}. (31) \citet{2012ApJ...754..140B}. (32) \citet{2004AN....325...92F}. (33) \citet{2018A&A...616A.120M}. (34) \citet{2004MNRAS.350...35F}}
\end{table*}

\section{Measuring the outer shape of edge-on galaxies}
\label{sec:shape}

\begin{figure}
\label{fig:inner_mask}
\centering
\includegraphics[width=\columnwidth]{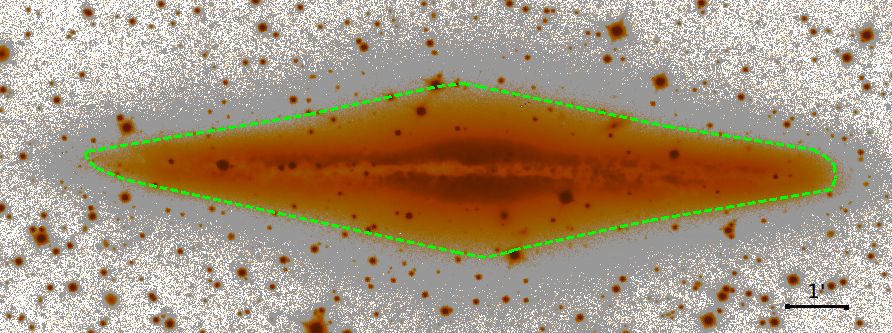}
\caption{An example of masking the inner region with $\mu_r<24$~mag/arcsec$^2$ for NGC\,891. The green dashed line shows the borders of this inner masking.}
\end{figure}

To measure the outer shape of the sample galaxies, we apply the following approach. For each galaxy image, we masked out the region inside of the isophote at 24~mag/arcsec$^2$ (all pixels with SB less than this value, see Fig.~\ref{fig:inner_mask}). With this mask (plus the mask of foreground Galaxy stars and overlapping galaxies) we fitted a S\'ersic function to the galaxy image. For this purpose we use the {\sc galfit} code \citep{2010AJ....139.2097P}, a S\'ersic component with generalised ellipses described by the free parameter $C_0$, which controls the discyness/boxyness of the isophotes. The radial pixel coordinate is defined by:

\begin {equation}
R(X,Y) = \left(\left|X-X_0\right|^{C_0+2}+\left| \frac{Y-Y_0}{b/a}\right|^{C_0+2}\right)^{\frac{1}{C_0+2}}\,,
\end{equation}
where $(X_0,Y_0)$ is the centre of the generalised ellipse, the axes of which are aligned with the coordinate axes.
This functionality appears very useful if one wants to analyse the average shape of the isophotes along with a S\'ersic parametrization. If $C_0=0$, the isophotes are described by pure ellipses. When $C_0<0$, the isophotes become more discy, and, vice versa, they become boxy if $C_0$ is positive. This approach has several advantages over a simple isophote fitting \citep[e.g. ][]{1987MNRAS.226..747J}. First, it allows one to take into account the PSF, which can severely affect the outer shape of the 2D galaxy profile. Second, we derive a single, average value of the discyness/boxyness parameter, which can be immediately interpreted. Third, the azimuthal average fitting often struggles to find optimal values for the isophote ellipticity and higher-order Fourier coefficients at low surface brightness (LSB) levels, when going from the inner to the outer galaxy region, so it actually does not fit the required parameters but makes them fixed to values found at a lower radius. 

We should point out that in our analysis we are interested in retrieving the parameters that describe the outer shape of edge-on galaxies, which are the $C_0$ parameter and the projected axes ratio $b/a$, where $b$ and $a$ are the minor and major axis, respectively. {\sc galfit} uses a Levenberg-Marquardt algorithm for $\chi^2$ minimization. In our fitting the S\'ersic function has six free parameters (the position is made fixed at the galaxy centre): position angle, axis ratio, total magnitude, effective (half-light) radius, S\'ersic index, and $C_0$ parameter. The first-guess values for these free parameters were taken from our single S\'ersic fitting of the whole galaxy from \citetalias{2019MNRAS.490.1539R}, whereas the initial value for $C_0$ is set to 0 (pure ellipse isophotes). The output parameters of the axis ratio and $C_0$ are listed in Table~\ref{tab:table1}.Unfortunately, the {\sc galfit} fitting method does not return a statistically correct error on the fitted parameters. However, in Sect.~\ref{sec:correlations} we apply this method to a sample of mock galaxy images and find an average uncertainty of this parameter to be 15-20\%.

Our method might potentially suffer from one relevant drawback: here we do not take into account the extended wings due to the PSF scattering from the inner part of the galaxy, which is masked out. As shown in many studies \citep{2008MNRAS.388.1521D,2014A&A...567A..97S,2015A&A...577A.106S,2016ApJ...823..123T,2017A&A...601A..86K}, the scattered light of an inner component can severely contaminate the light from an outer structure. Nevertheless, \citetalias{2019MNRAS.490.1539R} modelled the influence of the scattered PSF light on the measured radius $R_{28}$ for the whole {\sl HERON} sample and found only small overestimation (less than 10\% at most) of the true envelope radius. Also, we should emphasise that by our fitting we quantify a rather bright (though with SB$>24$~mag/arcsec$^2$) structure where regions with a higher signal-to-noise ratio have more weight than a very faint glow around the galaxy. In addition to that, in Sect.~\ref{sec:correlations} we do not find a significant influence of the extended PSF on the measured shape of the outer structure in mock galaxy images. We should, however, point out that if we had deeper observations and wanted to measure an outer structure beyond $\approx28-29$~mag/arcsec$^2$ (by masking out the inner region), where a stellar halo should dominate (see Sect.~\ref{sec:discussion}), the influence of the scattered light on the shape of this structure would not be neglected by any means \citep{2016ApJ...823..123T}.

Also, we note that the influence of the dust attenuation on the fitting results in the case of the outer structure is practically small. The dust is concentrated towards the galactic plane and its distribution in the vertical direction is very narrow (the average ratio of the dust disc scaleheight to the thick disc scaleheight for 7 nearby edge-on galaxies from \citealt{2018A&A...616A.120M} is $0.22\pm0.10$) and, thus, should not affect the observed disc outer structure at SB $\mu_r\geq24$\,mag/arcsec$^2$.

\section{Results}
\label{sec:results}

In this section we consider the results from applying our method to our sample of galaxies. First, we provide a brief description for each galaxy, with a useful information on its structural properties collected from the literature. In Sect.~\ref{sec:correlations} we discuss the general properties of the outer shape of edge-on galaxies. In the Appendix~A, Fig.~\ref{Images}, we provide deep images for the sample galaxies.

\subsection{Description of the individual galaxies}
\label{sec:description}

\textbf{NGC\,128}. This lenticular galaxy is famous for its prominent peanut-shaped bulge, which is thought to be associated with a buckling instability of a stellar bar \citep{1991Natur.352..411R}. \citet{2016MNRAS.459.1276C} discovered that there are two such nested structures, possibly associated with two stellar bars. Also, it has a strongly warped stellar disc. A colour image\footnote{Adam Block/Mount Lemmon SkyCenter/University of Arizona in collaboration with the Chart32 team for additional color data: \url{http://www.caelumobservatory.com/gallery/n125.shtml}.} of this galaxy clearly shows a bridge (both in the stellar, as well as dusty material, see also \citealt{1994cag..book.....S}) between this galaxy and a neighbour, NGC\,127, the spiral arms of which might be induced by the tidal interaction with NGC\,128 \citep{1979ApJ...233..539K}. We also note a faint bridge between NGC\,128 and NGC\,130. The outer shape of NGC\,128 is highly asymmetric: the north side of the disc is boxy and the disc is truncated at smaller radii, whereas the south side is oval/discy and more extended. The overall shape is close to oval. We suspect that because of the poor resolution more LSB details near the galaxy are hidden: the ``fluffy'' filamentary outer structure clearly points to that.

\textbf{NGC\,509}. This is a S0 galaxy with a lens \citep{2012ARep...56..578I}. It is a group member \citep{1993A&AS..100...47G} with no bright close neighbours. The envelope looks quite regular with oval isophotes at the periphery. We do not notice LSB features near this galaxy. However, we note that its outer structure is slightly twisted with respect to its inner structure, similar to NGC\,4469.

\textbf{NGC\,518}. This is a Sa galaxy with possible spiral arms and belongs to the loose NGC\,524 group \citep{2006MNRAS.369..360S,2012ARep...56..578I}. It has a faint inclined dust lane. The outer shape is boxy, with a LSB feature, possibly a stellar stream, on the west side of the disc in the South-West (SW) direction. This feature has not been mentioned in the literature. By its outer structure it resembles NGC\,3628, a galaxy with a very thick, disturbed disc. 

\textbf{NGC\,530}. This is a lenticular galaxy in the NGC\,545 group, which belongs to the galaxy cluster Abell~194. By visual inspection, it has a B/PS bulge and a ring (the inclination is $\approx85\degr$). We did not identify bright interacting companions near this galaxy. Its outer shape is pure elliptical without LSB features.

\textbf{NGC\,891}. This galaxy has been well-studied for decades because it is similar to our Galaxy in many regards \citep{1984A&A...140..470V,1992A&A...266...21G}. It belongs to the NGC\,1023 group, although we do not see signatures of interactions in our deep image. Nevertheless, \citet{2010ApJ...714L..12M} detected overdensities of RGB stars in the halo of NGC\,891, which define streams and loops. These features might appear due to one or more accretion events in the present or past. \citet{2007AJ....134.1019O} detected a cold gaseous halo, which might be created due to a galactic fountain and/or accretion of gas form intergalactic space. An extended dust component was also found around NGC\,891 \citep{2016A&A...586A...8B}, which is likely to be embedded in an atomic/molecular gas and heated by a thick stellar disc. The outer shape of this galaxy appears as a diamond, making this galaxy the best example of diamond-shape galaxies in our sample. We consider its profile in detail in Sect.~\ref{sec:examples}. 

\textbf{NGC\,1055}. This galaxy is located in the NGC\,1068 group \citep{1993A&AS...97..887B}. Its remarkable bulge was classified as a ``thick boxy bulge'' \citep{2004A&A...417..527L}; in \citet{2010AJ....140..962M} it is called a boxy-shaped inner halo  with multiple streams that seem to emerge from the galaxy's disc. The comparison with the deep image from \citet{2010AJ....140..962M} (see their figure~1) does not add new information about the LSB structure of this galaxy. The slightly asymmetric structure with the debris is evidence for a past/current multiple minor merging. We should notice here that such thick boxy bulges are quite rare: \citet{2004A&A...417..527L} found only 20 such objects out of about 1300 galaxies based on RC3 and NIR images.

\textbf{NGC\,2481}. This is a giant spiral flocculent galaxy comparable to the Andromeda galaxy by its diameter of $D\approx50$~kpc \citep{2001ApJ...559..243M}. It is in the process of colliding with another edge-on galaxy NGC\,2480: we can see an apparent bridge in our image. Also, we can see a distorted structure of NGC\,2480 and some other signatures of their close interaction: tidal streams, plums, a highly flared disc of NGC\,2480 and a U-warped stellar disc of NGC\,2481. The outer shape of NGC\,2481 is oval, but the orientation is far from pure edge-on.

\textbf{NGC\,2549}. This lenticular galaxy has a complex structure. Apart from a spheroid and a disc component, this object shows the signatures of two nested B/PS structures associated with two bars \citep{2016MNRAS.459.1276C,2016PASA...33...62C}. The outer shape is oval, although it is viewed almost exactly edge-on. In our image we do not observe LSB features related to this galaxy.

\textbf{NGC\,2683}. This is an isolated \citep{1979AN....300..181K}, nearly edge-on Sb galaxy with a B/PS bulge \citep{2009AJ....138.1082K}. The spirals look flocculent; the GALEX image shows two possible ring-like structures. The optical data do not show a disc warp or strong flaring of the disc, which is not the case for the H{\sc i} disc \citep{2016A&A...586A..98V}. The outer structure of the disc shows discy isophotes. However, the central region is very extended in the vertical direction -- the ratio of the minor to major axis for the isophote 27~mag/arcsec$^2$ is 0.62. This outstanding central feature makes this galaxy looking as a four-point star.
% Surprisingly, the \textit{WISE}~W3 image of this galaxy shows a similar extended structure which resembles the bipolar outflow (superwind) from M\,82 (NGC\,3034). The \textit{WISE}~W4 image is too shallow and does not show any extraplanar features. However, in the \textit{Spitzer} MIPS~24~$\mu$m band we can discern the same high-latitude structure.
This enigmatic structure will be considered in a future paper (Mosenkov et al. 2020 in prep.)

\textbf{NGC\,3034}. This is the famous starburst galaxy in the M\,81 group. It has spiral arms and a bar \citep{2005ApJ...628L..33M}. The starburst at the centre is assumed to be driven by interaction with M\,81 \citep[see e.g. ][and references therein]{2019arXiv191014672S}. Interestingly, the shape of the outer isophotes truly resembles a diamond but with the cut-off edges (tips). This shape can be partially explained by the very bright orthogonal filaments due to enormous starburst activity in the central region. 

\textbf{NGC\,3079}. This is a starburst/Seyfert galaxy with an extended and lopsided H{\sc i} disc and prominent gas streams due to strong interactions with its neighbours \citep{2015MNRAS.454.1404S}. Our observations do not reveal any LSB details near the galaxy, perhaps, because of the severe light contamination from bright foreground stars. The outer shape is boxy/oval, which can be explained by the strong interactions with its neighbours.

\textbf{NGC\,3115}. This is the closest S0 galaxy to the Milky Way (9.7 Mpc), with very little dust and gas \citep{2011ApJ...737...41L,2013MNRAS.428.2085L}. It represents an object with a complex stellar structure: a nuclear and outer disc \citep[see e.g.][]{1987AJ.....94.1519C,1991A&A...244L..25N}, rings and spiral arms \citep{2007A&A...464..507M,2016A&A...591A.143G}, a possible bar \citep{2016A&A...591A.143G} and a bright fast-rotating spheroid \citep[e.g.][]{2011ApJ...736L..26A}. Our data show that the outer shape has pure elliptical isophotes ($C_0\approx0$), with some apparent asymmetry at the faintest isolevels. \citet{2016A&A...591A.143G} discussed different possible scenarios to explain the observed photometric and kinematical properties of NGC\,3115 (see also references therein). They suggest that most of its mass in the inner part formed at redshifts $\textsc{z}>3$, while the outer flattened spheroid was then formed due to the long growth of the outer parts via minor mergers \citep{2011ApJ...736L..26A}. The ring, the spirals, and the (possible) bar point to secular evolution through dynamical processes. The slightly irregular outer shape, which we observe in our deep image, may confirm that the flattened spheroid formed via multiple minor mergers.

\textbf{NGC\,3556}. This isolated galaxy in the Ursa Major cluster is viewed not perfectly edge-on, so its spiral structure and a bar can be seen. This galaxy shows strong evidence for extraplanar radio emission \citep{2003ApJ...598..969W}, as well as large amounts of extraplanar diffuse X-ray emission, important for studying disc--halo interaction. Our deep observations do not reveal any interesting LSB details; the outer isophotes have a pure elliptical shape.  

\textbf{NGC\,3628}. This galaxy has an outstanding disc, puffed up in the vertical direction in the shape of a ``dog bone'', which makes this edge-on galaxy unique among all galaxies in our sample (we consider its profile in detail in Sect.~\ref{sec:examples}). The disc is warped as also seen by the broken dust lane dissecting the galaxy body. The X-shaped structure at the centre signifies that this is a barred galaxy \citep{2016MNRAS.459.1276C}. The spectacular tidal stream stretches for about 100~kpc which can be explained by the accretion of a dwarf galaxy \citep{2015ApJ...812L..10J} or tidal interaction between NGC\,3627 and NGC\,3628 \citep{1978AJ.....83..219R} (this galaxy belongs to the Leo Triplet, along with NGC\,3623 and NGC\,3627). The strong boxy/X-shape of the outermost isophotes in our observation suggests that this galaxy has been indeed significantly influenced by its local environment. We consider this galaxy in detail in Sect.~\ref{sec:examples} as a prototype for boxy outer structures in disc galaxies.

\textbf{NGC\,4206}. This galaxy is not seen purely edge-on. Our observations do not reveal anything interesting regarding LSB structures: the outer shape is oval, as we would expect for a galaxy with $i\approx80\degr$.

\textbf{NGC\,4216}. This galaxy is located in the Virgo Cluster \citep{1985AJ.....90.1681B} and exhibits multiple faint structures. We can clearly observe a tidal stream with an angular extent of $\sim12.5\arcmin$ in the sky (50.2~kpc), which seems to turn towards and join the second largest filament ejecting from the galaxy \citep{2013ApJ...767..133P}. Also, other LSB filaments, loops, and plums in the extended boxy envelope are well seen in our observation. As noted by \citet{2013ApJ...767..133P}, possible progenitors of these faint structures are dwarf satellites ``caught in the act of being destroyed'', which bombard the host galaxy. The strong boxy, or `emerald-cut' \citep{2012ApJ...750..121G} shape suggests that these interactions are indeed very intensive.

\textbf{NGC\,4222}. This galaxy is a member of the Virgo Cluster \citep{1985AJ.....90.1681B} and is a companion of NGC\,4216 (at the projected distance of 56~kpc, \citealt{2009AJ....138.1741C}), which does not show signs of interaction with NGC\,4222 \citep{2009AJ....138.1741C,2013ApJ...767..133P}. This galaxy has a discy outer shape and does not look to be disturbed. No LSB features near this galaxy are detected. 

\textbf{NGC\,4244}. This galaxy belongs to the nearby M\,94 Group. It represents a flat galaxy without a bulge. Its thick disc is subtle as compared to other edge-on galaxies in the Local Universe \citep{2011ApJ...729...18C}. \citet{2007IAUS..241..523S} reported on a detection of a stellar halo in NGC\,4244 using number counts of red giant branch (RGB) stars along the minor axis of the galaxy out to 10~kpc, with a limiting SB of $\mu_R \sim 31$~mag/arcsec$^2$. \citet{2005Ap.....48..221T} discovered a faint extended halo with a transition from the thick disc to the halo at $z=2.7$~kpc. Our imaging shows discy isophotes at the periphery with a U-shape disc warp and no LSB signs of interaction. 

\textbf{NGC\,4302}. This galaxy is a member of the Virgo Cluster. It is oriented exactly edge-on, which is seen from the prominent, extended dust lane going across the plane of the galaxy. The galaxy has a prominent component of diffuse ionised gas out to a height of 2~kpc \citep{1996ApJ...462..712R}. The H{\sc i} disc is truncated to well within the stellar disc, indicating that this galaxy is affected by ram-pressure \citep{1972ApJ...176....1G}: it is falling into the center of the Virgo Cluster on a highly radial orbit \citep{2007ApJ...659L.115C}. \citet{2015ApJ...799...61Z} detected a bridge between NGC\,4302 and its companion, NGC\,4298. The apparent outer shape of NGC\,4302 is oval and asymmetric which can be explained by the interaction with NGC\,4298, although we do not detect other signatures of a tidal disturbance. We consider this galaxy in Sect.~\ref{sec:examples}.  

\textbf{NGC\,4469}. This galaxy also belongs to the Virgo Cluster. It has a prominent X-shape central structure and filamentary dust lanes. Its outer
shape is slightly twisted with respect to its inner structure consisting of a B/PS bulge and a ring \citep[see also ][]{2015ApJS..216....9C}. The outer shape is oval/boxy. No other LSB features are seen.

\textbf{NGC\,4517}. \citet{2014ApJ...782....4K} showed that this galaxy is located in front of the Virgo cluster and exhibits the infall toward the cluster. The outer shape is oval. No obvious signatures of a tidal disturbance are visible.

\textbf{NGC\,4550}. It is a famous counter-rotating S0 galaxy \citep{1992ApJ...394L...9R}. The outer isophotes are discy; no LSB traces of interactions are found.

\textbf{NGC\,4565}. This giant spiral galaxy is located in the Coma I Group \citep{1977ApJ...213..345G}. The galaxy has a pseudo-bulge within a B/PS bulge, which represents a bar viewed side-on \citep{2009ASPC..419..149B}. \citet{2019MNRAS.483..664M} studied NUV, optical and NIR observations of NGC\,4565 and found that its disc truncations are observed at the same spatial location up to a height of $z=3$~kpc at all wavelengths, which can be  associated with a star formation threshold and migration of stars to the outer galaxy regions. \citet{2019arXiv191005358G} studied an extremely deep image of NGC\,4565 and found asymmetric truncation of the disc and a fan-like feature which is proposed to be a tidal ribbon. From their analysis and simulations they suggested that the observed outer structure of NGC\,4565 can be explained by an interaction with a satellite galaxy, which must be accreted in the plane on a disc-like orbit. Also, IC\,3571 and NGC\,4562 might perturb the disc of NGC\,4565. Our image is not that deep, as of \citet{2019arXiv191005358G}, but in both images we do not discern long tidal streams, plums and other signatures, apart from those presented in \citet{2019arXiv191005358G}. This may signify that this galaxy does not experience a strong interaction. If we consider the outer isophotes, not fainter than 28~mag/arcsec$^2$ (the thick disc), than both the images show practically a similar diamond shape (see their figure~1). However, the deepest isophotes down to 30~mag/arcsec$^2$ are oval, which can be interpreted as a faint stellar halo.

\textbf{NGC\,4594}. The well-known Sombrero galaxy,  M\,104, shows a very complex structure \citep{2012MNRAS.423..877G}. The outer shape, at first glance, looks like a round elliptical galaxy. However, as noted by \citet{1997PASA...14...52M}, a faint loop at a projected distance of about 20$\arcmin$ to the SSW of the galaxy is found (see also their figure 6). Also, we confirm a faint extended detail, mentioned by \citet{1997PASA...14...52M} --  a diffuse faint structure to the NNE of the galaxy, diametrically opposed to the loop. We discuss in detail the outer shape of the Sombrero galaxy in Mosenkov et al. (in prep.). All these morphological characteristics point to a major merger in the past \citep{2012MNRAS.423..877G}. The Sombrero galaxy lies within a complex, filament-like cloud of galaxies that extends to the south of the Virgo Cluster \citep{1994yCat.7145....0T}. However, gravitational bond with other close galaxies is unclear (see different points of view on this question in \citealt{1993A&AS..100...47G,2000ApJ...543..178G}). 

\textbf{NGC\,4631}. \citet{2017ApJ...842..127T} studied the interacting galaxy system NGC\,4631 and NGC\,4656 and identified 11 dwarf galaxies and two tidal streams in its outer region based on the stellar density maps divided into resolved stellar populations. \citet{2005AJ....129.1331S} found somewhat thickened disc structures via HST/ACS observations. \citet{2015AJ....150..116M} discovered a giant stellar tidal stream based on the integrated surface light analysis. The observed faint features suggest that the disc of NGC\,4631 was perturbed by an interaction with multiple dwarf galaxies. Our observation shows an oval/boxy outer structure and a tidal stream in the SSE direction, which can again be explained by the tight environment of NGC\,4631.

\textbf{NGC\,4638}. This is another member of the Virgo Cluster. It has a tiny bulge and an edge-on disc embedded in a boxy spheroidal halo (the Sersic index $n=1.1$, \citealt{2012ApJS..198....2K}). It is located close to the elongated dwarf Sph galaxy NGC\,4637. The boxy halo, as proposed by \citet{2012ApJS..198....2K}, can be a heavily dynamically heated remnant of a disc. In our image we can clearly see two stellar streams: one in the west direction and the other in the NNE direction.

\textbf{NGC\,4710}. This galaxy is located in the Virgo cluster outskirts \citep{2007ApJ...655..144M}. The bulge of NGC\,4710 has a B/PS morphology with a pronounced X-shape, showing no indication of any additional spheroidally distributed bulge population \citep{2016A&A...591A...7G}. \citet{2016MNRAS.460L..89K} studied deep spectral observations of its thick disc from the SAO RAS 6-m telescope and found statistically significant differences in chemical abundances between the thin and thick discs with a flaring at the periphery of the disc. % They proposed that the galaxy has passed the cluster centre. The thick disc formed first, then the thin disc grew up due to fuelling by additional gas. Later, the ram pressure by hot intracluster gas near the cluster centre striped the outer part of the thin gaseous disc and quenched the star formation. The disc flaring would quickly occur beyond this radius because there will be no more gas and dynamically cold stars forming in the mid-plane.
Our observations show that this galaxy has a diamond-like outer shape with discy isophotes at all radii except for the central region where the B/PS bulge dominates. 

\textbf{NGC\,4762}. This is another member of the Virgo cluster. This lenticular galaxy has a very asymmetric, tidally distorted and warped disc. The structural composition of this galaxy is complex: a classical bulge, bar, lens, and outer ring \citep{2012ApJS..198....2K}. The outer disc structure is a result of an ongoing tidal encounter. Using galaxy scaling relations, \citet{2012ApJS..198....2K} show that NGC\,4762 is a link between earlier-type S0 galaxies and Sphs. Our observation shows that the outer shape is boxy which is also explained by harassment in action.

\textbf{NGC\,4866}. This spiral galaxy is not viewed exactly edge-on, so a ring structure is well seen. Despite the non edge-on orientation, it's quite discy in shape. \citet{2018A&A...614A.143M} noted an unclassifiable disc feature to the west of the galaxy, possibly formed by tidal interaction. However, we do not observed any LSB features near this galaxy.

\textbf{NGC\,5170}. This galaxy with a Hubble type of Sc (HyperLeda) is similar to the Milky Way by its morphology. It is relatively isolated and located behind but close to the Virgo cluster southern extension. \citet{2010MNRAS.403..429F} estimated the total number of globular clusters in this galaxy to be $600\pm100$,  much more than found in the Milky Way \citep{1996AJ....112.1487H}. They also detected an ultra compact dwarf, which is assumed to be physically associated with NGC\,5170. \citet{2009ApJ...697...79R} failed to detect any extraplanar X-ray emission. The outer shape shows discy isophotes. However, the U-warped tips of the disc are quite boxy. We also note a slight asymmetry of the outer envelope.

\textbf{NGC\,5746}. This is a massive, quiescent edge-on spiral of SB(r)bc type. \citet{2012ApJ...754..140B} studied profiles of this galaxy in the optical and NIR and detected no merger-built bulge. Instead the central ``boxy bulge'' represents a bar, seen end-on, and an inner secularly evolved pseudobulge. As in the case of NGC\,5170, \citet{2009ApJ...697...79R} failed to detect any diffuse X-ray emission outside the optical disc. The shape of the outer envelope is discy.

\textbf{NGC\,5866}. This galaxy belongs to the small NGC\,5866 group (along with NGC\,5879 and NGC\,5907). Its morphology is unclear: it is either lenticular or spiral \citep{2002AJ....123.3067B}. Its dust disc may contain a ring-like structure, although the shape of this structure is difficult to determine given the edge-on orientation of the galaxy \citep{2004A&A...416...41X}. The shape of the halo is oval/boxy. We suppose that its ``fluffy'' structure contains many faint filamentary structures, as there are some hints on that in our image.

\textbf{NGC\,5907}. This is a well-known Knife Edge Galaxy, a member of the NGC\,5866 group. It has a pronounced integral-shape H{\sc i} warp, although in the optical it looks quite small. \citet{1998ApJ...504L..23S} discovered a stellar tidal stream wrapping around the galaxy in the shape of a ring which is likely to be created due to a tidal disruption of a companion dwarf Sph galaxy. They discovered two interactions with companion dwarf galaxies. One of them is seen at the tip of the H{\sc i} warp and in the direction of the warp. Therefore, this warp might be excited by this dwarf galaxy. \citet{2008ApJ...689..184M} re-observed NGC\,5907 and found two complete loops, enveloping NGC\,5907, instead of one ring. However, recently \citet{2019ApJ...883L..32V} observed this galaxy with their Dragonfly Telephoto Array \citep{2014PASP..126...55A} down to $\mu_{g}=30.3$\,mag/arcsec$^2$ and re-considered the morphology of this tidal stream: it appeared to be a single curved stream with a total length of 220\,kpc, similar to the Sagittarius stream which wraps around the Milky Way. The Dragonfly image has been confirmed by new observations by \citep{2019A&A...632L..13M}. The structure of this galaxy looks simple, but in fact the presence of a thick disc or a halo in this galaxy is under question \citep{1994Natur.370..441S,2014A&A...567A..97S,2018A&A...616A.120M}. Our deep observation does reveal the stellar stream, although the deepness of our image is worse than that from the Dragonfly telescope, so we cannot trace in detail its morphology. No other LSB features are found. The outer shape is oval, but the galaxy is viewed not strictly edge-on ($i\approx85\degr$).

\textbf{NGC\,7332}. It is an edge-on peculiar lenticular galaxy with a bar. NGC\,7332 and NGC\,7339 form a dynamically isolated binary system \citep{1987MoIzN....T....K}. \citet{1996A&A...307..391P} found that this galaxy has two gaseous components with opposite angular momentum and have different inclination angles. The gas is distributed asymmetrically and displays non-circular motions indicating that it has not reached equilibrium. Also, \citet{2002ASPC..282..216F} found that this galaxy deviates from the fundamental plane, without being obviously interacting. Also, the stellar colours in this galaxy are uniform, with only a very small blueing towards the outer parts \citep{1994AJ....107..160F}. These observations support of a recent merger having occurred in NGC\,7332 within a relatively small time scale. The outer shape is contradictory: the overall shape looks diamond-like, but the disc tips are strongly boxy.

\textbf{UGC\,4872}. This galaxy does not show any obvious signatures of a tidal disturbance. Being not exactly edge-on, it has an oval outer shape. According to \citet{2012A&A...540A.106T}, this galaxy belongs to a group but it is relatively isolated from other group members.

\subsection{The outer shape of edge-on galaxies}
\label{sec:correlations}

%Pair galaxies:

%NGC\,4762:The galaxy's disc is asymmetric and warped, which could potentially be explained by NGC 4762 violently cannibalising a smaller galaxy in the past. The remains of this former companion may then have settled within NGC 4762's disc, redistributing the gas and stars and so changing the disc's morphology.

%Second, the outer disk of NGC 4452 is warped and thicker than the ``superthin'' edge-on, late-type galaxies seen in isolated environments \citep{2011ARA&A..49..301V}. Similarly, NGC 4762’s outer disk is thick, warped, and tidally distorted. Gravitational encounters may be at fault (NGC 4762 with NGC 4754; NGC 4452 with IC 3381). \citet{2012ApJS..198....2K} suggested that these are signs of environmental heating that helps to convert flat disks into less flat spheroidals. They have very flat inner disks but very thick outer disks, as expected from dynamical heating processes. The fat outer disk of NGC 4762 is still warped and irregular; this is one example among many of an ongoing tidal encounter; i.e., of harassment in action. But the outer disk is thicker, warped, and distorted symmetrically into an  structure that is a signature of tidal responses. 

%NGC\,4517 with NGC\,4517A
%NGC\,5746 has a pair with NGC\,5740

\begin{figure*}
\label{fig:C0_incl}
\centering
\includegraphics[height=7.2cm]{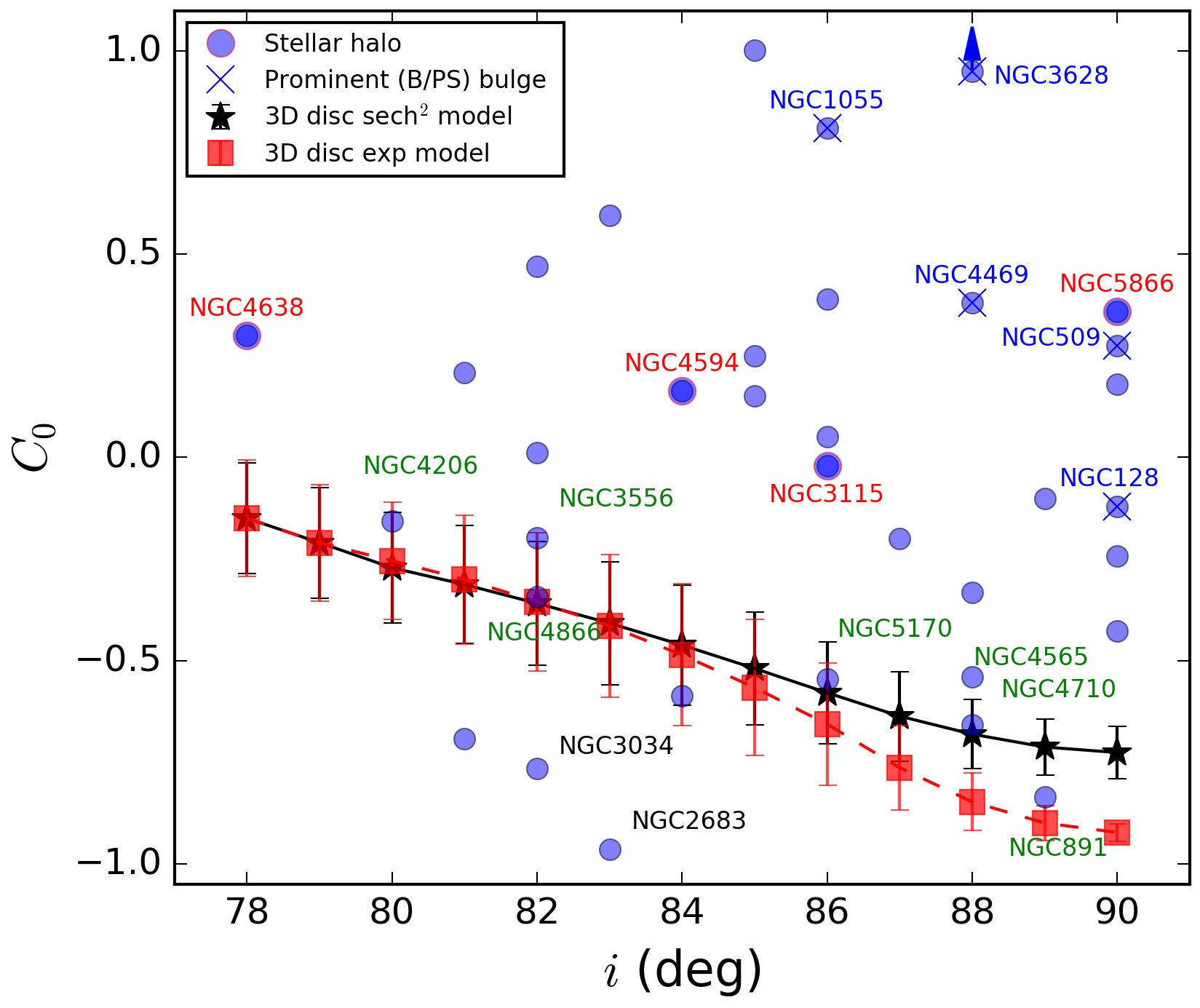}
\includegraphics[height=7.2cm]{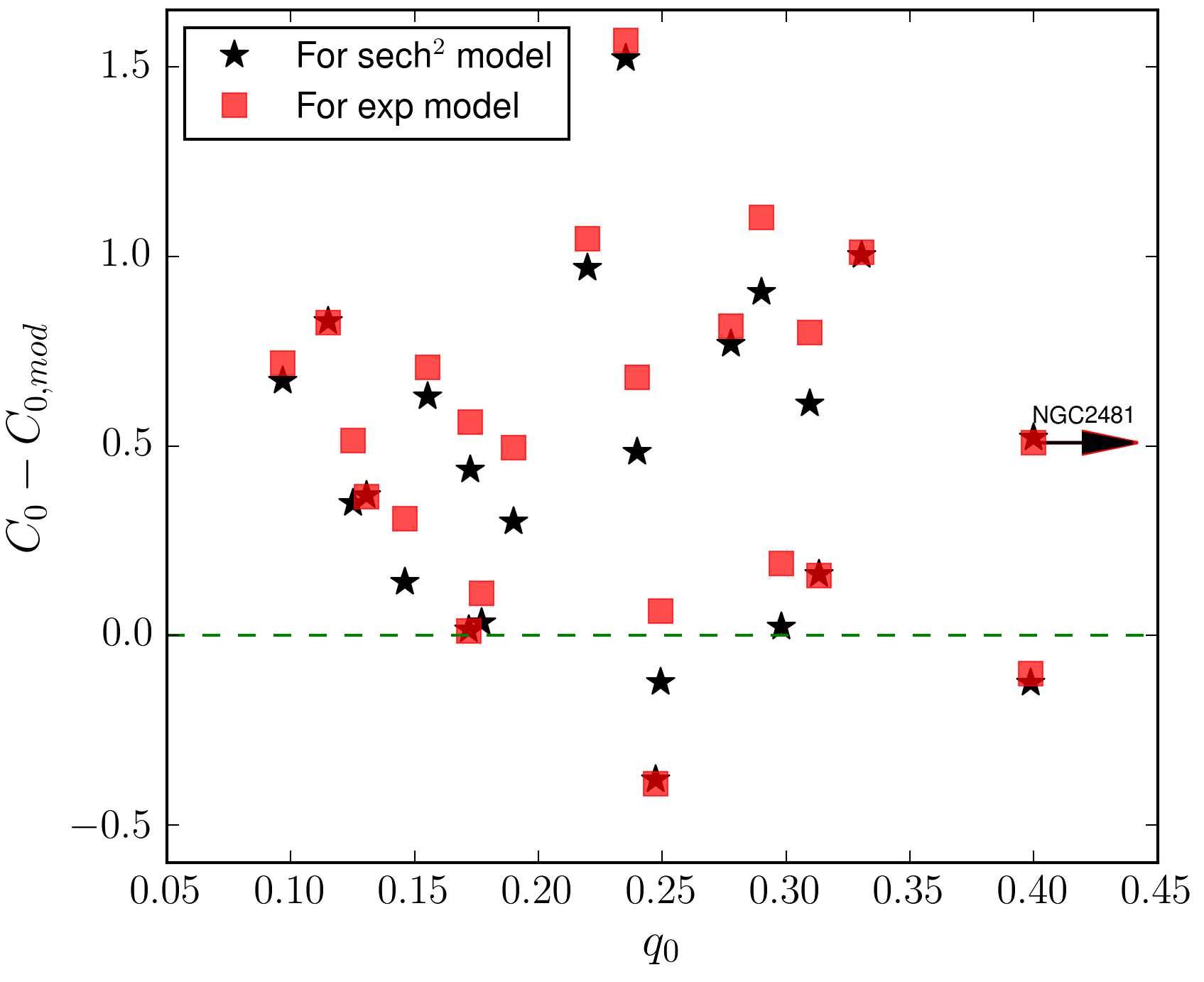}
\caption{\textit{Left plot:} Dependence of the discyness/boxyness parameter on galaxy inclination. The connected black stars and red squares represent 3D models with a sech$^2$ and exponential vertical distribution, respectively. The galaxies that are visually close to the model trends are marked by the green labels. Also, galaxies with a bright stellar halo (red labels), prominent B/PS bulges (blue labels) and orthogonal structures (black labels) are shown. \textit{Right plot:} Correlation between the outer disc intrinsic flattening (see text) and the difference between the observed and modelled $C_0$. The green line shows the perfect agreement between the model and observations.}
\end{figure*}

To ensure that our method for measuring the outer structure of galaxies is not affected by the PSF, we modelled mock galaxy images using 3D edge-on disc models, where the disc inclination is a free parameter, the radial distribution is described by an exponential law and the vertical distribution is described by a sech$^2$-law (isothermal disc) or exponential decline (both these laws are often used in the literature, see the review by \citealt{2011ARA&A..49..301V}). To model these images, we used the results of decompositions on a thin and thick disc for a sample of edge-on galaxies from the S$^4$G survey \citep{2010PASP..122.1397S,2015ApJS..219....4S}. For our modelling, we adopted fitted scalelengths, scaleheights and central surface brightnesses of the thin and thick discs for 142 edge-on galaxies in \citet{2015ApJS..219....4S}\footnote{We used Table 8 from \url{https://www.oulu.fi/astronomy/S4G\_PIPELINE4/MAIN/} to select galaxies with models which include thin and thick disc components.}. We used the {\sc imfit} code \citep{2015ApJ...799..226E}, which has a special function for modelling a 3D inclined disc with different profiles of the vertical distribution. To mimic real galaxies from our sample, we varied the inclinations of these mock galaxies from $78\degr$ to $90\degr$. No internal extinction was taken into account, as our primary goal here is to study the outer disc structure where the dust attenuation is significantly reduced \citep{2018A&A...616A.120M}. Also, we should note that the structural parameters of galaxies in the optical and NIR are different \citep[see e.g.][]{2016MNRAS.460.3458K}, but for our pedagogical purpose, where we study the shape of the outer structure for modelled galaxies irrelative of their structure, this is not important (broadly speaking, only co-existence of the thick and thin discs is relevant for our modelling). The produced mock images were scaled to 0.83\,arcsec/pix (a typical pixel scale for the {\sl HERON} survey) and convolved with an extended {\sl HERON} PSF. Also, we added Gaussian and Poisson noise to each image to mimic typical noise in our observations.

For measuring the shape of the outer structure in these mock galaxies we applied the same technique as for the sample galaxies in Sect.~\ref{sec:shape} (pixels with SB$>24$~mag/arcsec$^2$ were fitted with a S\'ersic function with a free parameter $C_0$). In the Appendix~\ref{Appendix:models} in Fig.~\ref{fig:compar_conv_unconv}, we show results of the fitting for the convolved and unconvolved mock galaxy images. As one can see, the disc apparent flattening naturally decreases with the inclination $i$. The exponential discs appear more flattened than the isothermal ones. For both the models, the flattening is slightly larger for the convolved images but this shift lies within the dispersion of the apparent disc flattening $q$. For the parameter $C_0$ we can also see no significant difference between the convolved and unconvolved models.

Here we show the results on measuring the galaxy outer shape for the whole sample. In Fig.~\ref{fig:C0_incl}, lefthand plot, we present the dependence of the  discyness/boxyness parameter $C_0$ on galaxy inclination. Also, we show the results for the mock galaxy images. As one can see, the exponential (in the vertical direction) discs look more discy at inclinations close to 90 than the isothermal discs (approximately -0.9 versus -0.7, see also Fig.~\ref{fig:compar_conv_unconv} and  fig.~5 in \citealt{2015ApJ...799..226E} which displays the difference between the isophote shapes for exponential and isothermal discs). However, this difference vanishes at lower inclinations: for $i\lesssim 85$ the $C_0$ parameter does not differ significantly for the exponential and sech$^2$ model.

It is interesting to compare how real galaxies follow the modelled ones. Only seven galaxies (marked by the green labels in Fig.~\ref{fig:C0_incl}, lefthand plot) out of 35 follow the model trends. We point out that four galaxies in our sample have bright haloes, the sizes of which are comparable to the discs (\citealt{2018JKAS...51...73A}, see also references in Sect.~\ref{sec:description}), therefore it is natural that they do not follow the discy models: their outer structures show oval or boxy isophotes ($C_0 \gtrsim 0$). The same applies to five galaxies with prominent B/PS or thick boxy bulges (labelled in Fig.~\ref{fig:C0_incl} by `X'), as noted in \cite{2017MNRAS.471.3261S}, \cite{2016MNRAS.459.1276C} and \cite{2004A&A...417..527L}, and galaxies with a central peculiar component (NGC\,2683 and M\,82). The vertical extent of these central components may contribute to the measured outer structure to some degree. If we exclude these 18 galaxies, the remaining 17 galaxies in our samples do not follow the general trends for the modelled thick discs. In almost all cases, except for NGC\,2683, NGC\,3034 and NGC\,4550, the $C_0$ parameter has a larger value than predicted ($C_0-C_\mathrm{0,mod}=0.47\pm0.50$ for the isothermal disc model versus $0.56\pm0.52$ for the exponential disc model). Edge-on discs look more oval or may even have a boxy outer shape: the average measured $C_0$ for the sample is $-0.06\pm0.47$ versus the predicted $C_0=-0.54\pm0.16$ for the isothermal disc and $C_0=-0.62\pm0.24$ for the exponential disc.

% delta(C0-C0_trend) Vs q
In the righthand plot of Fig.~\ref{fig:C0_incl}, we show the dependence of the difference between the observed and modelled $C_0$ (both for the exponential and isothermal discs) and the intrinsic disc flattening $q_0$, which was calculated using the formula for inclined oblate spheroids \citep{1958MeLuS.136....1H}: $q_0^2=(q^2-\cos^2i)/(1-\cos^2i)$, where $q=b/a$ and $i$ are listed in Table~\ref{tab:table1}. In this figure we do not include the above mentioned galaxies with bright halos and prominent central components (11 galaxies). As one can see, there is a trend for galaxies which lie above the line $C_0-C_\mathrm{0,mod}=0$: the less flattened the galaxy, the larger the difference between the observed and model discyness/boxyness. This means that less flattened outer discs look more oval than more flattened discs. Also, these galaxies often exhibit LSB features and their outer shape may look somewhat asymmetric. The galaxies, which are located close to the $C_0-C_\mathrm{0,mod}=0$ or slightly above or below it, are those which follow the model trends in the lefthand plot.

\section{Prototypes}
\label{sec:examples}

% The data: S4G, averaging, extended PSF
In this section we show most representative examples (prototypes) for each type of the outer shape: NGC\,891 (diamond-like/discy), NGC\,4302 (oval), and NGC\,3628 (boxy). We consider the dependence of the disc scaleheight and scalelength on galactocentric radius and height above the galactic plane, respectively. We naturally assume that the outer shape should be controlled by these dependencies. For example, if the disc scaleheight increases with radius, one would expect to see a flaring of the stellar disc, and, thus, the outer isophotes should look more puffed up at the periphery, than in the inner region.

\begin{figure*}
\label{fig:Example_galaxies}
\makebox[12.3cm][s]{\textbf{NGC\,891} \textbf{NGC\,4302} \textbf{NGC\,3628}} \par
\includegraphics[height=2cm]{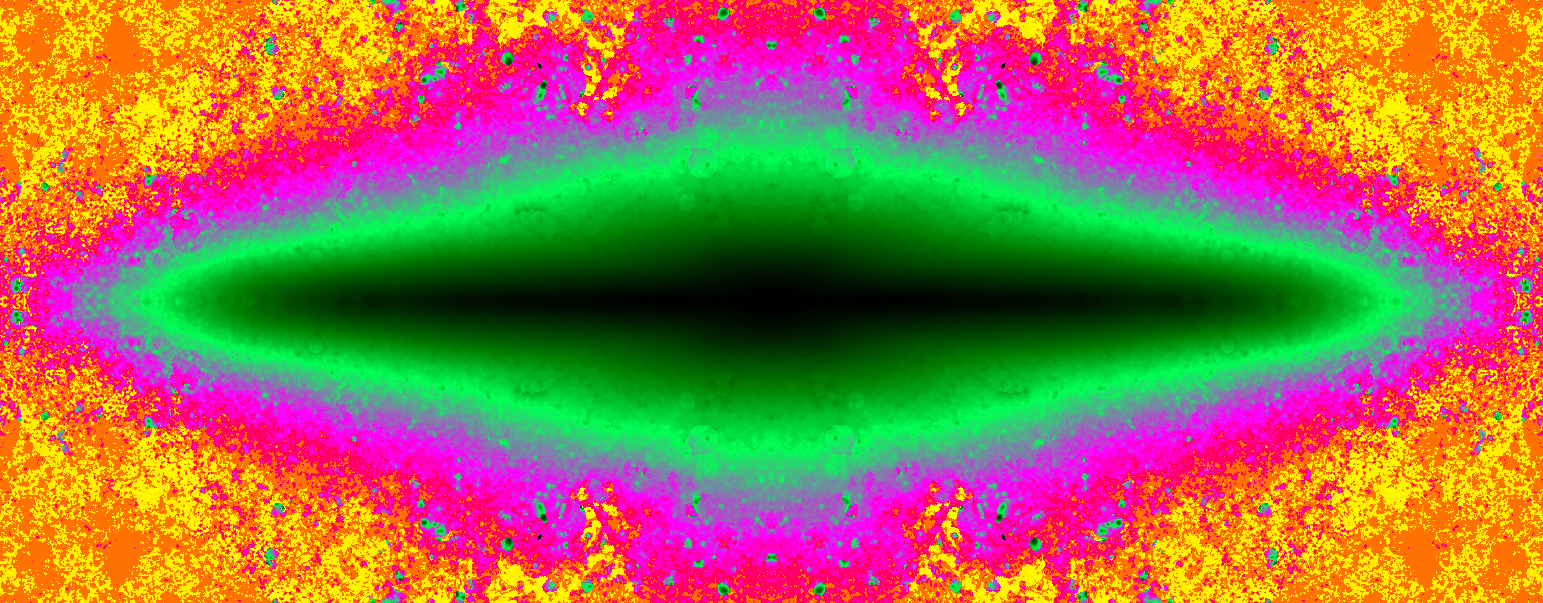}
\includegraphics[height=2cm]{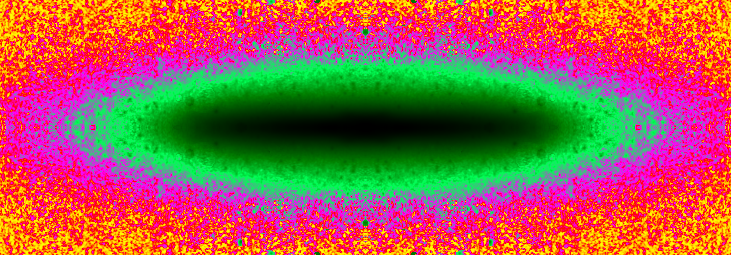}
\includegraphics[height=2cm]{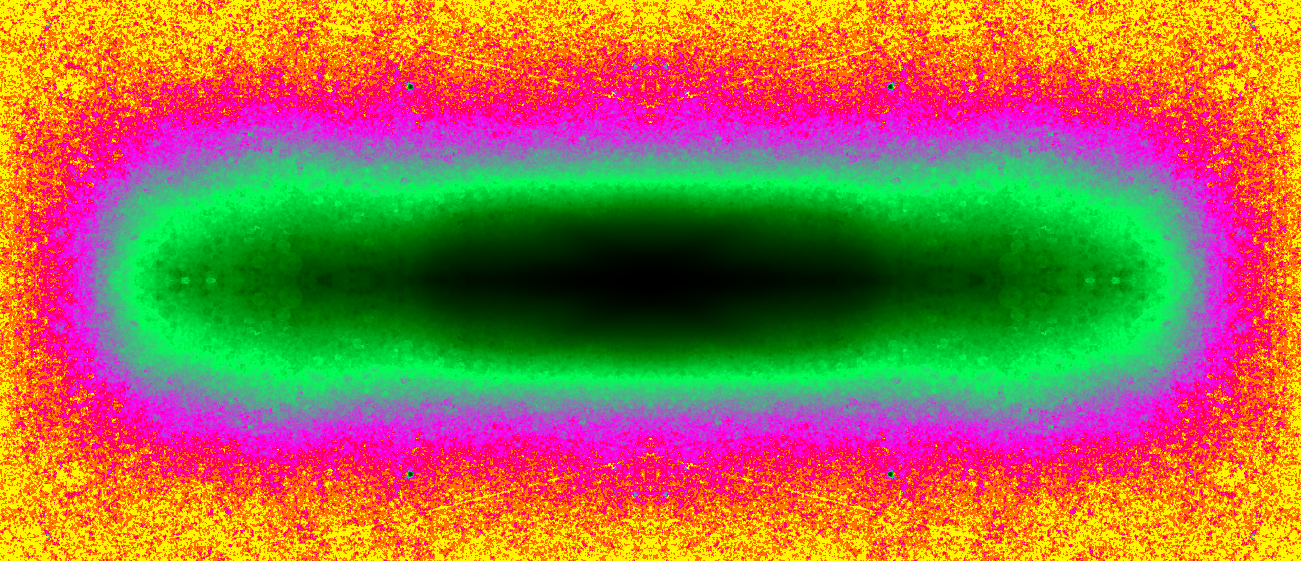}

\includegraphics[height=4.5cm]{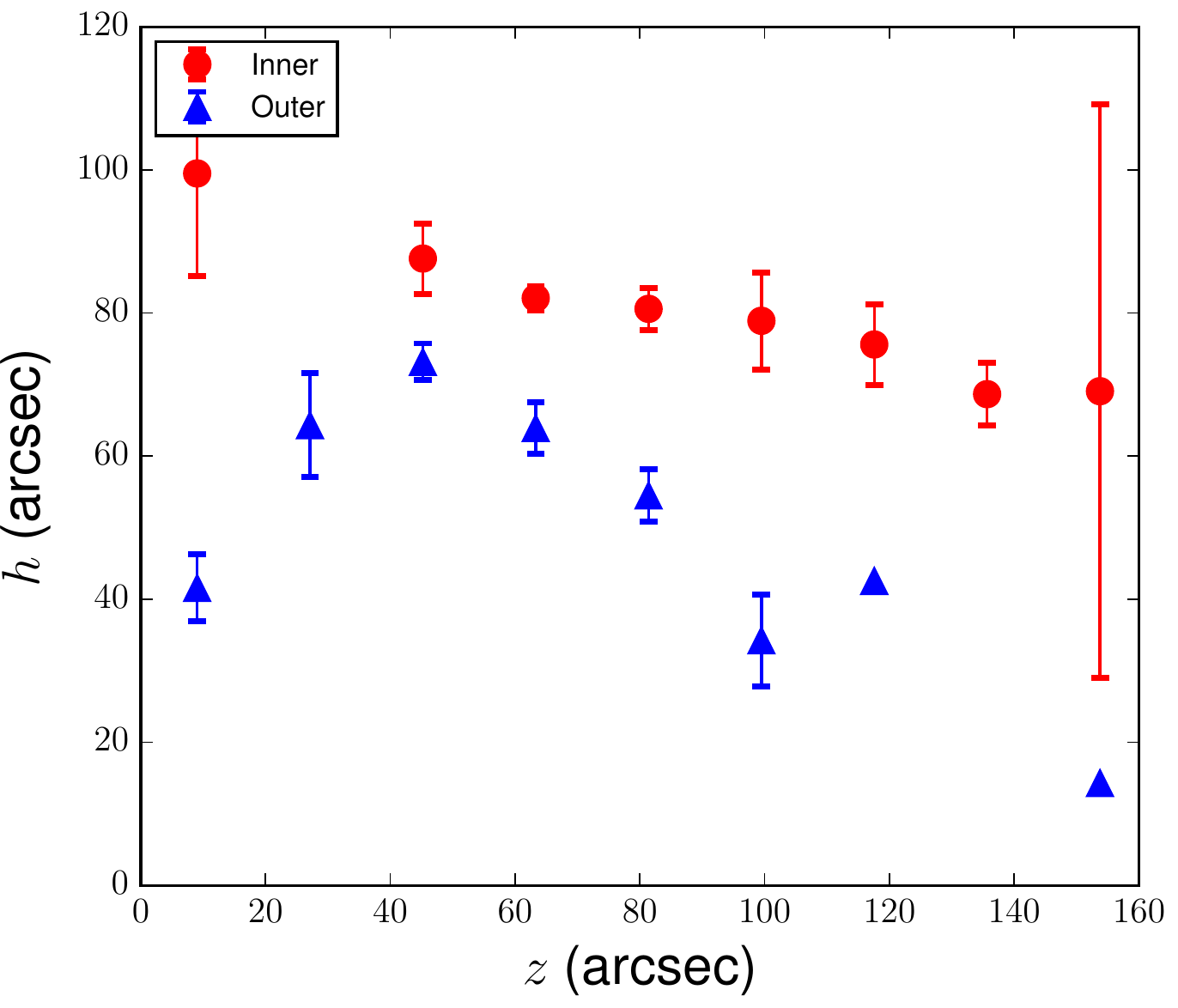}
\includegraphics[height=4.5cm]{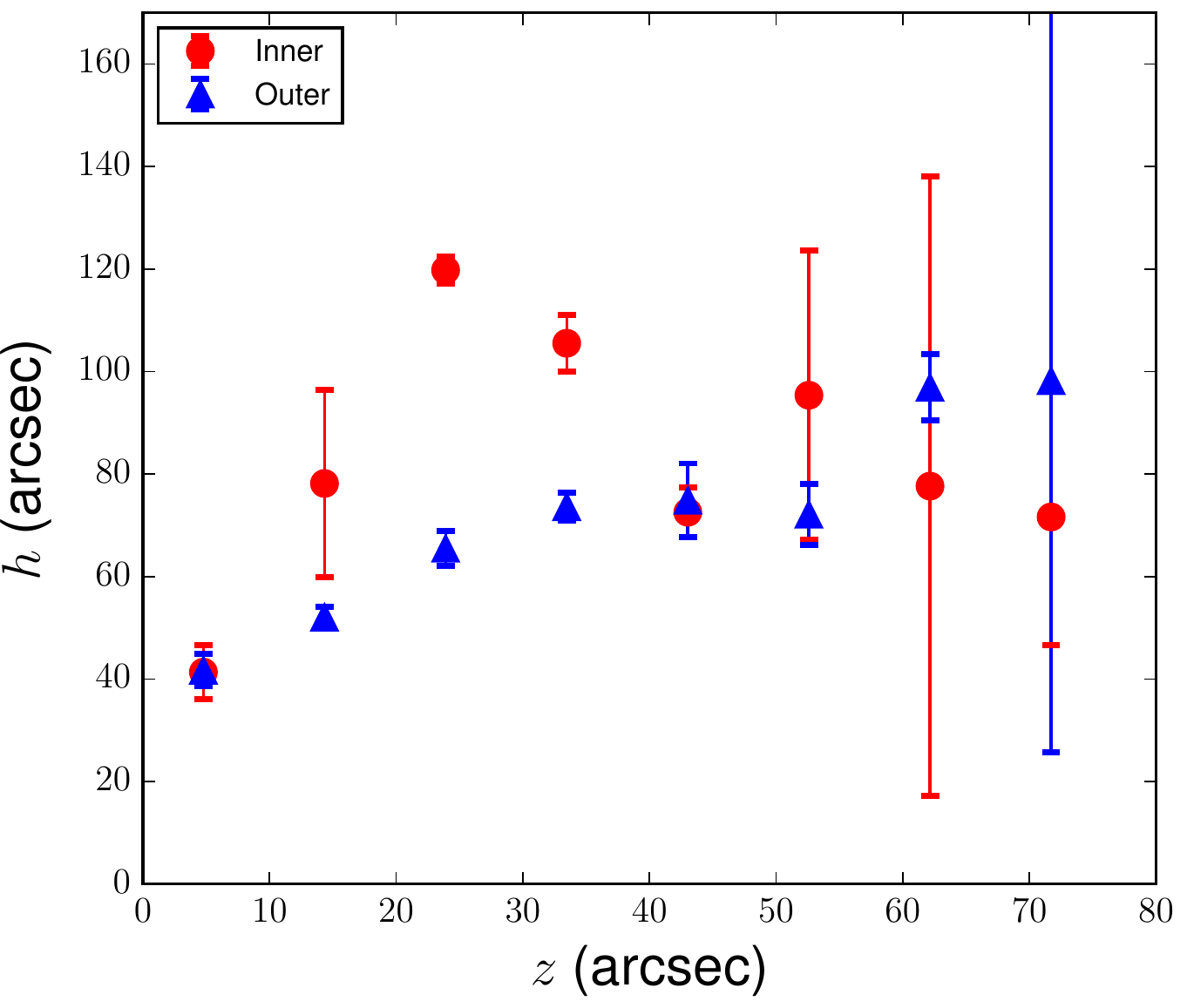}
\includegraphics[height=4.5cm]{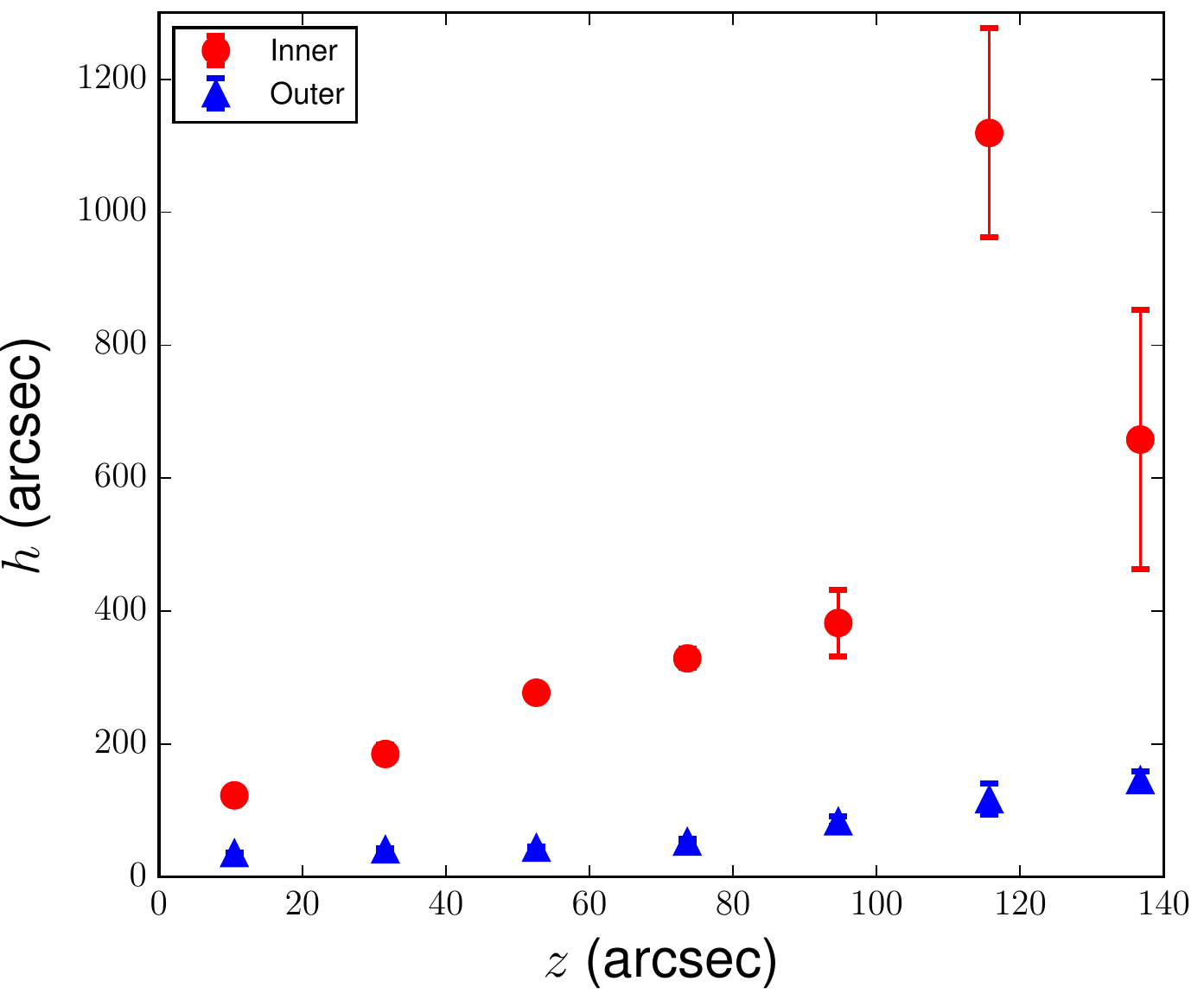}

\includegraphics[height=4.5cm]{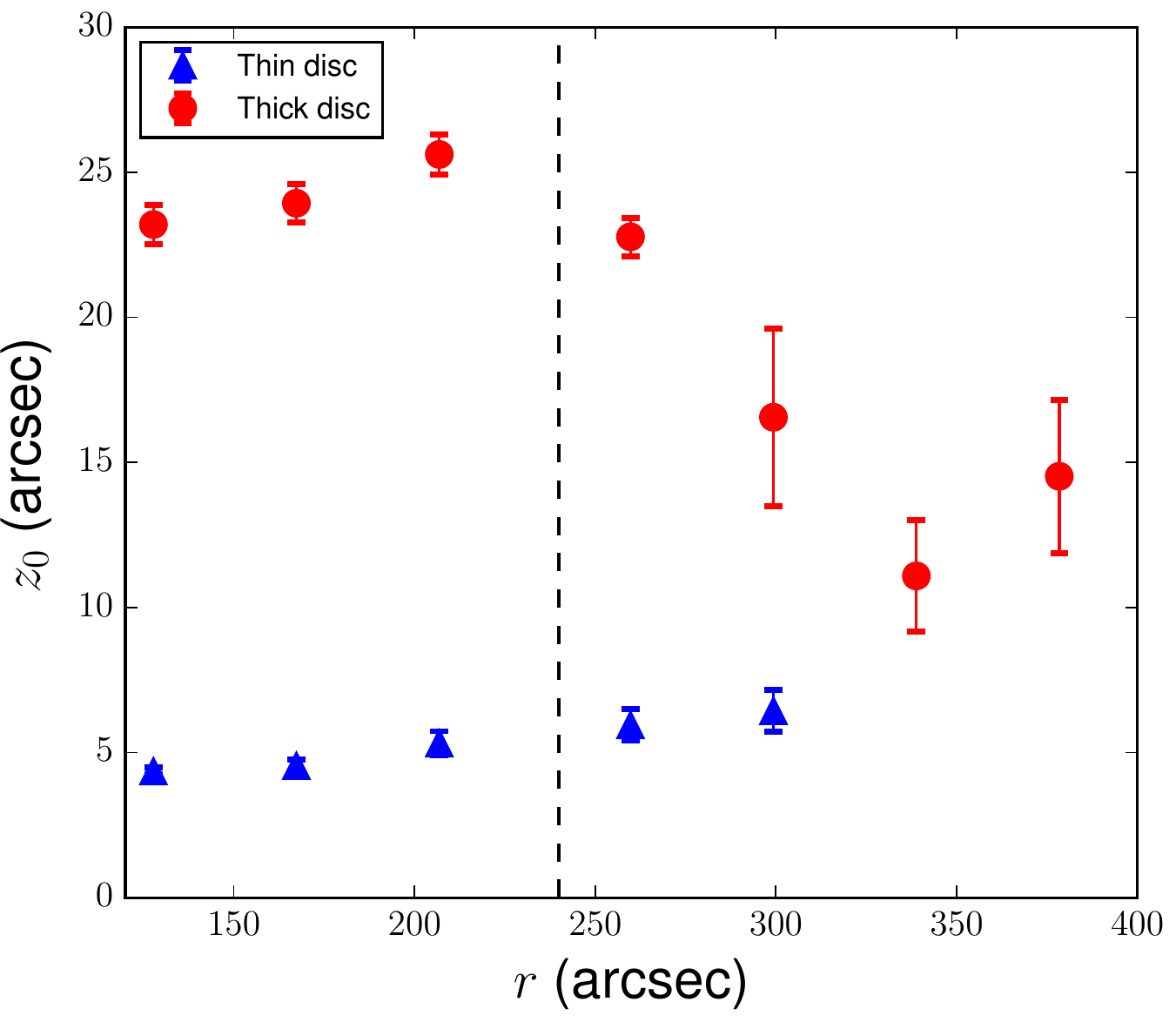}
\includegraphics[height=4.5cm]{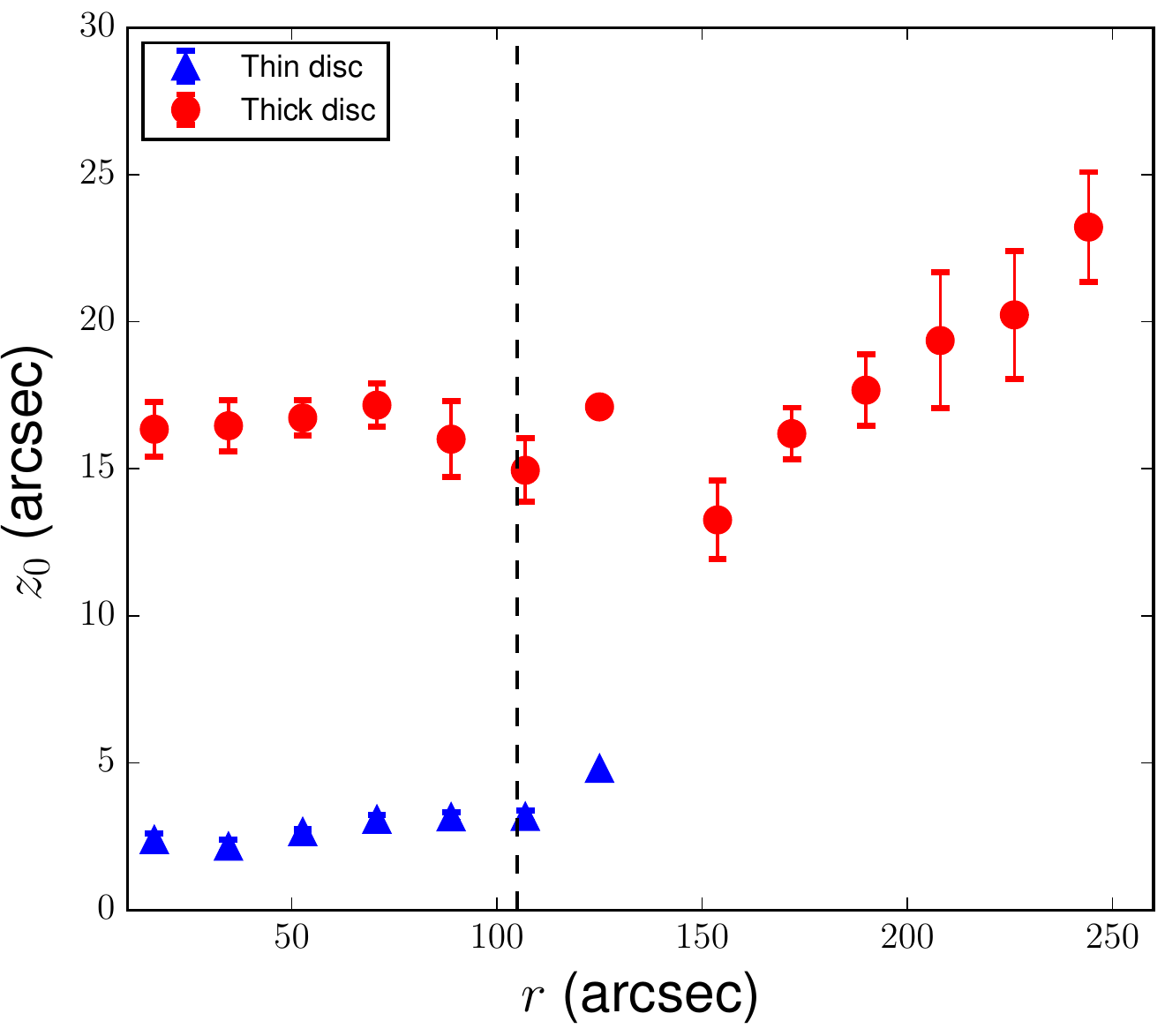}
\includegraphics[height=4.5cm]{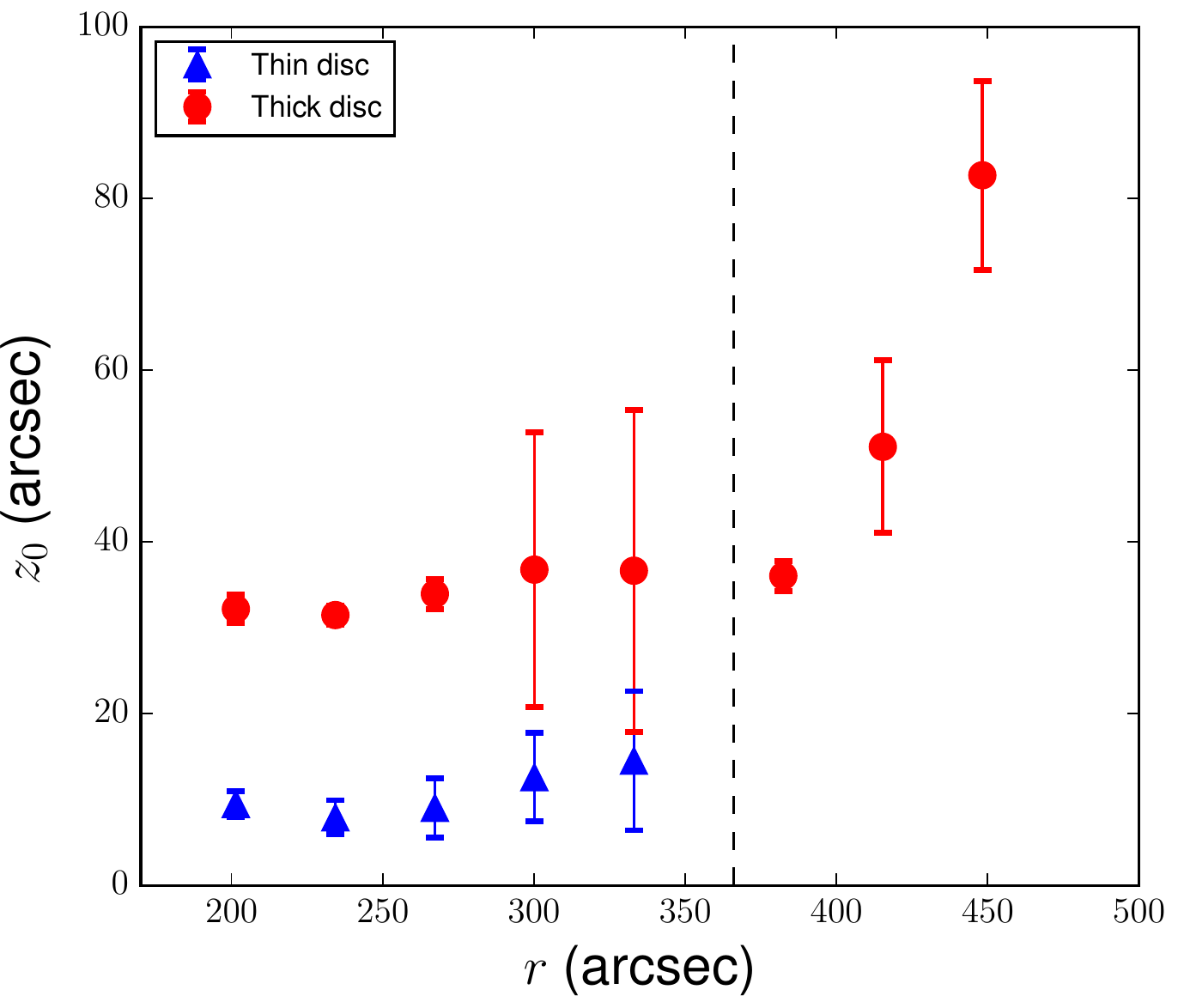}

\caption{\textit{The top panels.} Averaged IRAC~3.6~$\mu$m galaxy images for NGC\,891, NGC\,4302 and NGC\,3628. The darkest colours show maximum intensities. \textit{The middle panels.} Dependences of the disc scalelength on height above the plane. The inner disc (before the break) is shown by the red, whereas the outer disc (after the break radius) is shown by the blue colour. \textit{The bottom panels.} Dependences of the disc scaleheight on radius. The thin and thick discs are shown by the blue and red colour, respectively. The break radii are shown by the black dashed lines ($240\arcsec$, $105\arcsec$ and $366\arcsec$ for NGC\,891, NGC\,4302 and NGC\,3628, respectively).}
\end{figure*}

As our observations are made in an optical filter, due to dust the optical thickness of edge-on discs in galaxies is usually high \citep[$\tau_r>1$][]{2018A&A...616A.120M}, so to study the stellar distribution within and above the plane, one should use near- or mid-infrared data. Therefore, for these three galaxies we exploit \textit{Spitzer} \citep{2004ApJS..154....1W} observations in the Infrared Array Camera 3.6$\mu$m \citep[IRAC, ][]{2004ApJS..154...10F} band\footnote{The data were retrieved using the Spitzer Heritage Archive (the mosaic *maic.fits and uncertainties *munc.fits files from post-Basic Calibrated Data).}. 
For the retrieved images, we subtracted the sky background and masked out all contaminating foreground and background objects. Then we averaged their 2D profiles by averaging all four quadrants of the galaxy image (these quadrants can be obtained by dissecting the galaxy along the minor and major axis). This helped us to 1) increase the signal-to-noise ratio, 2) significantly reduce the number of masked objects (this is important in the case of 1D photometric cuts), and 3) average quite asymmetrical structure of NGC\,3628. Due to our averaging the information on the distribution of non-symmetric components (such as a bar and spirals arms) can be corrupted, but as we consider the smoothed disc component (see below), this averaging should not affect our results. The averaged images are given in Fig~\ref{fig:Example_galaxies}, bottom panels.

For this analysis, we retrieved an extended PSF from the IRSA website\footnote{\url{https://irsa.ipac.caltech.edu/data/SPITZER/docs/irac/calibrationfiles/psfprf/}}, which was then azimuthally averaged, so that we could apply it to all three images without modification \citep[the validity of this approach is discussed in ][]{2018A&A...610A...5C}. 

% Fitting method
To find the dependence of the disc scalelength and disc scaleheight on the vertical and radial coordinate, respectively, we do the following steps. 
%First, for each galaxy image we performed a 2D decomposition to estimate the parameters of the main galaxy components: thin and thick discs, viewed edge-on, and a central component, described by a S\'ersic function or a PSF. In fact, this central component may be consisting of a superposition of a stellar bar and a pseudobulge, but for our purposes the decomposition on the three components is sufficient). This step is done using the \textsc{galfit} code. The parameters of our decomposition model are listed in Table~\ref{tab:table3}. 
First, for each galaxy we find the truncation radius $R_\mathrm{tr}$, which is a radius at which the galaxy SB distribution along the major axis suddenly drops, making this position a sharp edge on the radial profile. 

Second, as each galaxy in our sample possesses a central component (a bar and, likely, a pseudo-bulge), we determine a region where the stellar disc starts to dominate over the central component ($108\arcsec$, $10\arcsec$, $185\arcsec$, for NGC\,891, NGC\,4302 and NGC\,3628, respectively). Here and further we only consider the region outside this inner radius and up to the truncation radius. If a break on the radial photometric cut is seen, we measure the disc scalelength up to the break, and beyond the break radius. We call them the inner and outer disc scalelengths, respectively. As the galaxies under consideration are nearby, we can clearly distinguish photometrically between their thin and thick discs. 
%All three galaxies show the extension of the thin disc along the radial direction up to the break radius, so beyond the break radius we only consider the thick disc. 

Third, once the inner break and truncation radius of the disc are determined, we use {\sc galfit} to fit the 2D profile of the galaxy image within the break radius but beyond the inner radius and within the truncation radius and beyond the break radius. These regions were fitted with two 2D edge-on profiles (thin and thick discs) to estimate their `average' disc scalelengths and disc scaleheights in these parts of the galaxy, using the following function for the intensity at $(R,z)$ for each disc:

\begin{equation}
\label{eon_disc}
I(R,z) \; = \; I_0 \; (R/h) \; K_{1}(R/h) \;\, {\rm sech}^{2} (z/z_{0})\,,
\end{equation}
where $I_0$ is the central surface brightness, $h$ is the disc scalelength, $z_{0}$ is
the disc scaleheight, $K_{1}$ is the modified Bessel function of the second kind.

However, to measure the dependence of these parameters on radius $R$ and height above the plane $z$, we dissect the galaxy image beyond the inner radius. We bin the radius extent to average vertical photometric cuts within each bin with a step of 0.1\,$R_\mathrm{tr}$. The same is done for horizontal (parallel to the major axis) photometric cuts -- we bin the vertical extent with a step of 0.1\,$b_{3.6}$, where $b_{3.6}$ is the minor axis of the isophote at 2\,$rms$ of the background in this image. The averaging within a bin is useful to make the signal-to-noise ratio higher, which is important for LSB levels. Then each 1D averaged cut, whether it is a radial or vertical one, is fitted using the \textsc{galfit} code (a weight map and the PSF convolution are both taken into account). The fitting of each averaged cut is done within the 2D galaxy image, at the specific radius $R$ (centre of the bin of vertical cuts) or height $z$ (centre of the bin of horizontal cuts) with the galaxy centre $(X_\mathrm{0},Y_\mathrm{0})$. For the radial distribution we use a single disc profile (within and beyond the disc break), whereas for the vertical distribution our model consists of an edge-on thick and thin disc (beyond some radius, however, we consider only one the single thick disc profile where the code cannot distinguish between the two discs). We note here that for fitting the vertical cuts we fix $h$ in eq.~(\ref{eon_disc}) and for fitting the horizontal cuts we fix $z_0$, which have been estimated from the 2D fitting above. This allows us to perform the correct 2D PSF convolution for all 1D photometric cuts. We tested this approach on a series of mock galaxy images and found it robust in retrieving parameters of 1D profiles. The results of the fitting are presented in Fig.~\ref{fig:Example_galaxies}, middle and bottom panels. The averaged profiles and corresponding models are given in the Appendix~B, Fig.~\ref{fig:R_profiles} and Fig.~\ref{fig:Z_profiles}, similar to fig.~3 in \citet{2019MNRAS.483..664M}.

%We provide the distributions of the fitted disc scalelength and scaleheight for all three galaxies in Fig.~\ref{}. The results of the fitting are presented in Table~\ref{}.

% Results: what we see, a possible formation scenario
Now let us briefly consider our results for each galaxy. 

\textbf{NGC\,891.} This galaxy shows prominent discy/diamond-like isophotes. The disc scalelength is decreasing with height $z$ for both the inner and outer parts of the disc. The disc scaleheight for the thin disc is increasing, whereas the thick discs is getting less vertically extended at the periphery. We do not see a significant flaring at the periphery for this galaxy. 

\textbf{NGC\,4302.} The outer isophotes of this galaxy look more oval, especially at the periphery. The scalelength of the inner disc shows a rather large variance along the $z$-axis, whereas the scalelength for the outer disc is increasing. The disc scaleheight is increasing at $R>150$~arcsec which demonstrates a disc flaring.

\textbf{NGC\,3628.} This galaxy has a prominent boxy outer shape illustrated in \citetalias{2019MNRAS.490.1539R}. The disc scalelength shows a tremendous increase by more than 5 times for the inner disc and 4 times for the outer disc. For the disc scaleheight we also observe a flaring at $R>400$~arcsec.

Comparing all three galaxies, we can see that their specific outer shape is controlled by two factors: an increase/decrease of the disc scalelength with the $z$-coordinate and a presence of the disc flaring at the periphery. The last factor can make the observed tips of an edge-on disc less sharp. However, the overall outer shape is more affected by the former factor, an increase of the disc scalelength with height.

\section{Discussion}
\label{sec:discussion}

In previous sections we have considered the outer shape of the sample galaxies. We found that the outer envelopes of edge-on galaxies fall into the three broad categories: discy/diamond-like, oval or boxy. However, in different edge-on galaxies their outer shape can be assigned to different galaxy components, which dominate the light in the periphery: an outer (thick) disc, bulge, or stellar halo. In the case of NGC\,2683 and M\,82 (NGC\,3034), the outer shape is affected by the presence of an inner orthogonal component (possibly, a remnant of minor merging in NGC\,2683 and superwinds in NGC\,3034). A massive bulge can outshine the disc, as, for example, in the galaxy NGC\,1055 with a thick boxy bulge or NGC\,1032 with its bright (classical?) bulge. The shape of bright stellar halos can be round, oval or boxy. Also, our deep observations show that these halos may have a slightly asymmetric shape of the outermost isophotes (NGC\,4594 and NGC\,3115). In its turn, the outer disc may have a discy, oval or boxy shape. Below we discuss this observed diversity of the galaxy outer shapes.

We consider the dependence of the observed outer shapes of the sample galaxies on different factors. First, as shown in \citet{2006AJ....131..226Y} low-mass ($V_\mathrm{max}<120$~km/s) galaxies have massive thick discs as compared to thin discs and dominate the stellar mass. On the contrary, massive galaxies are dominated by the thin disc. Thus, we might assume that the shape of the thick disc in massive and low-massive galaxies is different. However, this in not the case -- galaxies with different total mass may have quite different outer shapes. For example, NGC\,4762 ($V_\mathrm{max}=110$~km/s\footnote{The maximum rotation velocities are taken from the HyperLeda database, uncorrected for inclination.}) has a boxy outer shape, whereas NGC\,4222 ($V_\mathrm{max}=100$~km/s) and NGC\,4244 ($V_\mathrm{max}=89$~km/s) show pure discy outermost isophotes. Similarly, massive galaxies may also show different outer shapes.

We also compare the outer shape with the bulge type at the centre. Only 6 galaxies in our sample have confirmed classical bulges. Among them, four galaxies (NGC\,3115, NGC\,4594, NGC\,4638, and NGC\,5866) have bright oval or boxy halos. NGC\,1055 is a galaxy with a dominant thick boxy bulge and NGC\,4866 shows discy/oval isophotes. 

Most galaxies in our sample have confirmed pseudo-bulges, which grow due to internal secular evolution mostly associated with the disc potential \citep{2004ARA&A..42..603K}. According to the current study, the outer shape of these galaxies may be different. For example, NGC\,4762 shows a boxy outer structure, NGC\,5746 has oval outer isophotes, while NGC\,4565 is a diamond-like galaxy. We have several bulge-less galaxies which also exhibit different outer shapes. 

Also, we consider the dependence of the outer shape on the bar presence. 18 galaxies clearly show a B/PS bulge (a bar viewed side-on). Again, all these galaxies have different outer shapes (see, for example, the prototype galaxies in Sect.~\ref{sec:examples}). Three galaxies in our sample were confirmed as not hosting a bar structure, but those are not exactly edge-on.

In general, although the number of galaxies in our sample is rather modest, we did not find any correlation of the outer shape with the total galaxy mass and the presence (and the morphology) of the central component.

Let us now consider the environment of the sample galaxies. Three galaxies in our sample are marked in NED as isolated (`I' in column `Env' in Table~\ref{tab:table1}). All three galaxies have discy or discy/oval outer isophotes. 10 galaxies are marked as `in pair' and show some signatures of interaction. Almost all these galaxies have oval or boxy outer shapes. The remaining galaxies belong to groups of galaxies. Among these galaxies we see a diversity of types of the outer shape, with a large fraction of discy/diamond-like outer structures (7 out of 22).

To quantify the environment, we measured the surface and volume number density to the fifth nearest neighbour \citep[see e.g. ][for a review of different methods to estimate the local density]{2012MNRAS.419.2670M}. The HyperLeda database was used for this analysis. The results in the sense of the dependence of the outer envelope shape on the local density are similar for both approaches, therefore we provide in Table~\ref{tab:table1} only the volume number density, where the identified neighbours are indeed close to the target galaxy and not just show a small projection distance to it. The number of galaxies is small, so these results are very tentative. In general, we do not find any correlation between the local density and the outer shape (both for the $C_0$ and $C_0-C_\mathrm{0,mod}$, see Sect.~\ref{sec:correlations}). However, despite of the absence of this correlation, we strongly believe that it is galaxy interactions that are responsible for changing the observed outer shape of galaxies (see also a discussion below). All galaxies in our sample, which show LSB signatures in our deep images (column LSBF, `y' values in Table~\ref{tab:table1}), have an oval or boxy outer shape. For these galaxies $\langle C_0 \rangle =0.37\pm0.32$ versus $\langle C_0 \rangle =-0.29\pm0.40$ for the remaining galaxies. If we consider only galaxies which are close to edge-on ($i\geq86\degr$, see Table~\ref{tab:table2}), discy/diamond-like galaxies have, on average, a lower density environment than oval and boxy ones. Also, boxy galaxies have a lower flattening as compared to discy and oval shapes (see also Sect.~\ref{sec:correlations}). Four galaxies with bright stellar halos are oval or boxy and usually reside in very dense environments.

% Table with specific shape, C0 and ellipticity
\begin{table}
	\centering
	\caption{Average values of the discyness/boxyness parameter, flattening and volume density for different subsamples of galaxies. The number of galaxies in each sub-sample is provided in parenthesis.}
	\label{tab:table2}
	\begin{tabular}{cccc} % four columns, alignment for each
		\hline
		Shape of the outer disc & $C_0$       & $b/a$        & $\rho$        \\
                $i\geq86\degr$  &             &              &(Mpc$^{-3}$)  \\
		\hline
		Discy/diamond ($N=6$)  &$-0.6\pm0.2$ & $0.20\pm0.06$ & $0.62\pm0.61$ \\
		Oval ($N=5$)           &$0.0\pm0.3$  & $0.23\pm0.06$ & $1.36\pm2.53$ \\
		Boxy ($N=3$)	          &$2.0\pm3.0$  & $0.30\pm0.01$ & $1.60\pm0.74$  \\
		\hline
		\hline
		Halo ($N=4$)            &$0.2\pm0.2$  & $0.71\pm0.17$ & $342\pm625$   \\   
		\hline
	\end{tabular}
\end{table}

%Up-bending profiles are observed in high environments, so as boxy and oval shapes!
Now, let us turn to the literature where merging history of galaxies is studied.
One of the mechanisms to form stellar haloes in galaxies is related to \textit{in situ}
star formation \citep{2011MNRAS.416.2802F,2012MNRAS.420.2245M,2015MNRAS.454.3185C}.
On the other hand, as shown by \citet{2008MNRAS.389.1041R}, most massive mergers may heat the thin disc enough to produce a thick disc which can be as massive as that seen in the Milky Way (see the introduction of \citealt{2011ApJ...741...28C} where three other formation mechanisms of thick discs are briefly described). However, as shown by \citet{2009ApJ...691.1168H} disc heating caused by minor mergers is significantly reduced if a spiral galaxy has a high gas fraction \citep[see also][]{2009MNRAS.396..696S,2010MNRAS.403.1009M}. This can prevent the destruction of thin stellar discs by minor mergers. On the other hand, the accretion of satellite galaxies leads to a disc size increase from redshifts $\textsc{z}\sim3$ to the present of a factor of $\sim4$ \citep{2012ApJ...747...34B}. \citet{2016MNRAS.458.2371R} studied the stellar mass assembly of galaxies in the Illustris simulation and found that the fraction of accreted stars, formed within satellite galaxies (\textit{ex situ}) and found throughout galaxies at $\textsc{z}=0$, increases with galaxy stellar mass.

As shown by \citet{2010MNRAS.406..744C,2013MNRAS.434.3348C}, accreted stellar halos are assembled between redshifts $1<\textsc{z}<7$ by tidal disruption of accreted dwarf (spheroidal) galaxies with stellar masses $10^7-10^8\,M_{\sun}$ \citep[see also][]{1999MNRAS.307..495H,2005ApJ...635..931B,2006MNRAS.365..747A,2009ApJ...702.1058Z}. However, the diversity of the observed stellar halos (their masses, extent and complexity of the structure) is explained by the number of the progenitors and their infall time. According to their simulations, we should already see tidal streams, shells and other fine substructures at SB levels of $\approx 28$~mag/arcsec$^2$. This is what we see in many edge-on galaxies in our survey, but almost half of our sample galaxies (16 out of 35) do not show any LSB features near them. One reason for this discrepancy might be the different depth of the deep images of the sample galaxies ($27.8\pm0.5$~mag/arcsec$^2$, see Sect.~\ref{sec:sample}). However, we did not find any correlation between the image depth and the presence or absence of LSB features. For example, NGC\,4206 and NGC\,4222 are present in the same frame (depth 28.3~mag/arcsec$^2$) as NGC\,4216, but only the latter galaxy demonstrates LSB features. A further investigation of deep imaging (PI. Noah Brosch, see \citealt{2017IAUS..321..293B}) for a larger sample of edge-on galaxies should clarify this situation. The discrepancy between the observed and predicted LSB features near edge-on galaxies would challenge cosmological simulations, where interactions and mergers should usually produce readily detectable fine substructures at these LSB levels.

We should note, however, that as shown in several studies \citep[see e.g.][]{2008MNRAS.389.1041R, 2018MNRAS.479.4720D,2019MNRAS.487..318K}, the impact inclination is important for producing structures of different morphology. For example, at low inclination encounters ($\theta<20\degr$) a dwarf satellite is dragged into the disc plane by dynamical friction. The accreted stars of this satellite settle into a thick disc of the host galaxy. Recently, \citet{2019arXiv191005358G} found a fan-like feature in NGC\,4565 seen to wrap around the northwest disc limb. These ripples may be a tidal ribbon which formed due to a satellite accreted in the plane of the host galaxy disc. Taking this and other LSB features (asymmetry in disc truncations and many LSB satellite candidates) into account, they make a conclusion on the accretion scenario of the outer disc growth in NGC\,4565. However, according to the present study, this galaxy has a diamond-like shape, whereas we would expect to see quite boxy isophotes at its periphery if multiple ripples due to a satellite accretion were formed. In this galaxy, \citet{2019arXiv191005358G} found only one fan-like feature, so either this accretion has recently begun, or this fan-feature has a different origin. In either case, the co-existence of a large number of dwarf satellites near this galaxy and its quite undisturbed disc confirms a general observational problem of galaxy formation and evolution in the $\Lambda$CDM model. 

For higher inclination encounters $\theta<20\degr$ (which are twice as likely as low-inclination ones), structures, which are produced by minor merging, closely resemble inner and outer stellar halos observed in our Galaxy \citep{1993AJ....106..578M,2007Natur.450.1020C} and other galaxies \citep[see e.g. ][]{2014MNRAS.439.3128T}. The other by-products of these encounters are pronounced flares and warp of stellar discs and fine substructures.

Also, as shown in \citet{2011A&A...530A..10Q}, in galaxies disturbed by minor mergers, the scalelength of the thick disc should be larger than that of the thin disc. This is what we observe indirectly in NGC\,4302 and NGC\,3628 -- their disc scalelengths increase with height $z$. According to the 2D decomposition on thick and thick discs by \citet{2015ApJS..219....4S}, the thick disc scalelength is two and three times the thin disc scalelength for NGC\,4302 and NGC\,3628, respectively. In the physically motivated approach which assumes that galaxy discs are made of two stellar discs in hydrostatic equilibrium \citep{2018A&A...610A...5C}, the ratio of the thick disc to thin disc scalelength is even larger (for example, 3.6 for NGC\,4302). However, this is not the case for NGC\,891, where we can even see a slight decrease of the scalelength with $z$ for radii before the break radius (in the inner part of the disc) and much steeper decrease, starting at some $z$, for the outer part. \citet{2013ApJ...773...45S} used near-infrared imaging and identified a superthin disc for a model consisting of three stellar discs. The radial scalelengths for the thin and thick discs are very close: 4.11 versus 4.80~kpc for the thin and thick disc, respectively.  %Also, as follows from simulations by \citep{2011A&A...530A..10Q}, the thick disc scaleheight should increase with radius (disc flaring) if such a disc formed through minor merging. 
Also, as follows from cosmological simulations, the ubiquity of merger events inevitably produces strongly flaring discs \citep{2009ApJ...707L...1B,2011A&A...530A..10Q}. However, as shown in \citet{2013MNRAS.433..976R}, if the disc thickens only due to internal processes, this induces only a minor amount of flaring. 
We observe disc flaring, which starts at some radius, in NGC\,4302 and NGC\,3628. However, NGC\,891 does not show a significant flare of the disc. Moreover, we can see a decrease of the thick disc scaleheight by more then two times when going from the inner to outer part of the galaxy, although we should admit that some scattered light from the bar may be still present at these radii and be erroneously attributed to the thick disc. 
Therefore, taking into account the absence of bright tidal structures near NGC\,891 (although by overdensities of RGB stars in the halo of NGC\,891, \citealt{2010ApJ...714L..12M} detected extended ultrafaint substructures which may point to one or more accretion events in the present or past), the merging history and its intensity for NGC\,891 should be rather quiet as compared to NGC\,4302 and NGC\,3628.

%Interestingly, using resolved stellar populations, \citet{2016A&A...585A..97S} studied three nearby low-mass edge-on galaxies (one of them is NGC\,4244 from our sample) and did not detect a thick disc in any of them and no or little disc flaring. These observational facts, as conclude the authors should pose strong constraints on galaxy simulations.

%As such, taking into account all of the above, we can propose that the observed outer shape of galaxies depends on their merging history.

\section{Conclusions}
\label{sec:conlusions}
In this paper we have studied the outer shape of the 35 edge-on galaxies from the {\sl HERON} survey. Also, we identified LSB features near these galaxies, such as tidal tails, stellar streams, shells, and bridges. The conclusions of this work can be summarised as follows:
\begin{enumerate}
\item We classified the observed outer shapes of the sample galaxies into several types: discy/diamond-like, oval, and boxy (emerald-cut). The bright stellar halo may have boxy (NGC\,4638, NGC\,5866), oval (NGC\,3115), or round isophotes (NGC\,4594, but at very faint SB levels we observe a prolate spheroid!). Separately, we classify a boxy outer structure due to a prominent thick boxy bulge (NGC\,1055) and two galaxies with an orthogonal component: NGC\,2683 (a remnant of past merging or a polar-ring galaxy?) and NGC\,3034 (superwinds in the inner region).
\item At least for 17 galaxies in our sample their outer discs deviate from the 3D models for exponential or isothermal discs. Only 7 galaxies (NGC\,891, NGC\,3556, NGC\,4206, NGC\,4565, NGC\,4710, NGC\,4866, and NGC\,5170) follow these models, that is they show discy isophotes. 
\item We find that the less flattened the outer structure is, the more it deviates from the 3D disc models and has more oval or even boxy isophotes. 
\item  We considered three examples of most representative galaxies in our sample by their prominent outer shape: NGC\,891 (diamond-like/discy), NGC\,4302 (discy/oval), and NGC\,3628 (boxy). We showed that the shape of the outer disc is controlled by two dependencies: the dependence of the radial scalelength on height and of the vertical height on galactocentric radius. Galaxies, which show an increasing disc scalelength with height show more oval or even boxy isophotes. The second factor, which may affect the outer shape, is the disc flaring. However, flaring without the first factor cannot  produce boxy isophotes alone.
\item We propose that the outer shape of galaxies is controlled by recent or past merger activity. If a merging event occurred not long ago, the disturbed outer disc shows boxy isophotes. If no recent merger took place, the outer shape remains undisturbed and shows discy/diamond-like isophotes. If we observe an interacting (in a pair) edge-on galaxy, its shape often shows oval isophotes. Therefore, the view of the outer galaxy shape gives us a hint on the merging history of the galaxy.
\item We report new LSB features, which have not been mentioned in the literature, near the following galaxies: NGC\,518 (stellar stream), NGC\,2481 (tidal streams, plums, bridge to NGC\,2480), and NGC\,4638 (multiple stellar streams going out of the galaxy).
\end{enumerate}

We should point out that our sample is by no means complete, therefore the conclusions of this paper should be taken with caution and valid only to this small sample. We are about to apply the analysis of the outer galaxy shape to a larger sample of edge-on galaxies, where we will be able to study the outer shape and the galaxy environment using a better statistics.

\newpage
\section*{Acknowledgements}
We thank the anonymous reviewer for helpful comments on the paper.

This research has made use of the NASA/IPAC Infrared Science Archive (IRSA; \url{http://irsa.ipac.caltech.edu/frontpage/}), and the NASA/IPAC Extragalactic Database (NED; \url{https://ned.ipac.caltech.edu/}), both of which are operated by the Jet Propulsion Laboratory, California Institute of Technology, under contract with the National Aeronautics and Space Administration.  This research has made use of the HyperLeda database (\url{http://leda.univ-lyon1.fr/}; \citealp{2014A&A...570A..13M}). 
This work is based in part on observations made with the {\it Spitzer} Space Telescope, which is operated by the Jet Propulsion Laboratory, California Institute of Technology under a contract with NASA. 
RMR acknowledges financial support from his late father Jay Baum Rich.
JR acknowledges financial support from the State Agency for Research  
of the Spanish MCIU through the ``Center of Excellence Severo Ochoa''  
award to the Instituto de Astrof\'isica de Andaluc\'ia (SEV-2017-0709) and  
support from the grant RTI2018-096228-B-C31 (MICIU/FEDER, EU).
PA acknowledges financial support from the Russian Science
Foundation (grant no. 19-12-00145).

%%%%%%%%%%%%%%%%%%%%%%%%%%%%%%%%%%%%%%%%%%%%%%%%%%

%%%%%%%%%%%%%%%%%%%% REFERENCES %%%%%%%%%%%%%%%%%%

% The best way to enter references is to use BibTeX:

\bibliographystyle{mnras}
\bibliography{art} % if your bibtex file is called example.bib

\begin{thebibliography}{}
\makeatletter
\relax
\def\mn@urlcharsother{\let\do\@makeother \do\$\do\&\do\#\do\^\do\_\do\%\do\~}
\def\mn@doi{\begingroup\mn@urlcharsother \@ifnextchar [ {\mn@doi@}
  {\mn@doi@[]}}
\def\mn@doi@[#1]#2{\def\@tempa{#1}\ifx\@tempa\@empty \href
  {http://dx.doi.org/#2} {doi:#2}\else \href {http://dx.doi.org/#2} {#1}\fi
  \endgroup}
\def\mn@eprint#1#2{\mn@eprint@#1:#2::\@nil}
\def\mn@eprint@arXiv#1{\href {http://arxiv.org/abs/#1} {{\tt arXiv:#1}}}
\def\mn@eprint@dblp#1{\href {http://dblp.uni-trier.de/rec/bibtex/#1.xml}
  {dblp:#1}}
\def\mn@eprint@#1:#2:#3:#4\@nil{\def\@tempa {#1}\def\@tempb {#2}\def\@tempc
  {#3}\ifx \@tempc \@empty \let \@tempc \@tempb \let \@tempb \@tempa \fi \ifx
  \@tempb \@empty \def\@tempb {arXiv}\fi \@ifundefined
  {mn@eprint@\@tempb}{\@tempb:\@tempc}{\expandafter \expandafter \csname
  mn@eprint@\@tempb\endcsname \expandafter{\@tempc}}}

\bibitem[\protect\citeauthoryear{{Abadi}, {Navarro}  \& {Steinmetz}}{{Abadi}
  et~al.}{2006}]{2006MNRAS.365..747A}
{Abadi} M.~G.,  {Navarro} J.~F.,   {Steinmetz} M.,  2006, \mn@doi [\mnras]
  {10.1111/j.1365-2966.2005.09789.x}, \href
  {https://ui.adsabs.harvard.edu/abs/2006MNRAS.365..747A} {365, 747}

\bibitem[\protect\citeauthoryear{{Abraham} \& {van Dokkum}}{{Abraham} \& {van
  Dokkum}}{2014}]{2014PASP..126...55A}
{Abraham} R.~G.,  {van Dokkum} P.~G.,  2014, \mn@doi [\pasp] {10.1086/674875},
  \href {https://ui.adsabs.harvard.edu/abs/2014PASP..126...55A} {126, 55}

\bibitem[\protect\citeauthoryear{{Afanasiev} \& {Sil'chenko}}{{Afanasiev} \&
  {Sil'chenko}}{2002}]{2002AJ....124..706A}
{Afanasiev} V.~L.,  {Sil'chenko} O.~K.,  2002, \mn@doi [\aj] {10.1086/341387},
  \href {https://ui.adsabs.harvard.edu/abs/2002AJ....124..706A} {124, 706}

\bibitem[\protect\citeauthoryear{{Ahn} et~al.,}{{Ahn}
  et~al.}{2014}]{2014ApJS..211...17A}
{Ahn} C.~P.,  et~al., 2014, \mn@doi [\apjs] {10.1088/0067-0049/211/2/17}, \href
  {https://ui.adsabs.harvard.edu/abs/2014ApJS..211...17A} {211, 17}

\bibitem[\protect\citeauthoryear{{Ann} \& {Park}}{{Ann} \&
  {Park}}{2018}]{2018JKAS...51...73A}
{Ann} H.~B.,  {Park} H.~W.,  2018, \mn@doi [Journal of Korean Astronomical
  Society] {10.5303/JKAS.2018.51.4.73}, \href
  {https://ui.adsabs.harvard.edu/abs/2018JKAS...51...73A} {51, 73}

\bibitem[\protect\citeauthoryear{{Arnold}, {Romanowsky}, {Brodie}, {Chomiuk},
  {Spitler}, {Strader}, {Benson}  \& {Forbes}}{{Arnold}
  et~al.}{2011}]{2011ApJ...736L..26A}
{Arnold} J.~A.,  {Romanowsky} A.~J.,  {Brodie} J.~P.,  {Chomiuk} L.,  {Spitler}
  L.~R.,  {Strader} J.,  {Benson} A.~J.,   {Forbes} D.~A.,  2011, \mn@doi
  [\apjl] {10.1088/2041-8205/736/2/L26}, \href
  {https://ui.adsabs.harvard.edu/abs/2011ApJ...736L..26A} {736, L26}

\bibitem[\protect\citeauthoryear{{Barentine} \& {Kormendy}}{{Barentine} \&
  {Kormendy}}{2009}]{2009ASPC..419..149B}
{Barentine} J.~C.,  {Kormendy} J.,  2009, {Detection of a Distinct Pseudobulge
  Hidden Inside the ``Box-Shaped Bulge'' of NGC 4565}.
p.~149

\bibitem[\protect\citeauthoryear{{Barentine} \& {Kormendy}}{{Barentine} \&
  {Kormendy}}{2012}]{2012ApJ...754..140B}
{Barentine} J.~C.,  {Kormendy} J.,  2012, \mn@doi [\apj]
  {10.1088/0004-637X/754/2/140}, \href
  {https://ui.adsabs.harvard.edu/abs/2012ApJ...754..140B} {754, 140}

\bibitem[\protect\citeauthoryear{{Barteldrees} \& {Dettmar}}{{Barteldrees} \&
  {Dettmar}}{1994}]{1994A&AS..103..475B}
{Barteldrees} A.,  {Dettmar} R.~J.,  1994, \aaps, \href
  {https://ui.adsabs.harvard.edu/abs/1994A&AS..103..475B} {103, 475}

\bibitem[\protect\citeauthoryear{{Bellstedt}, {Forbes}, {Foster}, {Romanowsky},
  {Brodie}, {Pastorello}, {Alabi}  \& {Villaume}}{{Bellstedt}
  et~al.}{2017}]{2017MNRAS.467.4540B}
{Bellstedt} S.,  {Forbes} D.~A.,  {Foster} C.,  {Romanowsky} A.~J.,  {Brodie}
  J.~P.,  {Pastorello} N.,  {Alabi} A.,   {Villaume} A.,  2017, \mn@doi
  [\mnras] {10.1093/mnras/stx418}, \href
  {https://ui.adsabs.harvard.edu/abs/2017MNRAS.467.4540B} {467, 4540}

\bibitem[\protect\citeauthoryear{{Bender}}{{Bender}}{1988}]{1988A&A...193L...7B}
{Bender} R.,  1988, \aap, \href
  {https://ui.adsabs.harvard.edu/abs/1988A&A...193L...7B} {193, L7}

\bibitem[\protect\citeauthoryear{{Bendo} et~al.,}{{Bendo}
  et~al.}{2002}]{2002AJ....123.3067B}
{Bendo} G.~J.,  et~al., 2002, \mn@doi [\aj] {10.1086/340083}, \href
  {https://ui.adsabs.harvard.edu/abs/2002AJ....123.3067B} {123, 3067}

\bibitem[\protect\citeauthoryear{{Bertin} \& {Arnouts}}{{Bertin} \&
  {Arnouts}}{1996}]{1996A&AS..117..393B}
{Bertin} E.,  {Arnouts} S.,  1996, \mn@doi [\aaps] {10.1051/aas:1996164}, \href
  {https://ui.adsabs.harvard.edu/abs/1996A&AS..117..393B} {117, 393}

\bibitem[\protect\citeauthoryear{{Binggeli}, {Sandage}  \&
  {Tammann}}{{Binggeli} et~al.}{1985}]{1985AJ.....90.1681B}
{Binggeli} B.,  {Sandage} A.,   {Tammann} G.~A.,  1985, \mn@doi [\aj]
  {10.1086/113874}, \href
  {https://ui.adsabs.harvard.edu/abs/1985AJ.....90.1681B} {90, 1681}

\bibitem[\protect\citeauthoryear{{Bizyaev} \& {Mitronova}}{{Bizyaev} \&
  {Mitronova}}{2002}]{2002A&A...389..795B}
{Bizyaev} D.,  {Mitronova} S.,  2002, \mn@doi [\aap]
  {10.1051/0004-6361:20020633}, \href
  {https://ui.adsabs.harvard.edu/abs/2002A&A...389..795B} {389, 795}

\bibitem[\protect\citeauthoryear{{Bizyaev}, {Kautsch}, {Mosenkov},
  {Reshetnikov}, {Sotnikova}, {Yablokova}  \& {Hillyer}}{{Bizyaev}
  et~al.}{2014}]{2014ApJ...787...24B}
{Bizyaev} D.~V.,  {Kautsch} S.~J.,  {Mosenkov} A.~V.,  {Reshetnikov} V.~P.,
  {Sotnikova} N.~Y.,  {Yablokova} N.~V.,   {Hillyer} R.~W.,  2014, \mn@doi
  [\apj] {10.1088/0004-637X/787/1/24}, \href
  {https://ui.adsabs.harvard.edu/abs/2014ApJ...787...24B} {787, 24}

\bibitem[\protect\citeauthoryear{{Bluck}, {Conselice}, {Buitrago},
  {Gr{\"u}tzbauch}, {Hoyos}, {Mortlock}  \& {Bauer}}{{Bluck}
  et~al.}{2012}]{2012ApJ...747...34B}
{Bluck} A. F.~L.,  {Conselice} C.~J.,  {Buitrago} F.~o.,  {Gr{\"u}tzbauch} R.,
  {Hoyos} C.,  {Mortlock} A.,   {Bauer} A.~E.,  2012, \mn@doi [\apj]
  {10.1088/0004-637X/747/1/34}, \href
  {https://ui.adsabs.harvard.edu/abs/2012ApJ...747...34B} {747, 34}

\bibitem[\protect\citeauthoryear{{Bocchio}, {Bianchi}, {Hunt}  \&
  {Schneider}}{{Bocchio} et~al.}{2016}]{2016A&A...586A...8B}
{Bocchio} M.,  {Bianchi} S.,  {Hunt} L.~K.,   {Schneider} R.,  2016, \mn@doi
  [\aap] {10.1051/0004-6361/201526950}, \href
  {https://ui.adsabs.harvard.edu/abs/2016A&A...586A...8B} {586, A8}

\bibitem[\protect\citeauthoryear{{Bottema}, {van der Kruit}  \&
  {Freeman}}{{Bottema} et~al.}{1987}]{1987A&A...178...77B}
{Bottema} R.,  {van der Kruit} P.~C.,   {Freeman} K.~C.,  1987, \aap, \href
  {https://ui.adsabs.harvard.edu/abs/1987A&A...178...77B} {178, 77}

\bibitem[\protect\citeauthoryear{{Bournaud}, {Elmegreen}  \&
  {Martig}}{{Bournaud} et~al.}{2009}]{2009ApJ...707L...1B}
{Bournaud} F.,  {Elmegreen} B.~G.,   {Martig} M.,  2009, \mn@doi [\apjl]
  {10.1088/0004-637X/707/1/L1}, \href
  {https://ui.adsabs.harvard.edu/abs/2009ApJ...707L...1B} {707, L1}

\bibitem[\protect\citeauthoryear{{Braine}, {Combes}, {Casoli}, {Dupraz},
  {Gerin}, {Klein}, {Wielebinski}  \& {Brouillet}}{{Braine}
  et~al.}{1993}]{1993A&AS...97..887B}
{Braine} J.,  {Combes} F.,  {Casoli} F.,  {Dupraz} C.,  {Gerin} M.,  {Klein}
  U.,  {Wielebinski} R.,   {Brouillet} N.,  1993, \aaps, \href
  {https://ui.adsabs.harvard.edu/abs/1993A&AS...97..887B} {97, 887}

\bibitem[\protect\citeauthoryear{{Bregman}}{{Bregman}}{1980}]{1980ApJ...236..577B}
{Bregman} J.~N.,  1980, \mn@doi [\apj] {10.1086/157776}, \href
  {https://ui.adsabs.harvard.edu/abs/1980ApJ...236..577B} {236, 577}

\bibitem[\protect\citeauthoryear{{Bregman} \& {Houck}}{{Bregman} \&
  {Houck}}{1997}]{1997ApJ...485..159B}
{Bregman} J.~N.,  {Houck} J.~C.,  1997, \mn@doi [\apj] {10.1086/304397}, \href
  {https://ui.adsabs.harvard.edu/abs/1997ApJ...485..159B} {485, 159}

\bibitem[\protect\citeauthoryear{{Brosch}, {Mosenkov}  \& {Rich}}{{Brosch}
  et~al.}{2017}]{2017IAUS..321..293B}
{Brosch} N.,  {Mosenkov} A.,   {Rich} R.~M.,  2017, in {Gil de Paz} A.,
  {Knapen} J.~H.,   {Lee} J.~C.,  eds,  IAU Symposium Vol. 321, Formation and
  Evolution of Galaxy Outskirts. pp 293--293,
  \mn@doi{10.1017/S1743921316011820}

\bibitem[\protect\citeauthoryear{{Bullock} \& {Johnston}}{{Bullock} \&
  {Johnston}}{2005}]{2005ApJ...635..931B}
{Bullock} J.~S.,  {Johnston} K.~V.,  2005, \mn@doi [\apj] {10.1086/497422},
  \href {https://ui.adsabs.harvard.edu/abs/2005ApJ...635..931B} {635, 931}

\bibitem[\protect\citeauthoryear{{Bureau}, {Aronica}, {Athanassoula},
  {Dettmar}, {Bosma}  \& {Freeman}}{{Bureau}
  et~al.}{2006}]{2006MNRAS.370..753B}
{Bureau} M.,  {Aronica} G.,  {Athanassoula} E.,  {Dettmar} R.~J.,  {Bosma} A.,
   {Freeman} K.~C.,  2006, \mn@doi [\mnras] {10.1111/j.1365-2966.2006.10471.x},
  \href {https://ui.adsabs.harvard.edu/abs/2006MNRAS.370..753B} {370, 753}

\bibitem[\protect\citeauthoryear{{Buta}}{{Buta}}{2011}]{2011arXiv1102.0550B}
{Buta} R.~J.,  2011, arXiv e-prints, \href
  {https://ui.adsabs.harvard.edu/abs/2011arXiv1102.0550B} {p. arXiv:1102.0550}

\bibitem[\protect\citeauthoryear{{Camm}}{{Camm}}{1950}]{1950MNRAS.110..305C}
{Camm} G.~L.,  1950, \mn@doi [\mnras] {10.1093/mnras/110.4.305}, \href
  {https://ui.adsabs.harvard.edu/abs/1950MNRAS.110..305C} {110, 305}

\bibitem[\protect\citeauthoryear{{Capaccioli}, {Held}  \& {Nieto}}{{Capaccioli}
  et~al.}{1987}]{1987AJ.....94.1519C}
{Capaccioli} M.,  {Held} E.~V.,   {Nieto} J.-L.,  1987, \mn@doi [\aj]
  {10.1086/114585}, \href
  {https://ui.adsabs.harvard.edu/abs/1987AJ.....94.1519C} {94, 1519}

\bibitem[\protect\citeauthoryear{{Cappellari} et~al.,}{{Cappellari}
  et~al.}{2013}]{2013MNRAS.432.1709C}
{Cappellari} M.,  et~al., 2013, \mn@doi [\mnras] {10.1093/mnras/stt562}, \href
  {https://ui.adsabs.harvard.edu/abs/2013MNRAS.432.1709C} {432, 1709}

\bibitem[\protect\citeauthoryear{{Carollo} et~al.,}{{Carollo}
  et~al.}{2007}]{2007Natur.450.1020C}
{Carollo} D.,  et~al., 2007, \mn@doi [\nat] {10.1038/nature06460}, \href
  {https://ui.adsabs.harvard.edu/abs/2007Natur.450.1020C} {450, 1020}

\bibitem[\protect\citeauthoryear{{Chung}, {van Gorkom}, {Kenney}  \&
  {Vollmer}}{{Chung} et~al.}{2007}]{2007ApJ...659L.115C}
{Chung} A.,  {van Gorkom} J.~H.,  {Kenney} J. D.~P.,   {Vollmer} B.,  2007,
  \mn@doi [\apjl] {10.1086/518034}, \href
  {https://ui.adsabs.harvard.edu/abs/2007ApJ...659L.115C} {659, L115}

\bibitem[\protect\citeauthoryear{{Chung}, {van Gorkom}, {Kenney}, {Crowl}  \&
  {Vollmer}}{{Chung} et~al.}{2009}]{2009AJ....138.1741C}
{Chung} A.,  {van Gorkom} J.~H.,  {Kenney} J. D.~P.,  {Crowl} H.,   {Vollmer}
  B.,  2009, \mn@doi [\aj] {10.1088/0004-6256/138/6/1741}, \href
  {https://ui.adsabs.harvard.edu/abs/2009AJ....138.1741C} {138, 1741}

\bibitem[\protect\citeauthoryear{{Ciambur}}{{Ciambur}}{2016}]{2016PASA...33...62C}
{Ciambur} B.~C.,  2016, \mn@doi [\pasa] {10.1017/pasa.2016.60}, \href
  {https://ui.adsabs.harvard.edu/abs/2016PASA...33...62C} {33, e062}

\bibitem[\protect\citeauthoryear{{Ciambur} \& {Graham}}{{Ciambur} \&
  {Graham}}{2016}]{2016MNRAS.459.1276C}
{Ciambur} B.~C.,  {Graham} A.~W.,  2016, \mn@doi [\mnras]
  {10.1093/mnras/stw759}, \href
  {https://ui.adsabs.harvard.edu/abs/2016MNRAS.459.1276C} {459, 1276}

\bibitem[\protect\citeauthoryear{{Combes}, {Debbasch}, {Friedli}  \&
  {Pfenniger}}{{Combes} et~al.}{1990}]{1990A&A...233...82C}
{Combes} F.,  {Debbasch} F.,  {Friedli} D.,   {Pfenniger} D.,  1990, \aap,
  \href {https://ui.adsabs.harvard.edu/abs/1990A&A...233...82C} {233, 82}

\bibitem[\protect\citeauthoryear{{Comer{\'o}n} et~al.,}{{Comer{\'o}n}
  et~al.}{2011a}]{2011ApJ...729...18C}
{Comer{\'o}n} S.,  et~al., 2011a, \mn@doi [\apj] {10.1088/0004-637X/729/1/18},
  \href {https://ui.adsabs.harvard.edu/abs/2011ApJ...729...18C} {729, 18}

\bibitem[\protect\citeauthoryear{{Comer{\'o}n} et~al.,}{{Comer{\'o}n}
  et~al.}{2011b}]{2011ApJ...738L..17C}
{Comer{\'o}n} S.,  et~al., 2011b, \mn@doi [\apjl]
  {10.1088/2041-8205/738/2/L17}, \href
  {https://ui.adsabs.harvard.edu/abs/2011ApJ...738L..17C} {738, L17}

\bibitem[\protect\citeauthoryear{{Comer{\'o}n} et~al.,}{{Comer{\'o}n}
  et~al.}{2011c}]{2011ApJ...741...28C}
{Comer{\'o}n} S.,  et~al., 2011c, \mn@doi [\apj] {10.1088/0004-637X/741/1/28},
  \href {https://ui.adsabs.harvard.edu/abs/2011ApJ...741...28C} {741, 28}

\bibitem[\protect\citeauthoryear{{Comer{\'o}n} et~al.,}{{Comer{\'o}n}
  et~al.}{2012}]{2012ApJ...759...98C}
{Comer{\'o}n} S.,  et~al., 2012, \mn@doi [\apj] {10.1088/0004-637X/759/2/98},
  \href {https://ui.adsabs.harvard.edu/abs/2012ApJ...759...98C} {759, 98}

\bibitem[\protect\citeauthoryear{{Comer{\'o}n}, {Salo}  \&
  {Knapen}}{{Comer{\'o}n} et~al.}{2018}]{2018A&A...610A...5C}
{Comer{\'o}n} S.,  {Salo} H.,   {Knapen} J.~H.,  2018, \mn@doi [\aap]
  {10.1051/0004-6361/201731415}, \href
  {https://ui.adsabs.harvard.edu/abs/2018A&A...610A...5C} {610, A5}

\bibitem[\protect\citeauthoryear{{Cooper} et~al.,}{{Cooper}
  et~al.}{2010}]{2010MNRAS.406..744C}
{Cooper} A.~P.,  et~al., 2010, \mn@doi [\mnras]
  {10.1111/j.1365-2966.2010.16740.x}, \href
  {https://ui.adsabs.harvard.edu/abs/2010MNRAS.406..744C} {406, 744}

\bibitem[\protect\citeauthoryear{{Cooper}, {D'Souza}, {Kauffmann}, {Wang},
  {Boylan-Kolchin}, {Guo}, {Frenk}  \& {White}}{{Cooper}
  et~al.}{2013}]{2013MNRAS.434.3348C}
{Cooper} A.~P.,  {D'Souza} R.,  {Kauffmann} G.,  {Wang} J.,  {Boylan-Kolchin}
  M.,  {Guo} Q.,  {Frenk} C.~S.,   {White} S. D.~M.,  2013, \mn@doi [\mnras]
  {10.1093/mnras/stt1245}, \href
  {https://ui.adsabs.harvard.edu/abs/2013MNRAS.434.3348C} {434, 3348}

\bibitem[\protect\citeauthoryear{{Cooper}, {Parry}, {Lowing}, {Cole}  \&
  {Frenk}}{{Cooper} et~al.}{2015}]{2015MNRAS.454.3185C}
{Cooper} A.~P.,  {Parry} O.~H.,  {Lowing} B.,  {Cole} S.,   {Frenk} C.,  2015,
  \mn@doi [\mnras] {10.1093/mnras/stv2057}, \href
  {https://ui.adsabs.harvard.edu/abs/2015MNRAS.454.3185C} {454, 3185}

\bibitem[\protect\citeauthoryear{{Corsini}, {M{\'e}ndez-Abreu}, {Pastorello},
  {Dalla Bont{\`a}}, {Morelli}, {Beifiori}, {Pizzella}  \& {Bertola}}{{Corsini}
  et~al.}{2012}]{2012MNRAS.423L..79C}
{Corsini} E.~M.,  {M{\'e}ndez-Abreu} J.,  {Pastorello} N.,  {Dalla Bont{\`a}}
  E.,  {Morelli} L.,  {Beifiori} A.,  {Pizzella} A.,   {Bertola} F.,  2012,
  \mn@doi [\mnras] {10.1111/j.1745-3933.2012.01261.x}, \href
  {https://ui.adsabs.harvard.edu/abs/2012MNRAS.423L..79C} {423, L79}

\bibitem[\protect\citeauthoryear{{Cort{\'e}s}, {Kenney}  \&
  {Hardy}}{{Cort{\'e}s} et~al.}{2015}]{2015ApJS..216....9C}
{Cort{\'e}s} J.~R.,  {Kenney} J. D.~P.,   {Hardy} E.,  2015, \mn@doi [\apjs]
  {10.1088/0067-0049/216/1/9}, \href
  {https://ui.adsabs.harvard.edu/abs/2015ApJS..216....9C} {216, 9}

\bibitem[\protect\citeauthoryear{{Decleir}}{{Decleir}}{2015}]{MarjorieThesis}
{Decleir} M.,  2015, Master's thesis, Master of Science in Physics and
  Astronomy, Belgium

\bibitem[\protect\citeauthoryear{{Dehnen} \& {Hasanuddin}}{{Dehnen} \&
  {Hasanuddin}}{2018}]{2018MNRAS.479.4720D}
{Dehnen} W.,  {Hasanuddin} 2018, \mn@doi [\mnras] {10.1093/mnras/sty1726},
  \href {https://ui.adsabs.harvard.edu/abs/2018MNRAS.479.4720D} {479, 4720}

\bibitem[\protect\citeauthoryear{{Dumke}, {Krause}  \& {Wielebinski}}{{Dumke}
  et~al.}{2004}]{2004A&A...414..475D}
{Dumke} M.,  {Krause} M.,   {Wielebinski} R.,  2004, \mn@doi [\aap]
  {10.1051/0004-6361:20031636}, \href
  {https://ui.adsabs.harvard.edu/abs/2004A&A...414..475D} {414, 475}

\bibitem[\protect\citeauthoryear{{Emsellem}, {Bacon}, {Monnet}  \&
  {Poulain}}{{Emsellem} et~al.}{1996}]{1996A&A...312..777E}
{Emsellem} E.,  {Bacon} R.,  {Monnet} G.,   {Poulain} P.,  1996, \aap, \href
  {https://ui.adsabs.harvard.edu/abs/1996A&A...312..777E} {312, 777}

\bibitem[\protect\citeauthoryear{{Emsellem}, {Dejonghe}  \& {Bacon}}{{Emsellem}
  et~al.}{1999}]{1999MNRAS.303..495E}
{Emsellem} E.,  {Dejonghe} H.,   {Bacon} R.,  1999, \mn@doi [\mnras]
  {10.1046/j.1365-8711.1999.02210.x}, \href
  {https://ui.adsabs.harvard.edu/abs/1999MNRAS.303..495E} {303, 495}

\bibitem[\protect\citeauthoryear{{Erwin}}{{Erwin}}{2015}]{2015ApJ...799..226E}
{Erwin} P.,  2015, \mn@doi [\apj] {10.1088/0004-637X/799/2/226}, \href
  {https://ui.adsabs.harvard.edu/abs/2015ApJ...799..226E} {799, 226}

\bibitem[\protect\citeauthoryear{{Falc{\'o}n-Barroso} \&
  {Knapen}}{{Falc{\'o}n-Barroso} \& {Knapen}}{2013}]{2013seg..book.....F}
{Falc{\'o}n-Barroso} J.,  {Knapen} J.~H.,  2013, {Secular Evolution of
  Galaxies}

\bibitem[\protect\citeauthoryear{{Falc{\'o}n-Barroso} \&
  {Peletier}}{{Falc{\'o}n-Barroso} \& {Peletier}}{2002}]{2002ASPC..282..216F}
{Falc{\'o}n-Barroso} J.,  {Peletier} R.~F.,  2002, {The Peculiar Galaxy NGC
  7332}.
p.~216

\bibitem[\protect\citeauthoryear{{Falc{\'o}n-Barroso}
  et~al.,}{{Falc{\'o}n-Barroso} et~al.}{2004a}]{2004AN....325...92F}
{Falc{\'o}n-Barroso} J.,  et~al., 2004a, \mn@doi [Astronomische Nachrichten]
  {10.1002/asna.200310182}, \href
  {https://ui.adsabs.harvard.edu/abs/2004AN....325...92F} {325, 92}

\bibitem[\protect\citeauthoryear{{Falc{\'o}n-Barroso}
  et~al.,}{{Falc{\'o}n-Barroso} et~al.}{2004b}]{2004MNRAS.350...35F}
{Falc{\'o}n-Barroso} J.,  et~al., 2004b, \mn@doi [\mnras]
  {10.1111/j.1365-2966.2004.07704.x}, \href
  {https://ui.adsabs.harvard.edu/abs/2004MNRAS.350...35F} {350, 35}

\bibitem[\protect\citeauthoryear{{Fazio} et~al.,}{{Fazio}
  et~al.}{2004}]{2004ApJS..154...10F}
{Fazio} G.~G.,  et~al., 2004, \mn@doi [\apjs] {10.1086/422843}, \href
  {https://ui.adsabs.harvard.edu/abs/2004ApJS..154...10F} {154, 10}

\bibitem[\protect\citeauthoryear{{Fisher} \& {Drory}}{{Fisher} \&
  {Drory}}{2011}]{2011ApJ...733L..47F}
{Fisher} D.~B.,  {Drory} N.,  2011, \mn@doi [\apjl]
  {10.1088/2041-8205/733/2/L47}, \href
  {https://ui.adsabs.harvard.edu/abs/2011ApJ...733L..47F} {733, L47}

\bibitem[\protect\citeauthoryear{{Fisher}, {Illingworth}  \& {Franx}}{{Fisher}
  et~al.}{1994}]{1994AJ....107..160F}
{Fisher} D.,  {Illingworth} G.,   {Franx} M.,  1994, \mn@doi [\aj]
  {10.1086/116841}, \href
  {https://ui.adsabs.harvard.edu/abs/1994AJ....107..160F} {107, 160}

\bibitem[\protect\citeauthoryear{{Font}, {McCarthy}, {Crain}, {Theuns},
  {Schaye}, {Wiersma}  \& {Dalla Vecchia}}{{Font}
  et~al.}{2011}]{2011MNRAS.416.2802F}
{Font} A.~S.,  {McCarthy} I.~G.,  {Crain} R.~A.,  {Theuns} T.,  {Schaye} J.,
  {Wiersma} R.~P.~C.,   {Dalla Vecchia} C.,  2011, \mn@doi [\mnras]
  {10.1111/j.1365-2966.2011.19227.x}, \href
  {https://ui.adsabs.harvard.edu/abs/2011MNRAS.416.2802F} {416, 2802}

\bibitem[\protect\citeauthoryear{{Forbes}, {Spitler}, {Harris}, {Bailin},
  {Strader}, {Brodie}  \& {Larsen}}{{Forbes}
  et~al.}{2010}]{2010MNRAS.403..429F}
{Forbes} D.~A.,  {Spitler} L.~R.,  {Harris} W.~E.,  {Bailin} J.,  {Strader} J.,
   {Brodie} J.~P.,   {Larsen} S.~S.,  2010, \mn@doi [\mnras]
  {10.1111/j.1365-2966.2009.16130.x}, \href
  {https://ui.adsabs.harvard.edu/abs/2010MNRAS.403..429F} {403, 429}

\bibitem[\protect\citeauthoryear{{Freeman}}{{Freeman}}{1970}]{1970ApJ...160..811F}
{Freeman} K.~C.,  1970, \mn@doi [\apj] {10.1086/150474}, \href
  {https://ui.adsabs.harvard.edu/abs/1970ApJ...160..811F} {160, 811}

\bibitem[\protect\citeauthoryear{{Gadotti} \& {S{\'a}nchez-Janssen}}{{Gadotti}
  \& {S{\'a}nchez-Janssen}}{2012}]{2012MNRAS.423..877G}
{Gadotti} D.~A.,  {S{\'a}nchez-Janssen} R.,  2012, \mn@doi [\mnras]
  {10.1111/j.1365-2966.2012.20925.x}, \href
  {https://ui.adsabs.harvard.edu/abs/2012MNRAS.423..877G} {423, 877}

\bibitem[\protect\citeauthoryear{{Garcia}}{{Garcia}}{1993}]{1993A&AS..100...47G}
{Garcia} A.~M.,  1993, \aaps, \href
  {https://ui.adsabs.harvard.edu/abs/1993A&AS..100...47G} {100, 47}

\bibitem[\protect\citeauthoryear{{Garcia-Burillo}, {Guelin}, {Cernicharo}  \&
  {Dahlem}}{{Garcia-Burillo} et~al.}{1992}]{1992A&A...266...21G}
{Garcia-Burillo} S.,  {Guelin} M.,  {Cernicharo} J.,   {Dahlem} M.,  1992,
  \aap, \href {https://ui.adsabs.harvard.edu/abs/1992A&A...266...21G} {266, 21}

\bibitem[\protect\citeauthoryear{{Gilhuly} et~al.,}{{Gilhuly}
  et~al.}{2019}]{2019arXiv191005358G}
{Gilhuly} C.,  et~al., 2019, arXiv e-prints, \href
  {https://ui.adsabs.harvard.edu/abs/2019arXiv191005358G} {p. arXiv:1910.05358}

\bibitem[\protect\citeauthoryear{{Gilmore} \& {Reid}}{{Gilmore} \&
  {Reid}}{1983}]{1983MNRAS.202.1025G}
{Gilmore} G.,  {Reid} N.,  1983, \mn@doi [\mnras] {10.1093/mnras/202.4.1025},
  \href {https://ui.adsabs.harvard.edu/abs/1983MNRAS.202.1025G} {202, 1025}

\bibitem[\protect\citeauthoryear{{Giuricin}, {Marinoni}, {Ceriani}  \&
  {Pisani}}{{Giuricin} et~al.}{2000}]{2000ApJ...543..178G}
{Giuricin} G.,  {Marinoni} C.,  {Ceriani} L.,   {Pisani} A.,  2000, \mn@doi
  [\apj] {10.1086/317070}, \href
  {https://ui.adsabs.harvard.edu/abs/2000ApJ...543..178G} {543, 178}

\bibitem[\protect\citeauthoryear{{Gonzalez} et~al.,}{{Gonzalez}
  et~al.}{2016}]{2016A&A...591A...7G}
{Gonzalez} O.~A.,  et~al., 2016, \mn@doi [\aap] {10.1051/0004-6361/201527806},
  \href {https://ui.adsabs.harvard.edu/abs/2016A&A...591A...7G} {591, A7}

\bibitem[\protect\citeauthoryear{{Graham}, {Spitler}, {Forbes}, {Lisker},
  {Moore}  \& {Janz}}{{Graham} et~al.}{2012}]{2012ApJ...750..121G}
{Graham} A.~W.,  {Spitler} L.~R.,  {Forbes} D.~A.,  {Lisker} T.,  {Moore} B.,
  {Janz} J.,  2012, \mn@doi [\apj] {10.1088/0004-637X/750/2/121}, \href
  {https://ui.adsabs.harvard.edu/abs/2012ApJ...750..121G} {750, 121}

\bibitem[\protect\citeauthoryear{{Gregory} \& {Thompson}}{{Gregory} \&
  {Thompson}}{1977}]{1977ApJ...213..345G}
{Gregory} S.~A.,  {Thompson} L.~A.,  1977, \mn@doi [\apj] {10.1086/155160},
  \href {https://ui.adsabs.harvard.edu/abs/1977ApJ...213..345G} {213, 345}

\bibitem[\protect\citeauthoryear{{Gu{\'e}rou}, {Emsellem}, {Krajnovi{\'c}},
  {McDermid}, {Contini}  \& {Weilbacher}}{{Gu{\'e}rou}
  et~al.}{2016}]{2016A&A...591A.143G}
{Gu{\'e}rou} A.,  {Emsellem} E.,  {Krajnovi{\'c}} D.,  {McDermid} R.~M.,
  {Contini} T.,   {Weilbacher} P.~M.,  2016, \mn@doi [\aap]
  {10.1051/0004-6361/201628743}, \href
  {https://ui.adsabs.harvard.edu/abs/2016A&A...591A.143G} {591, A143}

\bibitem[\protect\citeauthoryear{{Gunn} \& {Gott}}{{Gunn} \&
  {Gott}}{1972}]{1972ApJ...176....1G}
{Gunn} J.~E.,  {Gott} J.~Richard I.,  1972, \mn@doi [\apj] {10.1086/151605},
  \href {https://ui.adsabs.harvard.edu/abs/1972ApJ...176....1G} {176, 1}

\bibitem[\protect\citeauthoryear{{Harris}}{{Harris}}{1996}]{1996AJ....112.1487H}
{Harris} W.~E.,  1996, \mn@doi [\aj] {10.1086/118116}, \href
  {https://ui.adsabs.harvard.edu/abs/1996AJ....112.1487H} {112, 1487}

\bibitem[\protect\citeauthoryear{{Helmi} \& {White}}{{Helmi} \&
  {White}}{1999}]{1999MNRAS.307..495H}
{Helmi} A.,  {White} S. D.~M.,  1999, \mn@doi [\mnras]
  {10.1046/j.1365-8711.1999.02616.x}, \href
  {https://ui.adsabs.harvard.edu/abs/1999MNRAS.307..495H} {307, 495}

\bibitem[\protect\citeauthoryear{{Holmberg}}{{Holmberg}}{1958}]{1958MeLuS.136....1H}
{Holmberg} E.,  1958, Meddelanden fran Lunds Astronomiska Observatorium Serie
  II, \href {https://ui.adsabs.harvard.edu/abs/1958MeLuS.136....1H} {136, 1}

\bibitem[\protect\citeauthoryear{{Hopkins}, {Cox}, {Younger}  \&
  {Hernquist}}{{Hopkins} et~al.}{2009}]{2009ApJ...691.1168H}
{Hopkins} P.~F.,  {Cox} T.~J.,  {Younger} J.~D.,   {Hernquist} L.,  2009,
  \mn@doi [\apj] {10.1088/0004-637X/691/2/1168}, \href
  {https://ui.adsabs.harvard.edu/abs/2009ApJ...691.1168H} {691, 1168}

\bibitem[\protect\citeauthoryear{{Howk} \& {Savage}}{{Howk} \&
  {Savage}}{1999}]{1999AJ....117.2077H}
{Howk} J.~C.,  {Savage} B.~D.,  1999, \mn@doi [\aj] {10.1086/300857}, \href
  {https://ui.adsabs.harvard.edu/abs/1999AJ....117.2077H} {117, 2077}

\bibitem[\protect\citeauthoryear{{Il'ina} \& {Sil'chenko}}{{Il'ina} \&
  {Sil'chenko}}{2012}]{2012ARep...56..578I}
{Il'ina} M.~A.,  {Sil'chenko} O.~K.,  2012, \mn@doi [Astronomy Reports]
  {10.1134/S1063772912080045}, \href
  {https://ui.adsabs.harvard.edu/abs/2012ARep...56..578I} {56, 578}

\bibitem[\protect\citeauthoryear{{Jedrzejewski}}{{Jedrzejewski}}{1987}]{1987MNRAS.226..747J}
{Jedrzejewski} R.~I.,  1987, \mn@doi [\mnras] {10.1093/mnras/226.4.747}, \href
  {https://ui.adsabs.harvard.edu/abs/1987MNRAS.226..747J} {226, 747}

\bibitem[\protect\citeauthoryear{{Jennings} et~al.,}{{Jennings}
  et~al.}{2015}]{2015ApJ...812L..10J}
{Jennings} Z.~G.,  et~al., 2015, \mn@doi [\apjl] {10.1088/2041-8205/812/1/L10},
  \href {https://ui.adsabs.harvard.edu/abs/2015ApJ...812L..10J} {812, L10}

\bibitem[\protect\citeauthoryear{{Juri{\'c}} et~al.,}{{Juri{\'c}}
  et~al.}{2008}]{2008ApJ...673..864J}
{Juri{\'c}} M.,  et~al., 2008, \mn@doi [\apj] {10.1086/523619}, \href
  {https://ui.adsabs.harvard.edu/abs/2008ApJ...673..864J} {673, 864}

\bibitem[\protect\citeauthoryear{{Karabal}, {Duc}, {Kuntschner}, {Chanial},
  {Cuillandre}  \& {Gwyn}}{{Karabal} et~al.}{2017}]{2017A&A...601A..86K}
{Karabal} E.,  {Duc} P.~A.,  {Kuntschner} H.,  {Chanial} P.,  {Cuillandre}
  J.~C.,   {Gwyn} S.,  2017, \mn@doi [\aap] {10.1051/0004-6361/201629974},
  \href {https://ui.adsabs.harvard.edu/abs/2017A&A...601A..86K} {601, A86}

\bibitem[\protect\citeauthoryear{{Karachentsev}}{{Karachentsev}}{1987}]{1987MoIzN....T....K}
{Karachentsev} I.,  1987, Moscow Izdatel Nauka, \href
  {https://ui.adsabs.harvard.edu/abs/1987MoIzN....T....K} {}

\bibitem[\protect\citeauthoryear{{Karachentsev}, {Tully}, {Wu}, {Shaya}  \&
  {Dolphin}}{{Karachentsev} et~al.}{2014}]{2014ApJ...782....4K}
{Karachentsev} I.~D.,  {Tully} R.~B.,  {Wu} P.-F.,  {Shaya} E.~J.,   {Dolphin}
  A.~E.,  2014, \mn@doi [\apj] {10.1088/0004-637X/782/1/4}, \href
  {https://ui.adsabs.harvard.edu/abs/2014ApJ...782....4K} {782, 4}

\bibitem[\protect\citeauthoryear{{Karademir}, {Remus}, {Burkert}, {Dolag},
  {Hoffmann}, {Moster}, {Steinwandel}  \& {Zhang}}{{Karademir}
  et~al.}{2019}]{2019MNRAS.487..318K}
{Karademir} G.~S.,  {Remus} R.-S.,  {Burkert} A.,  {Dolag} K.,  {Hoffmann}
  T.~L.,  {Moster} B.~P.,  {Steinwandel} U.~P.,   {Zhang} J.,  2019, \mn@doi
  [\mnras] {10.1093/mnras/stz1251}, \href
  {https://ui.adsabs.harvard.edu/abs/2019MNRAS.487..318K} {487, 318}

\bibitem[\protect\citeauthoryear{{Kasparova}, {Katkov}, {Chilingarian},
  {Silchenko}, {Moiseev}  \& {Borisov}}{{Kasparova}
  et~al.}{2016}]{2016MNRAS.460L..89K}
{Kasparova} A.~V.,  {Katkov} I.~Y.,  {Chilingarian} I.~V.,  {Silchenko} O.~K.,
  {Moiseev} A.~V.,   {Borisov} S.~B.,  2016, \mn@doi [\mnras]
  {10.1093/mnrasl/slw083}, \href
  {https://ui.adsabs.harvard.edu/abs/2016MNRAS.460L..89K} {460, L89}

\bibitem[\protect\citeauthoryear{{Kautsch}, {Grebel}, {Barazza}  \&
  {Gallagher}}{{Kautsch} et~al.}{2006}]{2006A&A...445..765K}
{Kautsch} S.~J.,  {Grebel} E.~K.,  {Barazza} F.~D.,   {Gallagher} J.~S. I.,
  2006, \mn@doi [\aap] {10.1051/0004-6361:20053981}, \href
  {https://ui.adsabs.harvard.edu/abs/2006A&A...445..765K} {445, 765}

\bibitem[\protect\citeauthoryear{{Kennedy} et~al.,}{{Kennedy}
  et~al.}{2016}]{2016MNRAS.460.3458K}
{Kennedy} R.,  et~al., 2016, \mn@doi [\mnras] {10.1093/mnras/stw1176}, \href
  {https://ui.adsabs.harvard.edu/abs/2016MNRAS.460.3458K} {460, 3458}

\bibitem[\protect\citeauthoryear{{Kodaira} \& {Yamashita}}{{Kodaira} \&
  {Yamashita}}{1996}]{1996PASJ...48..581K}
{Kodaira} K.,  {Yamashita} T.,  1996, \mn@doi [\pasj] {10.1093/pasj/48.4.581},
  \href {https://ui.adsabs.harvard.edu/abs/1996PASJ...48..581K} {48, 581}

\bibitem[\protect\citeauthoryear{{Kormendy}}{{Kormendy}}{2013}]{2013seg..book....1K}
{Kormendy} J.,  2013, {Secular Evolution in Disk Galaxies}.
p.~1

\bibitem[\protect\citeauthoryear{{Kormendy} \& {Barentine}}{{Kormendy} \&
  {Barentine}}{2010}]{2010ApJ...715L.176K}
{Kormendy} J.,  {Barentine} J.~C.,  2010, \mn@doi [\apjl]
  {10.1088/2041-8205/715/2/L176}, \href
  {https://ui.adsabs.harvard.edu/abs/2010ApJ...715L.176K} {715, L176}

\bibitem[\protect\citeauthoryear{{Kormendy} \& {Bender}}{{Kormendy} \&
  {Bender}}{2012}]{2012ApJS..198....2K}
{Kormendy} J.,  {Bender} R.,  2012, \mn@doi [\apjs]
  {10.1088/0067-0049/198/1/2}, \href
  {https://ui.adsabs.harvard.edu/abs/2012ApJS..198....2K} {198, 2}

\bibitem[\protect\citeauthoryear{{Kormendy} \& {Kennicutt}}{{Kormendy} \&
  {Kennicutt}}{2004}]{2004ARA&A..42..603K}
{Kormendy} J.,  {Kennicutt} Robert~C. J.,  2004, \mn@doi [\araa]
  {10.1146/annurev.astro.42.053102.134024}, \href
  {https://ui.adsabs.harvard.edu/abs/2004ARA&A..42..603K} {42, 603}

\bibitem[\protect\citeauthoryear{{Kormendy} \& {Norman}}{{Kormendy} \&
  {Norman}}{1979}]{1979ApJ...233..539K}
{Kormendy} J.,  {Norman} C.~A.,  1979, \mn@doi [\apj] {10.1086/157414}, \href
  {https://ui.adsabs.harvard.edu/abs/1979ApJ...233..539K} {233, 539}

\bibitem[\protect\citeauthoryear{{Kraan-Korteweg} \&
  {Tammann}}{{Kraan-Korteweg} \& {Tammann}}{1979}]{1979AN....300..181K}
{Kraan-Korteweg} R.~C.,  {Tammann} G.~A.,  1979, \mn@doi [Astronomische
  Nachrichten] {10.1002/asna.19793000403}, \href
  {https://ui.adsabs.harvard.edu/abs/1979AN....300..181K} {300, 181}

\bibitem[\protect\citeauthoryear{{Kregel}, {van der Kruit}  \& {de
  Grijs}}{{Kregel} et~al.}{2002}]{2002MNRAS.334..646K}
{Kregel} M.,  {van der Kruit} P.~C.,   {de Grijs} R.,  2002, \mn@doi [\mnras]
  {10.1046/j.1365-8711.2002.05556.x}, \href
  {https://ui.adsabs.harvard.edu/abs/2002MNRAS.334..646K} {334, 646}

\bibitem[\protect\citeauthoryear{{Kuzio de Naray}, {Zagursky}  \&
  {McGaugh}}{{Kuzio de Naray} et~al.}{2009}]{2009AJ....138.1082K}
{Kuzio de Naray} R.,  {Zagursky} M.~J.,   {McGaugh} S.~S.,  2009, \mn@doi [\aj]
  {10.1088/0004-6256/138/4/1082}, \href
  {https://ui.adsabs.harvard.edu/abs/2009AJ....138.1082K} {138, 1082}

\bibitem[\protect\citeauthoryear{{Laurikainen} \& {Salo}}{{Laurikainen} \&
  {Salo}}{2017}]{2017A&A...598A..10L}
{Laurikainen} E.,  {Salo} H.,  2017, \mn@doi [\aap]
  {10.1051/0004-6361/201628936}, \href
  {https://ui.adsabs.harvard.edu/abs/2017A&A...598A..10L} {598, A10}

\bibitem[\protect\citeauthoryear{{Laurikainen}, {Salo}, {Laine}  \&
  {Janz}}{{Laurikainen} et~al.}{2018}]{2018A&A...618A..34L}
{Laurikainen} E.,  {Salo} H.,  {Laine} J.,   {Janz} J.,  2018, \mn@doi [\aap]
  {10.1051/0004-6361/201833138}, \href
  {https://ui.adsabs.harvard.edu/abs/2018A&A...618A..34L} {618, A34}

\bibitem[\protect\citeauthoryear{{Lee} \& {Irwin}}{{Lee} \&
  {Irwin}}{1997}]{1997ApJ...490..247L}
{Lee} S.-W.,  {Irwin} J.~A.,  1997, \mn@doi [\apj] {10.1086/304840}, \href
  {https://ui.adsabs.harvard.edu/abs/1997ApJ...490..247L} {490, 247}

\bibitem[\protect\citeauthoryear{{Li} \& {Wang}}{{Li} \&
  {Wang}}{2013}]{2013MNRAS.428.2085L}
{Li} J.-T.,  {Wang} Q.~D.,  2013, \mn@doi [\mnras] {10.1093/mnras/sts183},
  \href {https://ui.adsabs.harvard.edu/abs/2013MNRAS.428.2085L} {428, 2085}

\bibitem[\protect\citeauthoryear{{Li}, {Wang}, {Li}  \& {Chen}}{{Li}
  et~al.}{2011}]{2011ApJ...737...41L}
{Li} J.-T.,  {Wang} Q.~D.,  {Li} Z.,   {Chen} Y.,  2011, \mn@doi [\apj]
  {10.1088/0004-637X/737/1/41}, \href
  {https://ui.adsabs.harvard.edu/abs/2011ApJ...737...41L} {737, 41}

\bibitem[\protect\citeauthoryear{{L{\'o}pez-Corredoira} \&
  {Molg{\'o}}}{{L{\'o}pez-Corredoira} \&
  {Molg{\'o}}}{2014}]{2014A&A...567A.106L}
{L{\'o}pez-Corredoira} M.,  {Molg{\'o}} J.,  2014, \mn@doi [\aap]
  {10.1051/0004-6361/201423706}, \href
  {https://ui.adsabs.harvard.edu/abs/2014A&A...567A.106L} {567, A106}

\bibitem[\protect\citeauthoryear{{L{\"u}tticke}, {Dettmar}  \&
  {Pohlen}}{{L{\"u}tticke} et~al.}{2000}]{2000A&A...362..435L}
{L{\"u}tticke} R.,  {Dettmar} R.~J.,   {Pohlen} M.,  2000, \aap, \href
  {https://ui.adsabs.harvard.edu/abs/2000A&A...362..435L} {362, 435}

\bibitem[\protect\citeauthoryear{{L{\"u}tticke}, {Pohlen}  \&
  {Dettmar}}{{L{\"u}tticke} et~al.}{2004}]{2004A&A...417..527L}
{L{\"u}tticke} R.,  {Pohlen} M.,   {Dettmar} R.~J.,  2004, \mn@doi [\aap]
  {10.1051/0004-6361:20031782}, \href
  {https://ui.adsabs.harvard.edu/abs/2004A&A...417..527L} {417, 527}

\bibitem[\protect\citeauthoryear{{Lynds} \& {Sandage}}{{Lynds} \&
  {Sandage}}{1963}]{1963ApJ...137.1005L}
{Lynds} C.~R.,  {Sandage} A.~R.,  1963, \mn@doi [\apj] {10.1086/147579}, \href
  {https://ui.adsabs.harvard.edu/abs/1963ApJ...137.1005L} {137, 1005}

\bibitem[\protect\citeauthoryear{{Macri}, {Stetson}, {Bothun}, {Freedman},
  {Garnavich}, {Jha}, {Madore}  \& {Richmond}}{{Macri}
  et~al.}{2001}]{2001ApJ...559..243M}
{Macri} L.~M.,  {Stetson} P.~B.,  {Bothun} G.~D.,  {Freedman} W.~L.,
  {Garnavich} P.~M.,  {Jha} S.,  {Madore} B.~F.,   {Richmond} M.~W.,  2001,
  \mn@doi [\apj] {10.1086/322395}, \href
  {https://ui.adsabs.harvard.edu/abs/2001ApJ...559..243M} {559, 243}

\bibitem[\protect\citeauthoryear{{Makarov}, {Prugniel}, {Terekhova}, {Courtois}
   \& {Vauglin}}{{Makarov} et~al.}{2014}]{2014A&A...570A..13M}
{Makarov} D.,  {Prugniel} P.,  {Terekhova} N.,  {Courtois} H.,   {Vauglin} I.,
  2014, \mn@doi [\aap] {10.1051/0004-6361/201423496}, \href
  {https://ui.adsabs.harvard.edu/abs/2014A&A...570A..13M} {570, A13}

\bibitem[\protect\citeauthoryear{{Malin} \& {Hadley}}{{Malin} \&
  {Hadley}}{1997}]{1997PASA...14...52M}
{Malin} D.,  {Hadley} B.,  1997, \mn@doi [\pasa] {10.1071/AS97052}, \href
  {https://ui.adsabs.harvard.edu/abs/1997PASA...14...52M} {14, 52}

\bibitem[\protect\citeauthoryear{{Marasco} et~al.,}{{Marasco}
  et~al.}{2019}]{2019A&A...631A..50M}
{Marasco} A.,  et~al., 2019, \mn@doi [\aap] {10.1051/0004-6361/201936338},
  \href {https://ui.adsabs.harvard.edu/abs/2019A&A...631A..50M} {631, A50}

\bibitem[\protect\citeauthoryear{{Mart{\'\i}nez-Delgado}, {Pe{\~n}arrubia},
  {Gabany}, {Trujillo}, {Majewski}  \& {Pohlen}}{{Mart{\'\i}nez-Delgado}
  et~al.}{2008}]{2008ApJ...689..184M}
{Mart{\'\i}nez-Delgado} D.,  {Pe{\~n}arrubia} J.,  {Gabany} R.~J.,  {Trujillo}
  I.,  {Majewski} S.~R.,   {Pohlen} M.,  2008, \mn@doi [\apj] {10.1086/592555},
  \href {https://ui.adsabs.harvard.edu/abs/2008ApJ...689..184M} {689, 184}

\bibitem[\protect\citeauthoryear{{Mart{\'\i}nez-Delgado}, {Pohlen}, {Gabany},
  {Majewski}, {Pe{\~n}arrubia}  \& {Palma}}{{Mart{\'\i}nez-Delgado}
  et~al.}{2009}]{2009ApJ...692..955M}
{Mart{\'\i}nez-Delgado} D.,  {Pohlen} M.,  {Gabany} R.~J.,  {Majewski} S.~R.,
  {Pe{\~n}arrubia} J.,   {Palma} C.,  2009, \mn@doi [\apj]
  {10.1088/0004-637X/692/2/955}, \href
  {https://ui.adsabs.harvard.edu/abs/2009ApJ...692..955M} {692, 955}

\bibitem[\protect\citeauthoryear{{Mart{\'\i}nez-Delgado}
  et~al.,}{{Mart{\'\i}nez-Delgado} et~al.}{2010}]{2010AJ....140..962M}
{Mart{\'\i}nez-Delgado} D.,  et~al., 2010, \mn@doi [\aj]
  {10.1088/0004-6256/140/4/962}, \href
  {https://ui.adsabs.harvard.edu/abs/2010AJ....140..962M} {140, 962}

\bibitem[\protect\citeauthoryear{{Mart{\'\i}nez-Delgado}, {D'Onghia}, {Chonis},
  {Beaton}, {Teuwen}, {GaBany}, {Grebel}  \& {Morales}}{{Mart{\'\i}nez-Delgado}
  et~al.}{2015}]{2015AJ....150..116M}
{Mart{\'\i}nez-Delgado} D.,  {D'Onghia} E.,  {Chonis} T.~S.,  {Beaton} R.~L.,
  {Teuwen} K.,  {GaBany} R.~J.,  {Grebel} E.~K.,   {Morales} G.,  2015, \mn@doi
  [\aj] {10.1088/0004-6256/150/4/116}, \href
  {https://ui.adsabs.harvard.edu/abs/2015AJ....150..116M} {150, 116}

\bibitem[\protect\citeauthoryear{{Mart{\'\i}nez-Lombilla}, {Trujillo}  \&
  {Knapen}}{{Mart{\'\i}nez-Lombilla} et~al.}{2019}]{2019MNRAS.483..664M}
{Mart{\'\i}nez-Lombilla} C.,  {Trujillo} I.,   {Knapen} J.~H.,  2019, \mn@doi
  [\mnras] {10.1093/mnras/sty2886}, \href
  {https://ui.adsabs.harvard.edu/abs/2019MNRAS.483..664M} {483, 664}

\bibitem[\protect\citeauthoryear{{Mayya}, {Carrasco}  \& {Luna}}{{Mayya}
  et~al.}{2005}]{2005ApJ...628L..33M}
{Mayya} Y.~D.,  {Carrasco} L.,   {Luna} A.,  2005, \mn@doi [\apjl]
  {10.1086/432644}, \href
  {https://ui.adsabs.harvard.edu/abs/2005ApJ...628L..33M} {628, L33}

\bibitem[\protect\citeauthoryear{{McCarthy}, {Font}, {Crain}, {Deason},
  {Schaye}  \& {Theuns}}{{McCarthy} et~al.}{2012}]{2012MNRAS.420.2245M}
{McCarthy} I.~G.,  {Font} A.~S.,  {Crain} R.~A.,  {Deason} A.~J.,  {Schaye} J.,
    {Theuns} T.,  2012, \mn@doi [\mnras] {10.1111/j.1365-2966.2011.20189.x},
  \href {https://ui.adsabs.harvard.edu/abs/2012MNRAS.420.2245M} {420, 2245}

\bibitem[\protect\citeauthoryear{{Mei} et~al.,}{{Mei}
  et~al.}{2007}]{2007ApJ...655..144M}
{Mei} S.,  et~al., 2007, \mn@doi [\apj] {10.1086/509598}, \href
  {https://ui.adsabs.harvard.edu/abs/2007ApJ...655..144M} {655, 144}

\bibitem[\protect\citeauthoryear{{Michard}}{{Michard}}{2007}]{2007A&A...464..507M}
{Michard} R.,  2007, \mn@doi [\aap] {10.1051/0004-6361:20066358}, \href
  {https://ui.adsabs.harvard.edu/abs/2007A&A...464..507M} {464, 507}

\bibitem[\protect\citeauthoryear{{Miskolczi}, {Bomans}  \&
  {Dettmar}}{{Miskolczi} et~al.}{2011}]{2011A&A...536A..66M}
{Miskolczi} A.,  {Bomans} D.~J.,   {Dettmar} R.~J.,  2011, \mn@doi [\aap]
  {10.1051/0004-6361/201116716}, \href
  {https://ui.adsabs.harvard.edu/abs/2011A&A...536A..66M} {536, A66}

\bibitem[\protect\citeauthoryear{{Morales}, {Mart{\'\i}nez-Delgado}, {Grebel},
  {Cooper}, {Javanmardi}  \& {Miskolczi}}{{Morales}
  et~al.}{2018}]{2018A&A...614A.143M}
{Morales} G.,  {Mart{\'\i}nez-Delgado} D.,  {Grebel} E.~K.,  {Cooper} A.~P.,
  {Javanmardi} B.,   {Miskolczi} A.,  2018, \mn@doi [\aap]
  {10.1051/0004-6361/201732271}, \href
  {https://ui.adsabs.harvard.edu/abs/2018A&A...614A.143M} {614, A143}

\bibitem[\protect\citeauthoryear{{Morrison}}{{Morrison}}{1993}]{1993AJ....106..578M}
{Morrison} H.~L.,  1993, \mn@doi [\aj] {10.1086/116662}, \href
  {https://ui.adsabs.harvard.edu/abs/1993AJ....106..578M} {106, 578}

\bibitem[\protect\citeauthoryear{{Mosenkov}, {Sotnikova}  \&
  {Reshetnikov}}{{Mosenkov} et~al.}{2010}]{2010MNRAS.401..559M}
{Mosenkov} A.~V.,  {Sotnikova} N.~Y.,   {Reshetnikov} V.~P.,  2010, \mn@doi
  [\mnras] {10.1111/j.1365-2966.2009.15671.x}, \href
  {https://ui.adsabs.harvard.edu/abs/2010MNRAS.401..559M} {401, 559}

\bibitem[\protect\citeauthoryear{{Mosenkov}, {Sotnikova}, {Reshetnikov},
  {Bizyaev}  \& {Kautsch}}{{Mosenkov} et~al.}{2015}]{2015MNRAS.451.2376M}
{Mosenkov} A.~V.,  {Sotnikova} N.~Y.,  {Reshetnikov} V.~P.,  {Bizyaev} D.~V.,
  {Kautsch} S.~J.,  2015, \mn@doi [\mnras] {10.1093/mnras/stv1085}, \href
  {https://ui.adsabs.harvard.edu/abs/2015MNRAS.451.2376M} {451, 2376}

\bibitem[\protect\citeauthoryear{{Mosenkov} et~al.,}{{Mosenkov}
  et~al.}{2018}]{2018A&A...616A.120M}
{Mosenkov} A.~V.,  et~al., 2018, \mn@doi [\aap] {10.1051/0004-6361/201832899},
  \href {https://ui.adsabs.harvard.edu/abs/2018A&A...616A.120M} {616, A120}

\bibitem[\protect\citeauthoryear{{Moster}, {Macci{\`o}}, {Somerville},
  {Johansson}  \& {Naab}}{{Moster} et~al.}{2010}]{2010MNRAS.403.1009M}
{Moster} B.~P.,  {Macci{\`o}} A.~V.,  {Somerville} R.~S.,  {Johansson} P.~H.,
  {Naab} T.,  2010, \mn@doi [\mnras] {10.1111/j.1365-2966.2009.16190.x}, \href
  {https://ui.adsabs.harvard.edu/abs/2010MNRAS.403.1009M} {403, 1009}

\bibitem[\protect\citeauthoryear{{Mouhcine}, {Ibata}  \& {Rejkuba}}{{Mouhcine}
  et~al.}{2010}]{2010ApJ...714L..12M}
{Mouhcine} M.,  {Ibata} R.,   {Rejkuba} M.,  2010, \mn@doi [\apjl]
  {10.1088/2041-8205/714/1/L12}, \href
  {https://ui.adsabs.harvard.edu/abs/2010ApJ...714L..12M} {714, L12}

\bibitem[\protect\citeauthoryear{{Muldrew} et~al.,}{{Muldrew}
  et~al.}{2012}]{2012MNRAS.419.2670M}
{Muldrew} S.~I.,  et~al., 2012, \mn@doi [\mnras]
  {10.1111/j.1365-2966.2011.19922.x}, \href
  {https://ui.adsabs.harvard.edu/abs/2012MNRAS.419.2670M} {419, 2670}

\bibitem[\protect\citeauthoryear{{M{\"u}ller}, {Vudragovi{\'c}}  \&
  {B{\'\i}lek}}{{M{\"u}ller} et~al.}{2019}]{2019A&A...632L..13M}
{M{\"u}ller} O.,  {Vudragovi{\'c}} A.,   {B{\'\i}lek} M.,  2019, \mn@doi [\aap]
  {10.1051/0004-6361/201937077}, \href
  {https://ui.adsabs.harvard.edu/abs/2019A&A...632L..13M} {632, L13}

\bibitem[\protect\citeauthoryear{{Nieto}, {Bender}, {Arnaud}  \&
  {Surma}}{{Nieto} et~al.}{1991}]{1991A&A...244L..25N}
{Nieto} J.~L.,  {Bender} R.,  {Arnaud} J.,   {Surma} P.,  1991, \aap, \href
  {https://ui.adsabs.harvard.edu/abs/1991A&A...244L..25N} {244, L25}

\bibitem[\protect\citeauthoryear{{Oosterloo}, {Fraternali}  \&
  {Sancisi}}{{Oosterloo} et~al.}{2007}]{2007AJ....134.1019O}
{Oosterloo} T.,  {Fraternali} F.,   {Sancisi} R.,  2007, \mn@doi [\aj]
  {10.1086/520332}, \href
  {https://ui.adsabs.harvard.edu/abs/2007AJ....134.1019O} {134, 1019}

\bibitem[\protect\citeauthoryear{{Patterson}}{{Patterson}}{1940}]{1940BHarO.914....9P}
{Patterson} F.~S.,  1940, Harvard College Observatory Bulletin, \href
  {https://ui.adsabs.harvard.edu/abs/1940BHarO.914....9P} {914, 9}

\bibitem[\protect\citeauthoryear{{Paudel} et~al.,}{{Paudel}
  et~al.}{2013}]{2013ApJ...767..133P}
{Paudel} S.,  et~al., 2013, \mn@doi [\apj] {10.1088/0004-637X/767/2/133}, \href
  {https://ui.adsabs.harvard.edu/abs/2013ApJ...767..133P} {767, 133}

\bibitem[\protect\citeauthoryear{{Peng}, {Ho}, {Impey}  \& {Rix}}{{Peng}
  et~al.}{2010}]{2010AJ....139.2097P}
{Peng} C.~Y.,  {Ho} L.~C.,  {Impey} C.~D.,   {Rix} H.-W.,  2010, \mn@doi [\aj]
  {10.1088/0004-6256/139/6/2097}, \href
  {https://ui.adsabs.harvard.edu/abs/2010AJ....139.2097P} {139, 2097}

\bibitem[\protect\citeauthoryear{{Plana} \& {Boulesteix}}{{Plana} \&
  {Boulesteix}}{1996}]{1996A&A...307..391P}
{Plana} H.,  {Boulesteix} J.,  1996, \aap, \href
  {https://ui.adsabs.harvard.edu/abs/1996A&A...307..391P} {307, 391}

\bibitem[\protect\citeauthoryear{{Pohlen}, {Dettmar}  \&
  {L{\"u}tticke}}{{Pohlen} et~al.}{2000}]{2000A&A...357L...1P}
{Pohlen} M.,  {Dettmar} R.~J.,   {L{\"u}tticke} R.,  2000, \aap, \href
  {https://ui.adsabs.harvard.edu/abs/2000A&A...357L...1P} {357, L1}

\bibitem[\protect\citeauthoryear{{Putman}, {Peek}  \& {Joung}}{{Putman}
  et~al.}{2012}]{2012ARA&A..50..491P}
{Putman} M.~E.,  {Peek} J.~E.~G.,   {Joung} M.~R.,  2012, \mn@doi [\araa]
  {10.1146/annurev-astro-081811-125612}, \href
  {https://ui.adsabs.harvard.edu/abs/2012ARA&A..50..491P} {50, 491}

\bibitem[\protect\citeauthoryear{{Qu}, {Di Matteo}, {Lehnert}  \& {van
  Driel}}{{Qu} et~al.}{2011}]{2011A&A...530A..10Q}
{Qu} Y.,  {Di Matteo} P.,  {Lehnert} M.~D.,   {van Driel} W.,  2011, \mn@doi
  [\aap] {10.1051/0004-6361/201015224}, \href
  {https://ui.adsabs.harvard.edu/abs/2011A&A...530A..10Q} {530, A10}

\bibitem[\protect\citeauthoryear{{Raha}, {Sellwood}, {James}  \& {Kahn}}{{Raha}
  et~al.}{1991}]{1991Natur.352..411R}
{Raha} N.,  {Sellwood} J.~A.,  {James} R.~A.,   {Kahn} F.~D.,  1991, \mn@doi
  [\nat] {10.1038/352411a0}, \href
  {https://ui.adsabs.harvard.edu/abs/1991Natur.352..411R} {352, 411}

\bibitem[\protect\citeauthoryear{{Rand}}{{Rand}}{1996}]{1996ApJ...462..712R}
{Rand} R.~J.,  1996, \mn@doi [\apj] {10.1086/177184}, \href
  {https://ui.adsabs.harvard.edu/abs/1996ApJ...462..712R} {462, 712}

\bibitem[\protect\citeauthoryear{{Rasmussen}, {Sommer-Larsen}, {Pedersen},
  {Toft}, {Benson}, {Bower}  \& {Grove}}{{Rasmussen}
  et~al.}{2009}]{2009ApJ...697...79R}
{Rasmussen} J.,  {Sommer-Larsen} J.,  {Pedersen} K.,  {Toft} S.,  {Benson} A.,
  {Bower} R.~G.,   {Grove} L.~F.,  2009, \mn@doi [\apj]
  {10.1088/0004-637X/697/1/79}, \href
  {https://ui.adsabs.harvard.edu/abs/2009ApJ...697...79R} {697, 79}

\bibitem[\protect\citeauthoryear{{Read}, {Lake}, {Agertz}  \&
  {Debattista}}{{Read} et~al.}{2008}]{2008MNRAS.389.1041R}
{Read} J.~I.,  {Lake} G.,  {Agertz} O.,   {Debattista} V.~P.,  2008, \mn@doi
  [\mnras] {10.1111/j.1365-2966.2008.13643.x}, \href
  {https://ui.adsabs.harvard.edu/abs/2008MNRAS.389.1041R} {389, 1041}

\bibitem[\protect\citeauthoryear{{Reshetnikov} \& {Combes}}{{Reshetnikov} \&
  {Combes}}{1998}]{1998A&A...337....9R}
{Reshetnikov} V.,  {Combes} F.,  1998, \aap, \href
  {https://ui.adsabs.harvard.edu/abs/1998A&A...337....9R} {337, 9}

\bibitem[\protect\citeauthoryear{{Reshetnikov} \& {Mosenkov}}{{Reshetnikov} \&
  {Mosenkov}}{2019}]{2019MNRAS.483.1470R}
{Reshetnikov} V.~P.,  {Mosenkov} A.~V.,  2019, \mn@doi [\mnras]
  {10.1093/mnras/sty3209}, \href
  {https://ui.adsabs.harvard.edu/abs/2019MNRAS.483.1470R} {483, 1470}

\bibitem[\protect\citeauthoryear{{Reshetnikov} \& {Sotnikova}}{{Reshetnikov} \&
  {Sotnikova}}{1997}]{1997A&A...325..933R}
{Reshetnikov} V.,  {Sotnikova} N.,  1997, \aap, \href
  {https://ui.adsabs.harvard.edu/abs/1997A&A...325..933R} {325, 933}

\bibitem[\protect\citeauthoryear{{Reshetnikov}, {Savchenko}, {Mosenkov},
  {Sotnikova}  \& {Bizyaev}}{{Reshetnikov} et~al.}{2015}]{2015AstL...41..748R}
{Reshetnikov} V.~P.,  {Savchenko} S.~S.,  {Mosenkov} A.~V.,  {Sotnikova} N.~Y.,
    {Bizyaev} D.~V.,  2015, \mn@doi [Astronomy Letters]
  {10.1134/S1063773715120117}, \href
  {https://ui.adsabs.harvard.edu/abs/2015AstL...41..748R} {41, 748}

\bibitem[\protect\citeauthoryear{{Reshetnikov}, {Mosenkov}, {Moiseev}, {Kotov}
  \& {Savchenko}}{{Reshetnikov} et~al.}{2016}]{2016MNRAS.461.4233R}
{Reshetnikov} V.~P.,  {Mosenkov} A.~V.,  {Moiseev} A.~V.,  {Kotov} S.~S.,
  {Savchenko} S.~S.,  2016, \mn@doi [\mnras] {10.1093/mnras/stw1623}, \href
  {https://ui.adsabs.harvard.edu/abs/2016MNRAS.461.4233R} {461, 4233}

\bibitem[\protect\citeauthoryear{{Rich} et~al.,}{{Rich}
  et~al.}{2019}]{2019MNRAS.490.1539R}
{Rich} R.~M.,  et~al., 2019, \mn@doi [\mnras] {10.1093/mnras/stz2106}, \href
  {https://ui.adsabs.harvard.edu/abs/2019MNRAS.490.1539R} {490, 1539}

\bibitem[\protect\citeauthoryear{{Rodriguez-Gomez} et~al.,}{{Rodriguez-Gomez}
  et~al.}{2016}]{2016MNRAS.458.2371R}
{Rodriguez-Gomez} V.,  et~al., 2016, \mn@doi [\mnras] {10.1093/mnras/stw456},
  \href {https://ui.adsabs.harvard.edu/abs/2016MNRAS.458.2371R} {458, 2371}

\bibitem[\protect\citeauthoryear{{Rots}}{{Rots}}{1978}]{1978AJ.....83..219R}
{Rots} A.~H.,  1978, \mn@doi [\aj] {10.1086/112195}, \href
  {https://ui.adsabs.harvard.edu/abs/1978AJ.....83..219R} {83, 219}

\bibitem[\protect\citeauthoryear{{Ro{\v{s}}kar}, {Debattista}  \&
  {Loebman}}{{Ro{\v{s}}kar} et~al.}{2013}]{2013MNRAS.433..976R}
{Ro{\v{s}}kar} R.,  {Debattista} V.~P.,   {Loebman} S.~R.,  2013, \mn@doi
  [\mnras] {10.1093/mnras/stt788}, \href
  {https://ui.adsabs.harvard.edu/abs/2013MNRAS.433..976R} {433, 976}

\bibitem[\protect\citeauthoryear{{Rubin}, {Graham}  \& {Kenney}}{{Rubin}
  et~al.}{1992}]{1992ApJ...394L...9R}
{Rubin} V.~C.,  {Graham} J.~A.,   {Kenney} J. D.~P.,  1992, \mn@doi [\apjl]
  {10.1086/186460}, \href
  {https://ui.adsabs.harvard.edu/abs/1992ApJ...394L...9R} {394, L9}

\bibitem[\protect\citeauthoryear{{Sackett}, {Morrisoni}, {Harding}  \&
  {Boroson}}{{Sackett} et~al.}{1994}]{1994Natur.370..441S}
{Sackett} P.~D.,  {Morrisoni} H.~L.,  {Harding} P.,   {Boroson} T.~A.,  1994,
  \mn@doi [\nat] {10.1038/370441a0}, \href
  {https://ui.adsabs.harvard.edu/abs/1994Natur.370..441S} {370, 441}

\bibitem[\protect\citeauthoryear{{Salo} et~al.,}{{Salo}
  et~al.}{2015}]{2015ApJS..219....4S}
{Salo} H.,  et~al., 2015, \mn@doi [\apjs] {10.1088/0067-0049/219/1/4}, \href
  {https://ui.adsabs.harvard.edu/abs/2015ApJS..219....4S} {219, 4}

\bibitem[\protect\citeauthoryear{{Sancisi}}{{Sancisi}}{1976}]{1976A&A....53..159S}
{Sancisi} R.,  1976, \aap, \href
  {https://ui.adsabs.harvard.edu/abs/1976A&A....53..159S} {53, 159}

\bibitem[\protect\citeauthoryear{{Sancisi}, {Fraternali}, {Oosterloo}  \& {van
  der Hulst}}{{Sancisi} et~al.}{2008}]{2008A&ARv..15..189S}
{Sancisi} R.,  {Fraternali} F.,  {Oosterloo} T.,   {van der Hulst} T.,  2008,
  \mn@doi [\aapr] {10.1007/s00159-008-0010-0}, \href
  {https://ui.adsabs.harvard.edu/abs/2008A&ARv..15..189S} {15, 189}

\bibitem[\protect\citeauthoryear{{Sandage} \& {Bedke}}{{Sandage} \&
  {Bedke}}{1994}]{1994cag..book.....S}
{Sandage} A.,  {Bedke} J.,  1994, {The Carnegie atlas of galaxies}.
 Vol. 638

\bibitem[\protect\citeauthoryear{{Sandin}}{{Sandin}}{2014}]{2014A&A...567A..97S}
{Sandin} C.,  2014, \mn@doi [\aap] {10.1051/0004-6361/201423429}, \href
  {https://ui.adsabs.harvard.edu/abs/2014A&A...567A..97S} {567, A97}

\bibitem[\protect\citeauthoryear{{Sandin}}{{Sandin}}{2015}]{2015A&A...577A.106S}
{Sandin} C.,  2015, \mn@doi [\aap] {10.1051/0004-6361/201425168}, \href
  {https://ui.adsabs.harvard.edu/abs/2015A&A...577A.106S} {577, A106}

\bibitem[\protect\citeauthoryear{{Savchenko}, {Sotnikova}, {Mosenkov},
  {Reshetnikov}  \& {Bizyaev}}{{Savchenko} et~al.}{2017}]{2017MNRAS.471.3261S}
{Savchenko} S.~S.,  {Sotnikova} N.~Y.,  {Mosenkov} A.~V.,  {Reshetnikov} V.~P.,
    {Bizyaev} D.~V.,  2017, \mn@doi [\mnras] {10.1093/mnras/stx1802}, \href
  {https://ui.adsabs.harvard.edu/abs/2017MNRAS.471.3261S} {471, 3261}

\bibitem[\protect\citeauthoryear{{Scannapieco}, {White}, {Springel}  \&
  {Tissera}}{{Scannapieco} et~al.}{2009}]{2009MNRAS.396..696S}
{Scannapieco} C.,  {White} S. D.~M.,  {Springel} V.,   {Tissera} P.~B.,  2009,
  \mn@doi [\mnras] {10.1111/j.1365-2966.2009.14764.x}, \href
  {https://ui.adsabs.harvard.edu/abs/2009MNRAS.396..696S} {396, 696}

\bibitem[\protect\citeauthoryear{{Schechtman-Rook} \&
  {Bershady}}{{Schechtman-Rook} \& {Bershady}}{2013}]{2013ApJ...773...45S}
{Schechtman-Rook} A.,  {Bershady} M.~A.,  2013, \mn@doi [\apj]
  {10.1088/0004-637X/773/1/45}, \href
  {https://ui.adsabs.harvard.edu/abs/2013ApJ...773...45S} {773, 45}

\bibitem[\protect\citeauthoryear{{Schechtman-Rook} \&
  {Bershady}}{{Schechtman-Rook} \& {Bershady}}{2014}]{2014ApJ...795..136S}
{Schechtman-Rook} A.,  {Bershady} M.~A.,  2014, \mn@doi [\apj]
  {10.1088/0004-637X/795/2/136}, \href
  {https://ui.adsabs.harvard.edu/abs/2014ApJ...795..136S} {795, 136}

\bibitem[\protect\citeauthoryear{{Sebastian}, {Kharb}, {O'Dea}, {Colbert}  \&
  {Baum}}{{Sebastian} et~al.}{2019}]{2019ApJ...883..189S}
{Sebastian} B.,  {Kharb} P.,  {O'Dea} C.~P.,  {Colbert} E.~J.~M.,   {Baum}
  S.~A.,  2019, \mn@doi [\apj] {10.3847/1538-4357/ab371a}, \href
  {https://ui.adsabs.harvard.edu/abs/2019ApJ...883..189S} {883, 189}

\bibitem[\protect\citeauthoryear{{Sengupta} \& {Balasubramanyam}}{{Sengupta} \&
  {Balasubramanyam}}{2006}]{2006MNRAS.369..360S}
{Sengupta} C.,  {Balasubramanyam} R.,  2006, \mn@doi [\mnras]
  {10.1111/j.1365-2966.2006.10307.x}, \href
  {https://ui.adsabs.harvard.edu/abs/2006MNRAS.369..360S} {369, 360}

\bibitem[\protect\citeauthoryear{{Seth}, {Dalcanton}  \& {de Jong}}{{Seth}
  et~al.}{2005}]{2005AJ....129.1331S}
{Seth} A.~C.,  {Dalcanton} J.~J.,   {de Jong} R.~S.,  2005, \mn@doi [\aj]
  {10.1086/427859}, \href
  {https://ui.adsabs.harvard.edu/abs/2005AJ....129.1331S} {129, 1331}

\bibitem[\protect\citeauthoryear{{Seth}, {de Jong}, {Dalcanton}  \& {GHOSTS
  Team}}{{Seth} et~al.}{2007}]{2007IAUS..241..523S}
{Seth} A.,  {de Jong} R.,  {Dalcanton} J.,   {GHOSTS Team} 2007, in {Vazdekis}
  A.,  {Peletier} R.,  eds,  IAU Symposium Vol. 241, Stellar Populations as
  Building Blocks of Galaxies. pp 523--524 (\mn@eprint {arXiv}
  {astro-ph/0701704}), \mn@doi{10.1017/S1743921307009003}

\bibitem[\protect\citeauthoryear{{Shafi}, {Oosterloo}, {Morganti},
  {Colafrancesco}  \& {Booth}}{{Shafi} et~al.}{2015}]{2015MNRAS.454.1404S}
{Shafi} N.,  {Oosterloo} T.~A.,  {Morganti} R.,  {Colafrancesco} S.,   {Booth}
  R.,  2015, \mn@doi [\mnras] {10.1093/mnras/stv2034}, \href
  {https://ui.adsabs.harvard.edu/abs/2015MNRAS.454.1404S} {454, 1404}

\bibitem[\protect\citeauthoryear{{Shang} et~al.,}{{Shang}
  et~al.}{1998}]{1998ApJ...504L..23S}
{Shang} Z.,  et~al., 1998, \mn@doi [\apjl] {10.1086/311563}, \href
  {https://ui.adsabs.harvard.edu/abs/1998ApJ...504L..23S} {504, L23}

\bibitem[\protect\citeauthoryear{{Sheth} et~al.,}{{Sheth}
  et~al.}{2010}]{2010PASP..122.1397S}
{Sheth} K.,  et~al., 2010, \mn@doi [\pasp] {10.1086/657638}, \href
  {https://ui.adsabs.harvard.edu/abs/2010PASP..122.1397S} {122, 1397}

\bibitem[\protect\citeauthoryear{{Shinn}}{{Shinn}}{2018}]{2018ApJS..239...21S}
{Shinn} J.-H.,  2018, \mn@doi [\apjs] {10.3847/1538-4365/aae3e5}, \href
  {https://ui.adsabs.harvard.edu/abs/2018ApJS..239...21S} {239, 21}

\bibitem[\protect\citeauthoryear{{Shinn} \& {Seon}}{{Shinn} \&
  {Seon}}{2015}]{2015ApJ...815..133S}
{Shinn} J.-H.,  {Seon} K.-I.,  2015, \mn@doi [\apj]
  {10.1088/0004-637X/815/2/133}, \href
  {https://ui.adsabs.harvard.edu/abs/2015ApJ...815..133S} {815, 133}

\bibitem[\protect\citeauthoryear{{Sil'chenko} \& {Afanasiev}}{{Sil'chenko} \&
  {Afanasiev}}{2012}]{2012AstBu..67..253S}
{Sil'chenko} O.~K.,  {Afanasiev} V.~L.,  2012, \mn@doi [Astrophysical Bulletin]
  {10.1134/S1990341312030029}, \href
  {https://ui.adsabs.harvard.edu/abs/2012AstBu..67..253S} {67, 253}

\bibitem[\protect\citeauthoryear{{Sil'chenko}, {Burenkov}  \&
  {Vlasyuk}}{{Sil'chenko} et~al.}{1999}]{1999AJ....117..826S}
{Sil'chenko} O.~K.,  {Burenkov} A.~N.,   {Vlasyuk} V.~V.,  1999, \mn@doi [\aj]
  {10.1086/300733}, \href
  {https://ui.adsabs.harvard.edu/abs/1999AJ....117..826S} {117, 826}

\bibitem[\protect\citeauthoryear{{Slater}, {Harding}  \& {Mihos}}{{Slater}
  et~al.}{2009}]{2009PASP..121.1267S}
{Slater} C.~T.,  {Harding} P.,   {Mihos} J.~C.,  2009, \mn@doi [\pasp]
  {10.1086/648457}, \href
  {https://ui.adsabs.harvard.edu/abs/2009PASP..121.1267S} {121, 1267}

\bibitem[\protect\citeauthoryear{{Smercina} et~al.,}{{Smercina}
  et~al.}{2019}]{2019arXiv191014672S}
{Smercina} A.,  et~al., 2019, arXiv e-prints, \href
  {https://ui.adsabs.harvard.edu/abs/2019arXiv191014672S} {p. arXiv:1910.14672}

\bibitem[\protect\citeauthoryear{{Sotnikova}, {Reshetnikov}  \&
  {Mosenkov}}{{Sotnikova} et~al.}{2012}]{2012A&AT...27..325S}
{Sotnikova} N.~Y.,  {Reshetnikov} V.~P.,   {Mosenkov} A.~V.,  2012,
  Astronomical and Astrophysical Transactions, \href
  {https://ui.adsabs.harvard.edu/abs/2012A&AT...27..325S} {27, 325}

\bibitem[\protect\citeauthoryear{{Spitzer}}{{Spitzer}}{1942}]{1942ApJ....95..329S}
{Spitzer} Lyman J.,  1942, \mn@doi [\apj] {10.1086/144407}, \href
  {https://ui.adsabs.harvard.edu/abs/1942ApJ....95..329S} {95, 329}

\bibitem[\protect\citeauthoryear{{Stein} et~al.,}{{Stein}
  et~al.}{2019}]{2019A&A...632A..13S}
{Stein} Y.,  et~al., 2019, \mn@doi [\aap] {10.1051/0004-6361/201935558}, \href
  {https://ui.adsabs.harvard.edu/abs/2019A&A...632A..13S} {632, A13}

\bibitem[\protect\citeauthoryear{{Tanaka}, {Chiba}  \& {Komiyama}}{{Tanaka}
  et~al.}{2017}]{2017ApJ...842..127T}
{Tanaka} M.,  {Chiba} M.,   {Komiyama} Y.,  2017, \mn@doi [\apj]
  {10.3847/1538-4357/aa6d11}, \href
  {https://ui.adsabs.harvard.edu/abs/2017ApJ...842..127T} {842, 127}

\bibitem[\protect\citeauthoryear{{Tempel}, {Tago}  \& {Liivam{\"a}gi}}{{Tempel}
  et~al.}{2012}]{2012A&A...540A.106T}
{Tempel} E.,  {Tago} E.,   {Liivam{\"a}gi} L.~J.,  2012, \mn@doi [\aap]
  {10.1051/0004-6361/201118687}, \href
  {https://ui.adsabs.harvard.edu/abs/2012A&A...540A.106T} {540, A106}

\bibitem[\protect\citeauthoryear{{Tikhonov} \& {Galazutdinova}}{{Tikhonov} \&
  {Galazutdinova}}{2005}]{2005Ap.....48..221T}
{Tikhonov} N.~A.,  {Galazutdinova} O.~A.,  2005, \mn@doi [Astrophysics]
  {10.1007/s10511-005-0021-8}, \href
  {https://ui.adsabs.harvard.edu/abs/2005Ap.....48..221T} {48, 221}

\bibitem[\protect\citeauthoryear{{Tissera}, {Beers}, {Carollo}  \&
  {Scannapieco}}{{Tissera} et~al.}{2014}]{2014MNRAS.439.3128T}
{Tissera} P.~B.,  {Beers} T.~C.,  {Carollo} D.,   {Scannapieco} C.,  2014,
  \mn@doi [\mnras] {10.1093/mnras/stu181}, \href
  {https://ui.adsabs.harvard.edu/abs/2014MNRAS.439.3128T} {439, 3128}

\bibitem[\protect\citeauthoryear{{Trujillo} \& {Fliri}}{{Trujillo} \&
  {Fliri}}{2016}]{2016ApJ...823..123T}
{Trujillo} I.,  {Fliri} J.,  2016, \mn@doi [\apj]
  {10.3847/0004-637X/823/2/123}, \href
  {https://ui.adsabs.harvard.edu/abs/2016ApJ...823..123T} {823, 123}

\bibitem[\protect\citeauthoryear{{Tully}}{{Tully}}{1994}]{1994yCat.7145....0T}
{Tully} R.~B.,  1994, VizieR Online Data Catalog, \href
  {https://ui.adsabs.harvard.edu/abs/1994yCat.7145....0T} {p. VII/145}

\bibitem[\protect\citeauthoryear{{Veilleux}, {Bland-Hawthorn}  \&
  {Cecil}}{{Veilleux} et~al.}{1999}]{1999AJ....118.2108V}
{Veilleux} S.,  {Bland-Hawthorn} J.,   {Cecil} G.,  1999, \mn@doi [\aj]
  {10.1086/301095}, \href
  {https://ui.adsabs.harvard.edu/abs/1999AJ....118.2108V} {118, 2108}

\bibitem[\protect\citeauthoryear{{Vollmer}, {Nehlig}  \& {Ibata}}{{Vollmer}
  et~al.}{2016}]{2016A&A...586A..98V}
{Vollmer} B.,  {Nehlig} F.,   {Ibata} R.,  2016, \mn@doi [\aap]
  {10.1051/0004-6361/201322899}, \href
  {https://ui.adsabs.harvard.edu/abs/2016A&A...586A..98V} {586, A98}

\bibitem[\protect\citeauthoryear{{Wainscoat}, {Freeman}  \&
  {Hyland}}{{Wainscoat} et~al.}{1989}]{1989ApJ...337..163W}
{Wainscoat} R.~J.,  {Freeman} K.~C.,   {Hyland} A.~R.,  1989, \mn@doi [\apj]
  {10.1086/167096}, \href
  {https://ui.adsabs.harvard.edu/abs/1989ApJ...337..163W} {337, 163}

\bibitem[\protect\citeauthoryear{{Wang}, {Chaves}  \& {Irwin}}{{Wang}
  et~al.}{2003}]{2003ApJ...598..969W}
{Wang} Q.~D.,  {Chaves} T.,   {Irwin} J.~A.,  2003, \mn@doi [\apj]
  {10.1086/379010}, \href
  {https://ui.adsabs.harvard.edu/abs/2003ApJ...598..969W} {598, 969}

\bibitem[\protect\citeauthoryear{{Werner} et~al.,}{{Werner}
  et~al.}{2004}]{2004ApJS..154....1W}
{Werner} M.~W.,  et~al., 2004, \mn@doi [\apjs] {10.1086/422992}, \href
  {https://ui.adsabs.harvard.edu/abs/2004ApJS..154....1W} {154, 1}

\bibitem[\protect\citeauthoryear{{Whitmore}, {Lucas}, {McElroy},
  {Steiman-Cameron}, {Sackett}  \& {Olling}}{{Whitmore}
  et~al.}{1990}]{1990AJ....100.1489W}
{Whitmore} B.~C.,  {Lucas} R.~A.,  {McElroy} D.~B.,  {Steiman-Cameron} T.~Y.,
  {Sackett} P.~D.,   {Olling} R.~P.,  1990, \mn@doi [\aj] {10.1086/115614},
  \href {https://ui.adsabs.harvard.edu/abs/1990AJ....100.1489W} {100, 1489}

\bibitem[\protect\citeauthoryear{{Xilouris}, {Alton}, {Davies}, {Kylafis},
  {Papamastorakis}  \& {Trewhella}}{{Xilouris}
  et~al.}{1998}]{1998A&A...331..894X}
{Xilouris} E.~M.,  {Alton} P.~B.,  {Davies} J.~I.,  {Kylafis} N.~D.,
  {Papamastorakis} J.,   {Trewhella} M.,  1998, \aap, \href
  {https://ui.adsabs.harvard.edu/abs/1998A&A...331..894X} {331, 894}

\bibitem[\protect\citeauthoryear{{Xilouris}, {Madden}, {Galliano}, {Vigroux}
  \& {Sauvage}}{{Xilouris} et~al.}{2004}]{2004A&A...416...41X}
{Xilouris} E.~M.,  {Madden} S.~C.,  {Galliano} F.,  {Vigroux} L.,   {Sauvage}
  M.,  2004, \mn@doi [\aap] {10.1051/0004-6361:20034020}, \href
  {https://ui.adsabs.harvard.edu/abs/2004A&A...416...41X} {416, 41}

\bibitem[\protect\citeauthoryear{{Yoachim} \& {Dalcanton}}{{Yoachim} \&
  {Dalcanton}}{2006}]{2006AJ....131..226Y}
{Yoachim} P.,  {Dalcanton} J.~J.,  2006, \mn@doi [\aj] {10.1086/497970}, \href
  {https://ui.adsabs.harvard.edu/abs/2006AJ....131..226Y} {131, 226}

\bibitem[\protect\citeauthoryear{{Yoshino} \& {Yamauchi}}{{Yoshino} \&
  {Yamauchi}}{2015}]{2015MNRAS.446.3749Y}
{Yoshino} A.,  {Yamauchi} C.,  2015, \mn@doi [\mnras] {10.1093/mnras/stu2249},
  \href {https://ui.adsabs.harvard.edu/abs/2015MNRAS.446.3749Y} {446, 3749}

\bibitem[\protect\citeauthoryear{{Zasov}, {Bizyaev}, {Makarov}  \&
  {Tyurina}}{{Zasov} et~al.}{2002}]{2002AstL...28..527Z}
{Zasov} A.~V.,  {Bizyaev} D.~V.,  {Makarov} D.~I.,   {Tyurina} N.~V.,  2002,
  \mn@doi [Astronomy Letters] {10.1134/1.1499176}, \href
  {https://ui.adsabs.harvard.edu/abs/2002AstL...28..527Z} {28, 527}

\bibitem[\protect\citeauthoryear{{Zolotov}, {Willman}, {Brooks}, {Governato},
  {Brook}, {Hogg}, {Quinn}  \& {Stinson}}{{Zolotov}
  et~al.}{2009}]{2009ApJ...702.1058Z}
{Zolotov} A.,  {Willman} B.,  {Brooks} A.~M.,  {Governato} F.,  {Brook} C.~B.,
  {Hogg} D.~W.,  {Quinn} T.,   {Stinson} G.,  2009, \mn@doi [\apj]
  {10.1088/0004-637X/702/2/1058}, \href
  {https://ui.adsabs.harvard.edu/abs/2009ApJ...702.1058Z} {702, 1058}

\bibitem[\protect\citeauthoryear{{Zschaechner}, {Rand}  \&
  {Walterbos}}{{Zschaechner} et~al.}{2015}]{2015ApJ...799...61Z}
{Zschaechner} L.~K.,  {Rand} R.~J.,   {Walterbos} R.,  2015, \mn@doi [\apj]
  {10.1088/0004-637X/799/1/61}, \href
  {https://ui.adsabs.harvard.edu/abs/2015ApJ...799...61Z} {799, 61}

\bibitem[\protect\citeauthoryear{{de Grijs} \& {van der Kruit}}{{de Grijs} \&
  {van der Kruit}}{1996}]{1996A&AS..117...19D}
{de Grijs} R.,  {van der Kruit} P.~C.,  1996, \aaps, \href
  {https://ui.adsabs.harvard.edu/abs/1996A&AS..117...19D} {117, 19}

\bibitem[\protect\citeauthoryear{{de Grijs}, {Peletier}  \& {van der
  Kruit}}{{de Grijs} et~al.}{1997}]{1997A&A...327..966D}
{de Grijs} R.,  {Peletier} R.~F.,   {van der Kruit} P.~C.,  1997, \aap, \href
  {https://ui.adsabs.harvard.edu/abs/1997A&A...327..966D} {327, 966}

\bibitem[\protect\citeauthoryear{{de Jong}}{{de
  Jong}}{2008}]{2008MNRAS.388.1521D}
{de Jong} R.~S.,  2008, \mn@doi [\mnras] {10.1111/j.1365-2966.2008.13505.x},
  \href {https://ui.adsabs.harvard.edu/abs/2008MNRAS.388.1521D} {388, 1521}

\bibitem[\protect\citeauthoryear{{van Dokkum} et~al.,}{{van Dokkum}
  et~al.}{2019}]{2019ApJ...883L..32V}
{van Dokkum} P.,  et~al., 2019, \mn@doi [\apjl] {10.3847/2041-8213/ab40c9},
  \href {https://ui.adsabs.harvard.edu/abs/2019ApJ...883L..32V} {883, L32}

\bibitem[\protect\citeauthoryear{{van der Kruit}}{{van der
  Kruit}}{1984}]{1984A&A...140..470V}
{van der Kruit} P.~C.,  1984, \aap, \href
  {https://ui.adsabs.harvard.edu/abs/1984A&A...140..470V} {140, 470}

\bibitem[\protect\citeauthoryear{{van der Kruit}}{{van der
  Kruit}}{1988}]{1988A&A...192..117V}
{van der Kruit} P.~C.,  1988, \aap, \href
  {https://ui.adsabs.harvard.edu/abs/1988A&A...192..117V} {192, 117}

\bibitem[\protect\citeauthoryear{{van der Kruit} \& {Freeman}}{{van der Kruit}
  \& {Freeman}}{2011}]{2011ARA&A..49..301V}
{van der Kruit} P.~C.,  {Freeman} K.~C.,  2011, \mn@doi [\araa]
  {10.1146/annurev-astro-083109-153241}, \href
  {https://ui.adsabs.harvard.edu/abs/2011ARA&A..49..301V} {49, 301}

\bibitem[\protect\citeauthoryear{{van der Kruit} \& {Searle}}{{van der Kruit}
  \& {Searle}}{1981a}]{1981A&A....95..105V}
{van der Kruit} P.~C.,  {Searle} L.,  1981a, \aap, \href
  {https://ui.adsabs.harvard.edu/abs/1981A&A....95..105V} {95, 105}

\bibitem[\protect\citeauthoryear{{van der Kruit} \& {Searle}}{{van der Kruit}
  \& {Searle}}{1981b}]{1981A&A....95..116V}
{van der Kruit} P.~C.,  {Searle} L.,  1981b, \aap, \href
  {https://ui.adsabs.harvard.edu/abs/1981A&A....95..116V} {95, 116}

\bibitem[\protect\citeauthoryear{{van der Kruit} \& {Searle}}{{van der Kruit}
  \& {Searle}}{1982}]{1982A&A...110...61V}
{van der Kruit} P.~C.,  {Searle} L.,  1982, \aap, \href
  {https://ui.adsabs.harvard.edu/abs/1982A&A...110...61V} {110, 61}

\makeatother
\end{thebibliography}

%%%%%%%%%%%%%%%%%%%%%%%%%%%%%%%%%%%%%%%%%%%%%%%%%%

%%%%%%%%%%%%%%%%%%%%%%%%%%%%%%%%%%%%%%%%%%%%%%%%%%

%%%%%%%%%%%%%%%%% APPENDICES %%%%%%%%%%%%%%%%%%%%%

\appendix

\section{{\sl HERON} images of edge-on galaxies}
\label{Appendix:figs}

\begin{figure*}
\centering
$\vcenter{\hbox{\includegraphics[width=16cm]{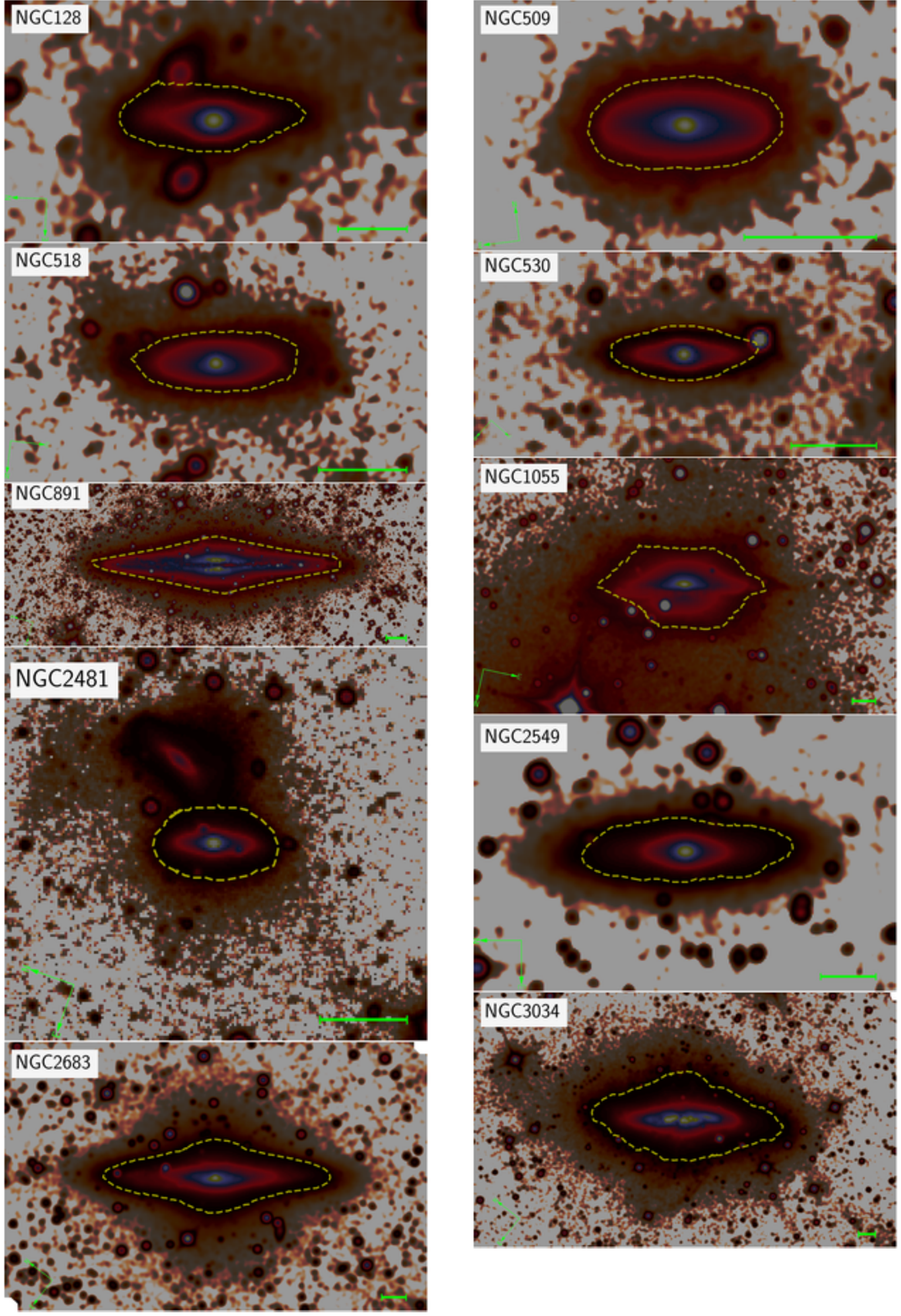}}}$
\caption{Smoothed images of the sample galaxies. A Gaussian filter with $\sigma=1$ was used ($\sigma=5$ for NGC\,2683, NGC\,3115, NGC\,4565, NGC\,4594, NGC\,4216, NGC\,5907). The green scale bar is $1\arcmin$. The yellow dashed contour represents an isophote of 24~mag/arcsec$^2$.} \label{Images}
\end{figure*}

\addtocounter{figure}{-1}
\begin{figure*}
\centering
$\vcenter{\hbox{\includegraphics[width=16cm]{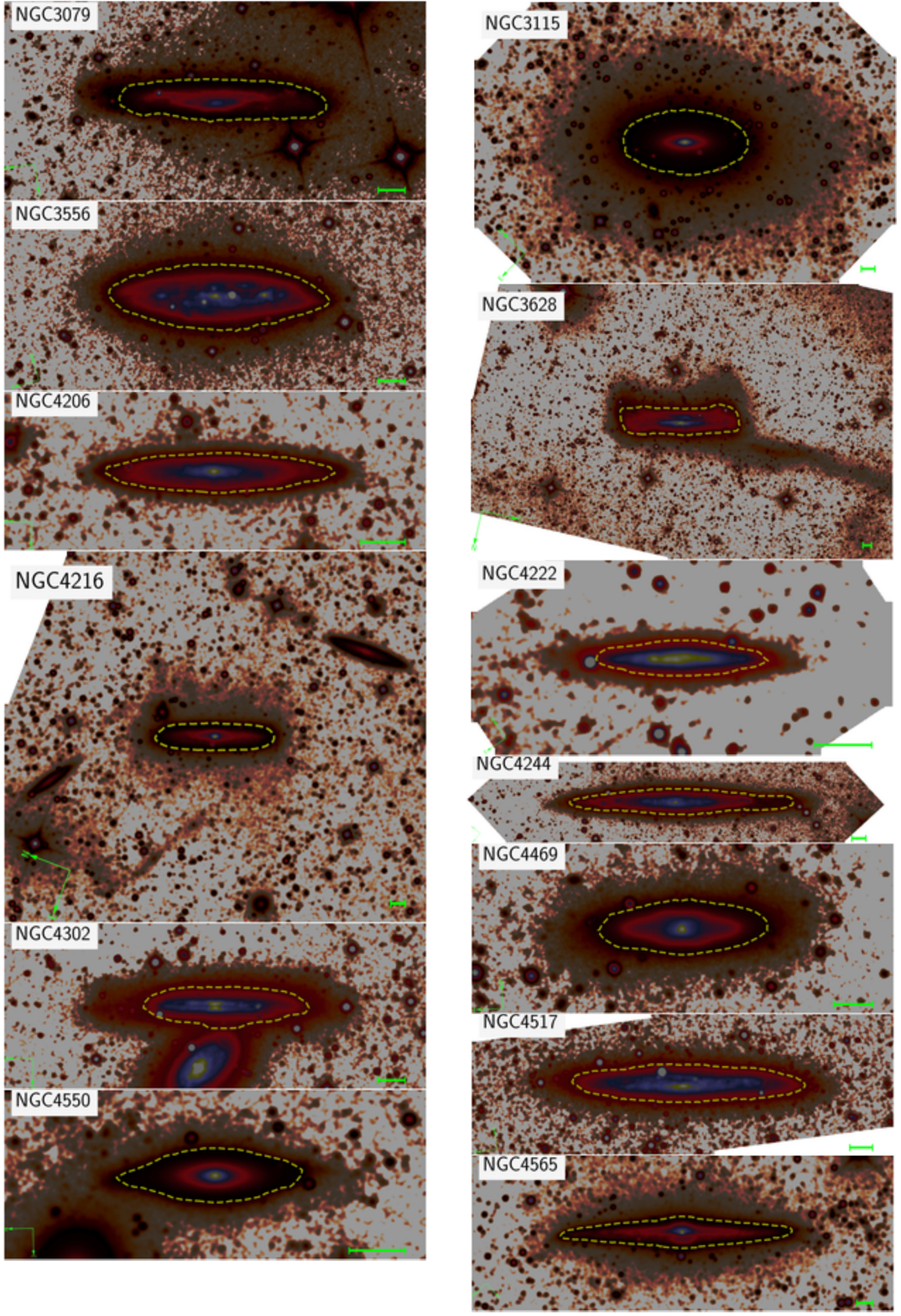}}}$

\caption{(continued)}
\end{figure*}

\addtocounter{figure}{-1}
\begin{figure*}
\centering
$\vcenter{\hbox{\includegraphics[width=16cm]{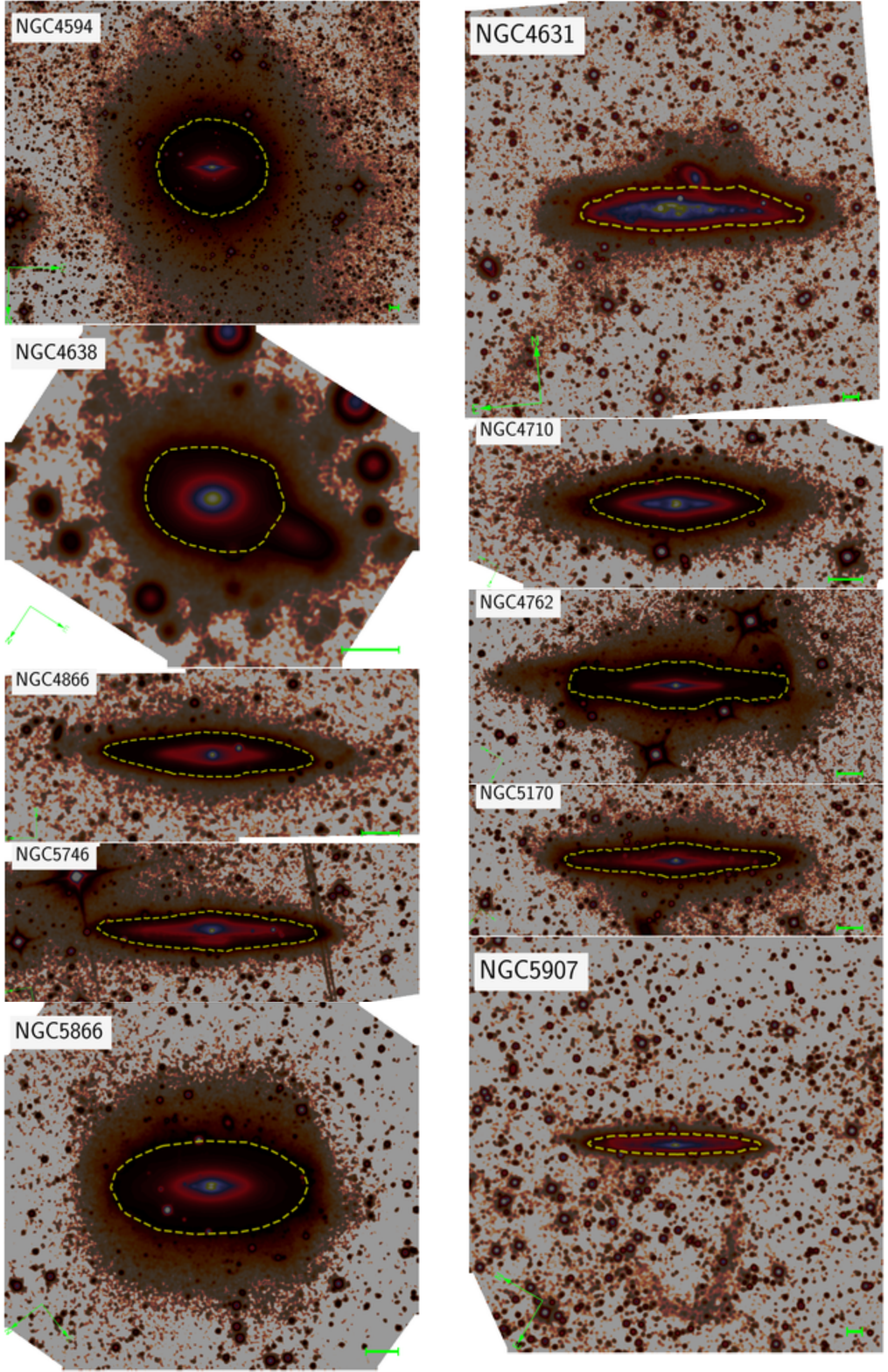}}}$

\caption{(continued)}
\end{figure*}

\addtocounter{figure}{-1}
\begin{figure*}
\centering
$\vcenter{\hbox{\includegraphics[width=8.5cm]{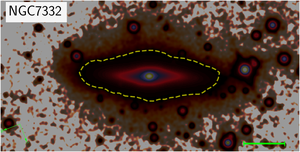}}}$
$\vcenter{\hbox{\includegraphics[width=8.5cm]{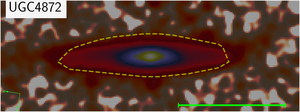}}}$
\caption{(continued)}
\end{figure*}

\section{Averaged profiles for the example galaxies}
\label{Appendix:figs}

\begin{figure*}
\label{fig:R_profiles}
\centering
\makebox[12.3cm][s]{\textbf{NGC\,891} \textbf{NGC\,4302} \textbf{NGC\,3628}} \par
$\vcenter{\hbox{\includegraphics[width=4.5cm]{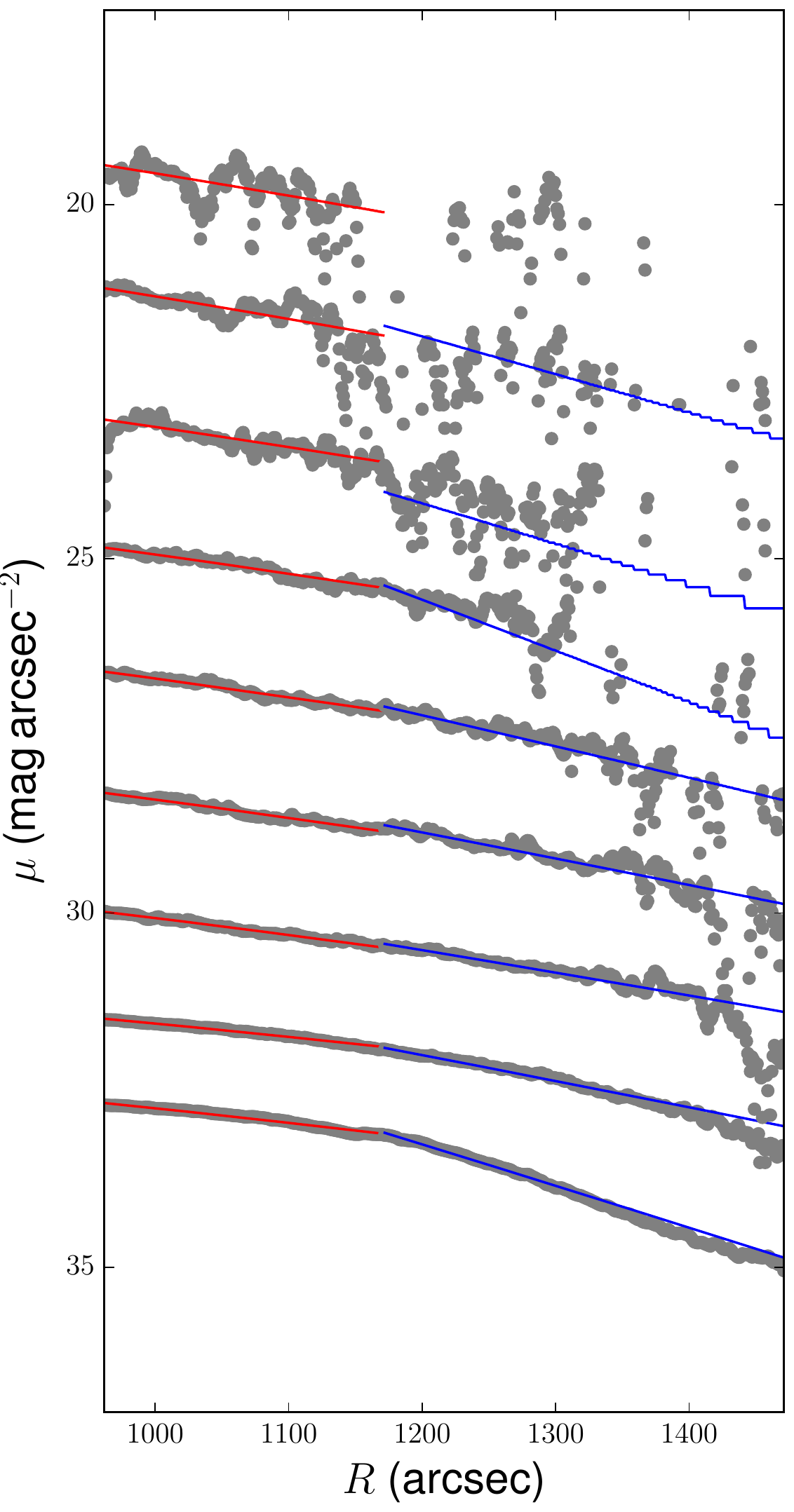}}}$
$\vcenter{\hbox{\includegraphics[width=4.5cm]{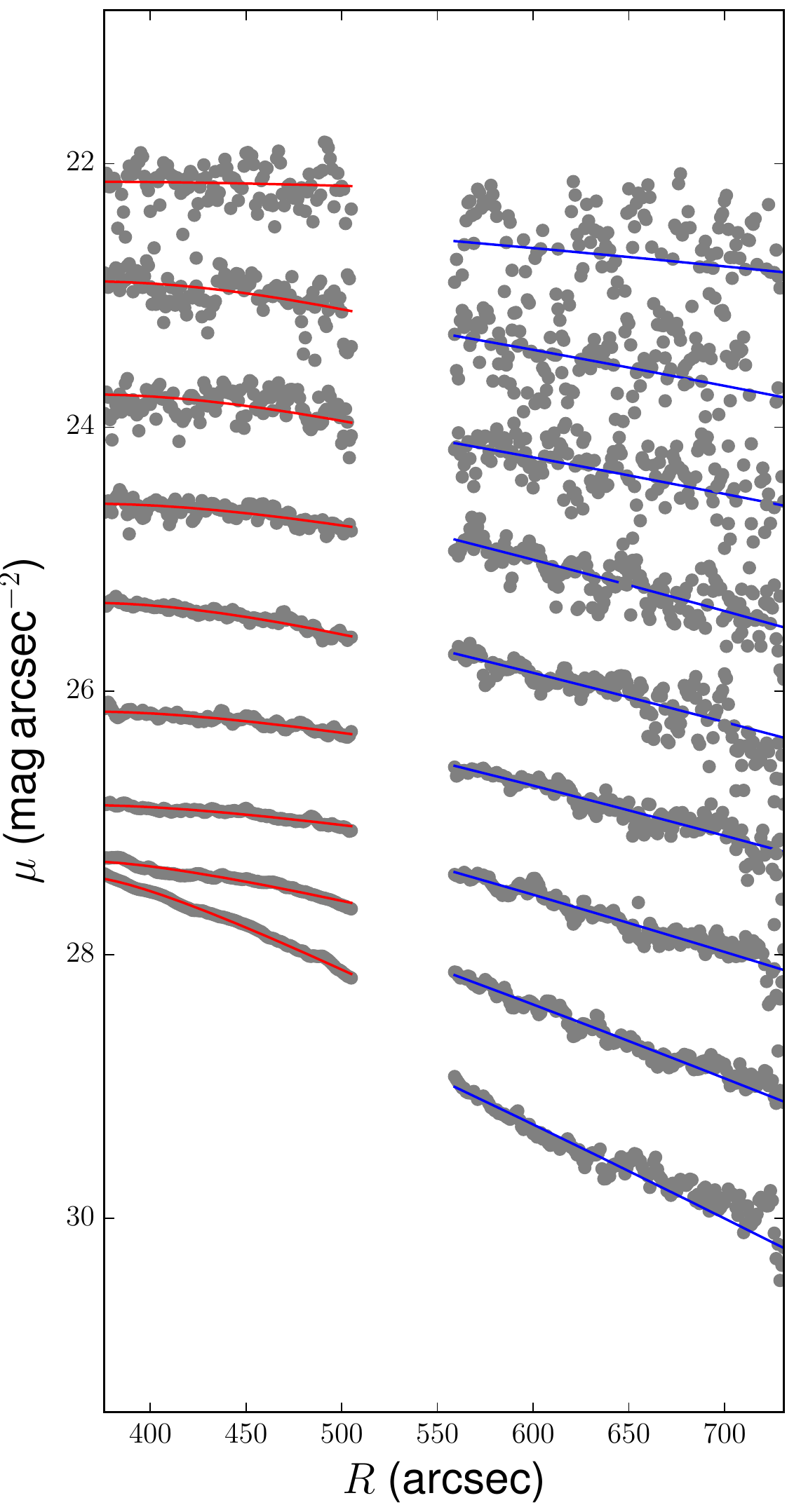}}}$
$\vcenter{\hbox{\includegraphics[width=4.5cm]{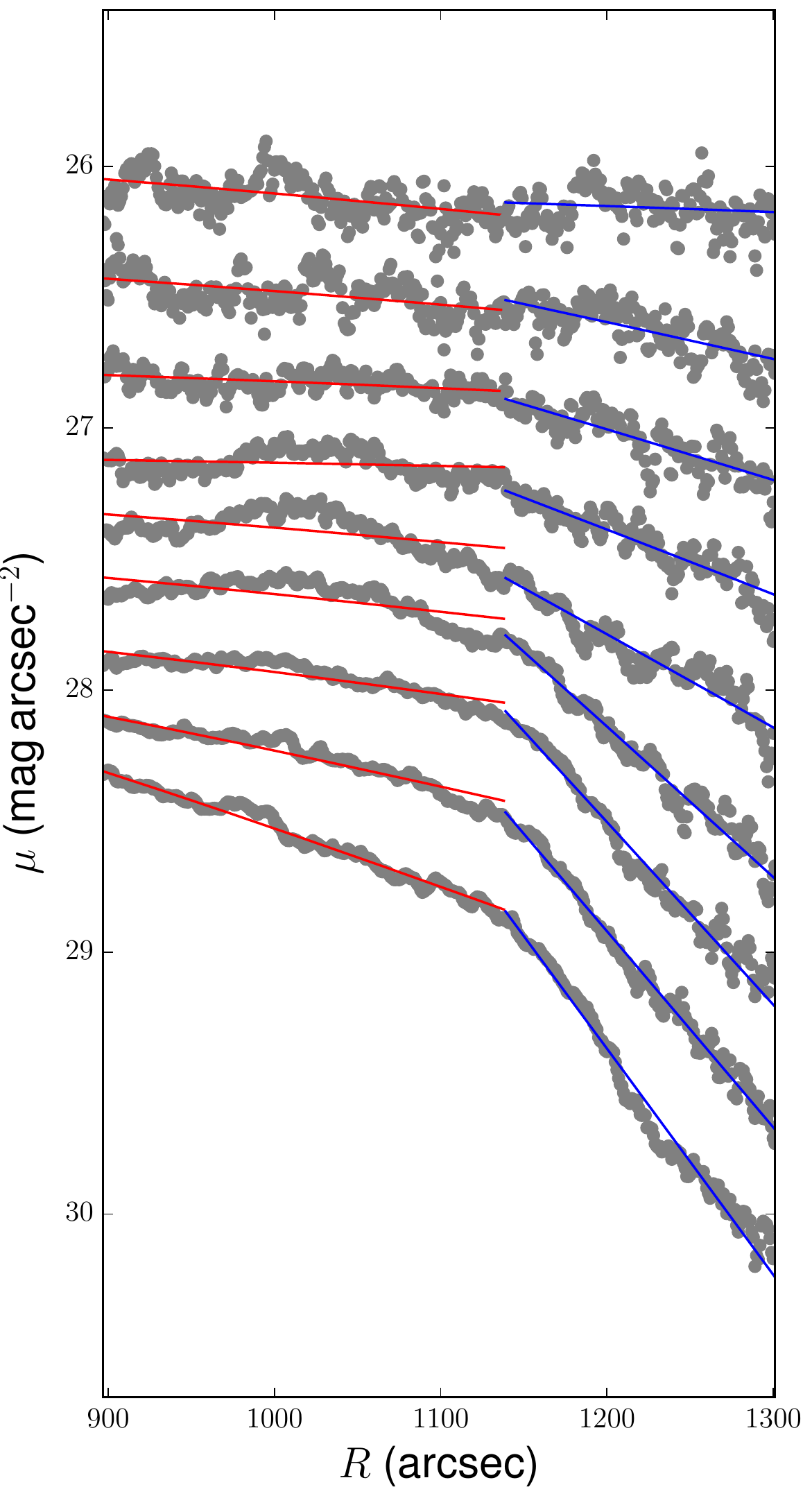}}}$
\caption{Averaged radial surface brightness profiles for NGC\,891 (left), NGC\,4302 (middle), and NGC\,3628 (right) in the 3.6~$\mu$m band, in relative units of surface brightnesses, starting from the midplane (bottom curves) up to the isophote of $2\,rms$ (top curves). The gray circles show the data, the red and blue lines depict the inner and outer parts of the discs, respectively.}
\end{figure*}

\begin{figure*}
\label{fig:Z_profiles}
\centering
\makebox[12.3cm][s]{\textbf{NGC\,891} \textbf{NGC\,4302} \textbf{NGC\,3628}} \par
$\vcenter{\hbox{\includegraphics[width=4.5cm]{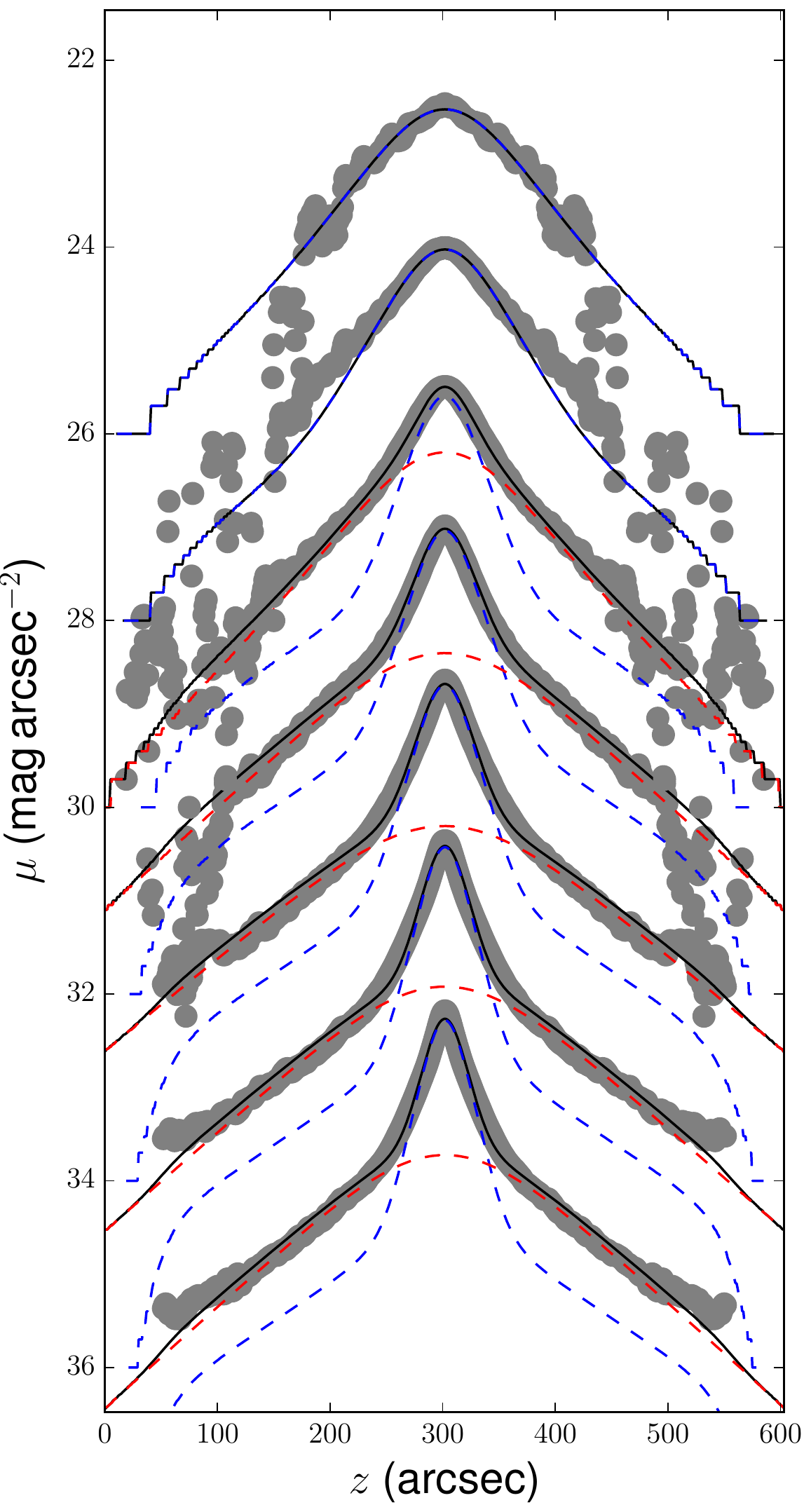}}}$
$\vcenter{\hbox{\includegraphics[width=4.5cm]{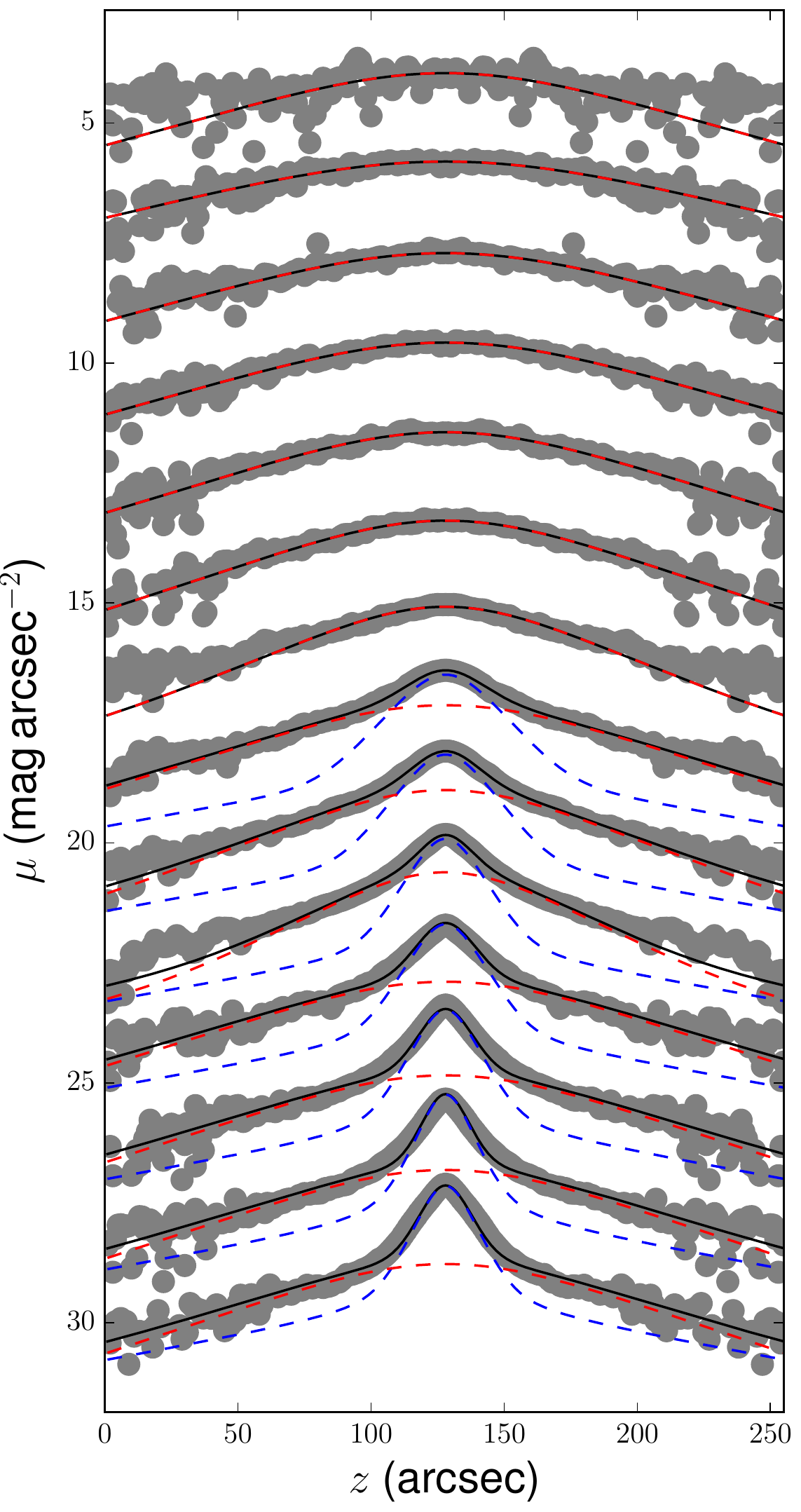}}}$
$\vcenter{\hbox{\includegraphics[width=4.5cm]{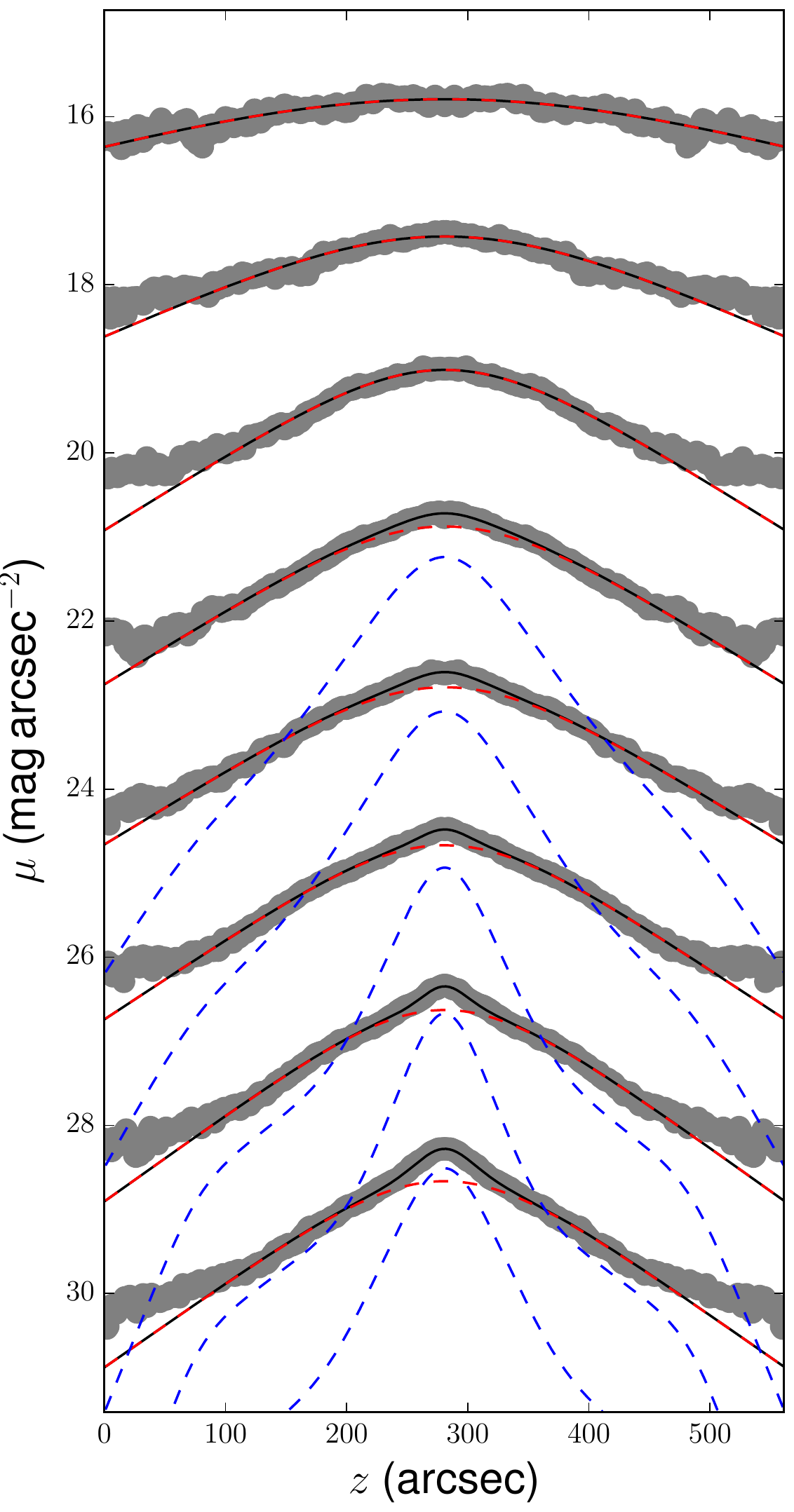}}}$
\caption{Averaged vertical surface brightness profiles for NGC\,891 (left), NGC\,4302 (middle), and NGC\,3628 (right) in the 3.6~$\mu$m band, in relative units of surface brightnesses, starting from the inner part (bottom curves) to the disc truncation (top curves). The gray circles show the data, the blue and red dashed lines depict the thin and thick discs, respectively, whereas the black solid lines show the total models.}
\end{figure*}

\section{Comparison of the fit parameters for convolved and unconvolved mock galaxy images}
\label{Appendix:models}
\begin{figure*}
\label{fig:compar_conv_unconv}
\centering
\includegraphics[width=\columnwidth]{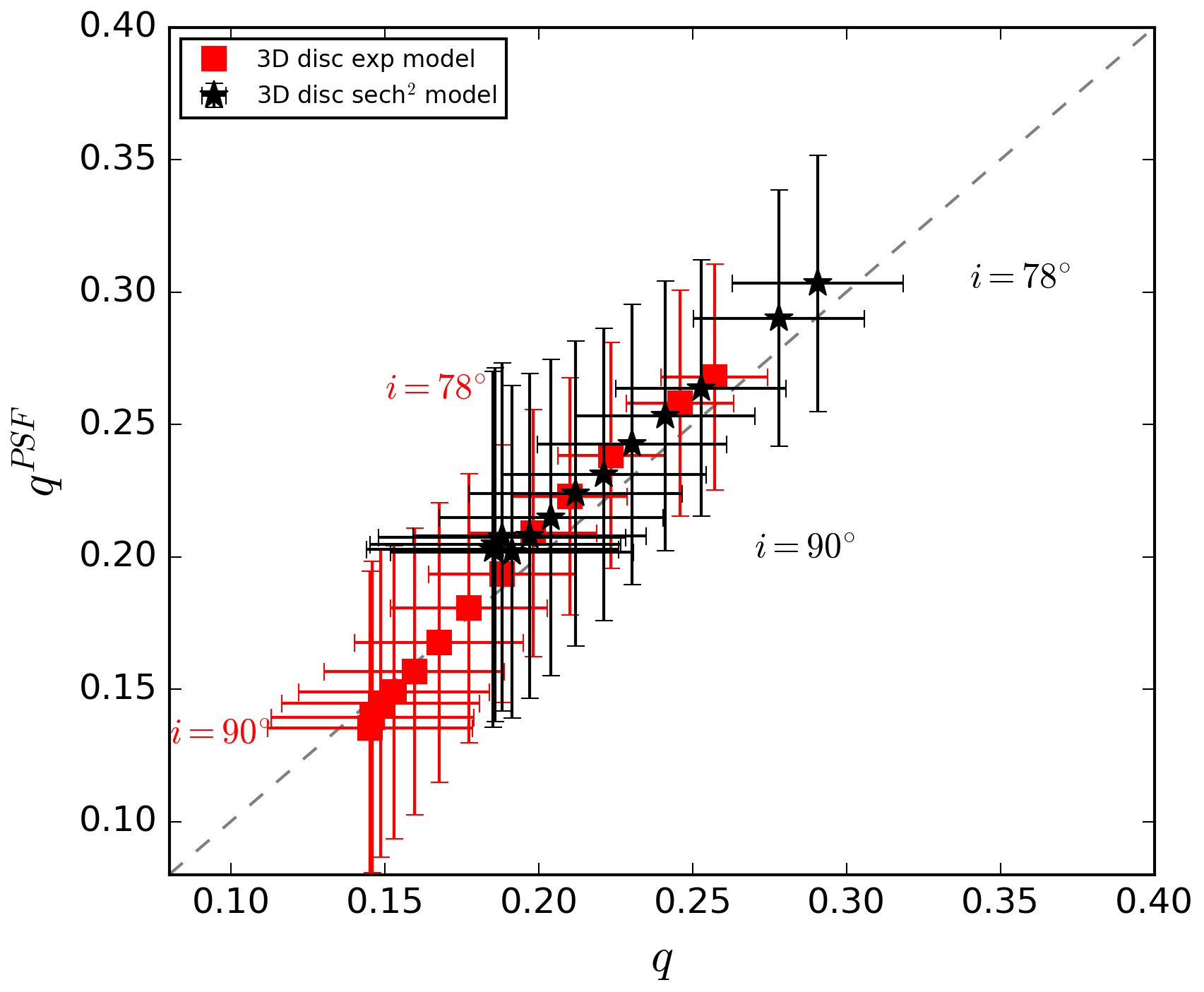}
\includegraphics[width=\columnwidth]{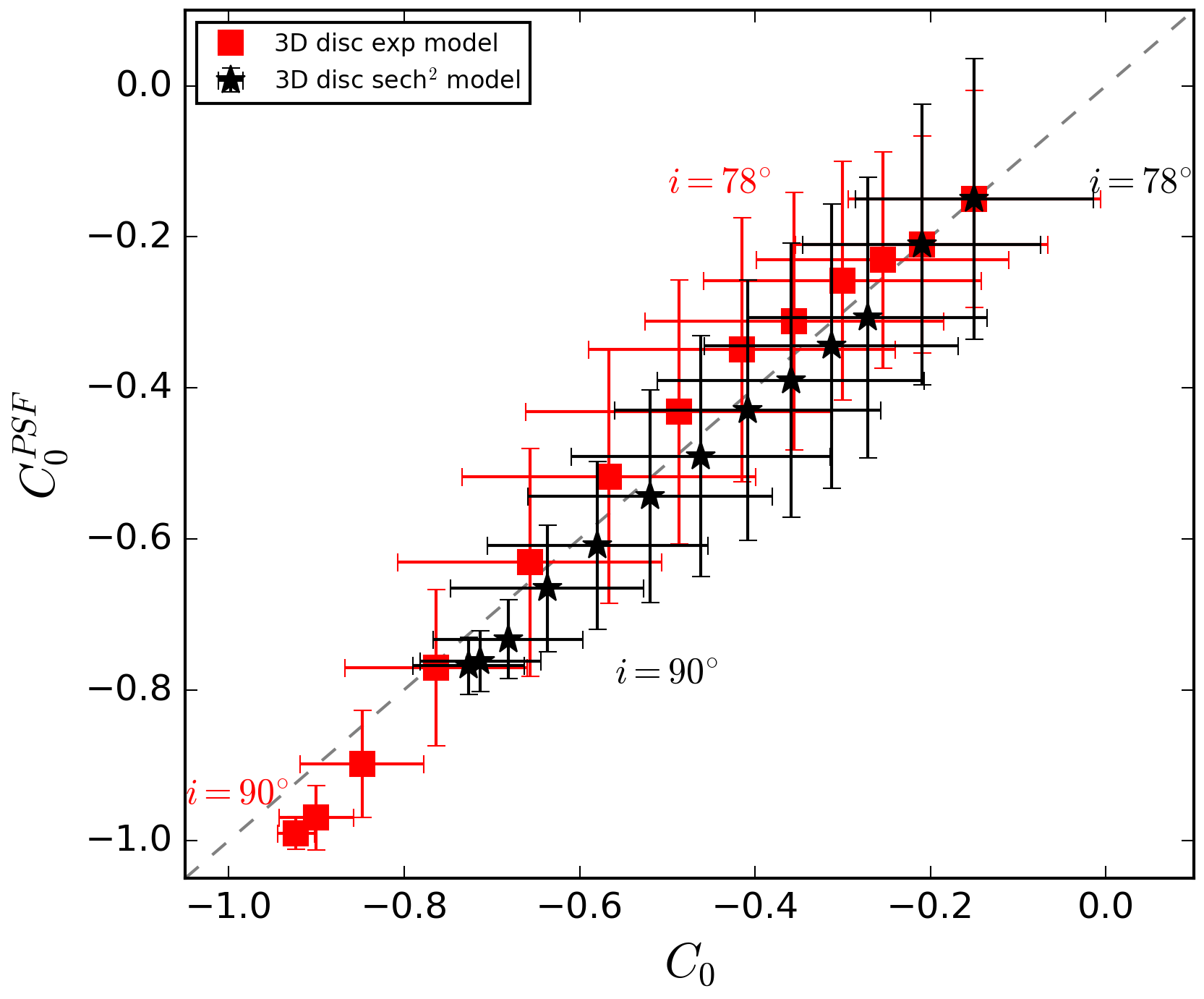}
\caption{Comparison for the two fit parameters of the outer galaxy structure, the apparent flattening $q$ (the lefthand plot) and the discyness/boxyness parameter $C_0$ (the righthand plot), for convolved ($y$-axis) and unconvolved ($x$-axis) images. The mock galaxy images consist of thin and thick discs parametrized using results from \citet{2015ApJS..219....4S} (see also Sect.~\ref{sec:shape}), but using two models of the vertical distribution in the discs: exponential (the red squares) and isothermal (the black stars). For the same mock galaxy, the parameters $C_0$ and $q$ decrease with the inclination $i$.}
\end{figure*}

%%%%%%%%%%%%%%%%%%%%%%%%%%%%%%%%%%%%%%%%%%%%%%%%%%

% Don't change these lines
\bsp	% typesetting comment
\label{lastpage}
\end{document}